\documentclass[11pt,a4paper]{article} 
\pdfoutput=1
\usepackage{graphicx}
\usepackage{color}
\usepackage{bbold}
\usepackage{style}
\usepackage{epstopdf}
\usepackage{enumerate}
\usepackage{epsfig}
\usepackage{amsmath,amssymb,bm,amsfonts,yfonts}
\newcommand{\be}{\begin{equation}}
\newcommand{\ee}{\end{equation}}
\newcommand{\bea}{\begin{eqnarray}}

\newcommand{\eea}{\end{eqnarray}}
\def\beq{\begin{equation}}
\def\eeq{\end{equation}}
\def\beeq{\begin{eqnarray}}
\def\eeeq{\end{eqnarray}}

\DeclareSymbolFont{mathscrUC}{U}{rsfs}{m}{n}  
\DeclareSymbolFont{mathscrLC}{OT1}{pzc}{m}{n} 

\makeatletter

\DeclareRobustCommand*{\mathscr}[1]{\gdef\F@ntPrefix{mathscr@char@}%
  \@EachCharacter #1\@EndEachCharacter}
\long\def\DoLongFutureLet #1#2#3#4{%
   \def\@FutureLetDecide{#1#2\@FutureLetToken
      \def\@FutureLetNext{#3}\else
      \def\@FutureLetNext{#4}\fi\@FutureLetNext}
   \futurelet\@FutureLetToken\@FutureLetDecide}
\def\DoFutureLet #1#2#3#4{\DoLongFutureLet{#1}{#2}{#3}{#4}}
\def\@EachCharacter{\DoFutureLet{\ifx}{\@EndEachCharacter}%
   {\@EachCharacterDone}{\@PickUpTheCharacter}}
\def\m@keCharacter#1{\csname\F@ntPrefix#1\endcsname}
\def\@PickUpTheCharacter#1{\m@keCharacter{#1}\@EachCharacter}
\def\@EachCharacterDone \@EndEachCharacter{}

\DeclareMathSymbol{\mathscr@char@A}{\mathord}{mathscrUC}{`A}
\DeclareMathSymbol{\mathscr@char@B}{\mathord}{mathscrUC}{`B}
\DeclareMathSymbol{\mathscr@char@C}{\mathord}{mathscrUC}{`C}
\DeclareMathSymbol{\mathscr@char@D}{\mathord}{mathscrUC}{`D}
\DeclareMathSymbol{\mathscr@char@E}{\mathord}{mathscrUC}{`E}
\DeclareMathSymbol{\mathscr@char@F}{\mathord}{mathscrUC}{`F}
\DeclareMathSymbol{\mathscr@char@G}{\mathord}{mathscrUC}{`G}
\DeclareMathSymbol{\mathscr@char@H}{\mathord}{mathscrUC}{`H}
\DeclareMathSymbol{\mathscr@char@I}{\mathord}{mathscrUC}{`I}
\DeclareMathSymbol{\mathscr@char@J}{\mathord}{mathscrUC}{`J}
\DeclareMathSymbol{\mathscr@char@K}{\mathord}{mathscrUC}{`K}
\DeclareMathSymbol{\mathscr@char@L}{\mathord}{mathscrUC}{`L}
\DeclareMathSymbol{\mathscr@char@M}{\mathord}{mathscrUC}{`M}
\DeclareMathSymbol{\mathscr@char@N}{\mathord}{mathscrUC}{`N}
\DeclareMathSymbol{\mathscr@char@O}{\mathord}{mathscrUC}{`O}
\DeclareMathSymbol{\mathscr@char@P}{\mathord}{mathscrUC}{`P}
\DeclareMathSymbol{\mathscr@char@Q}{\mathord}{mathscrUC}{`Q}
\DeclareMathSymbol{\mathscr@char@R}{\mathord}{mathscrUC}{`R}
\DeclareMathSymbol{\mathscr@char@S}{\mathord}{mathscrUC}{`S}
\DeclareMathSymbol{\mathscr@char@T}{\mathord}{mathscrUC}{`T}
\DeclareMathSymbol{\mathscr@char@U}{\mathord}{mathscrUC}{`U}
\DeclareMathSymbol{\mathscr@char@V}{\mathord}{mathscrUC}{`V}
\DeclareMathSymbol{\mathscr@char@W}{\mathord}{mathscrUC}{`W}
\DeclareMathSymbol{\mathscr@char@X}{\mathord}{mathscrUC}{`X}
\DeclareMathSymbol{\mathscr@char@Y}{\mathord}{mathscrUC}{`Y}
\DeclareMathSymbol{\mathscr@char@Z}{\mathord}{mathscrUC}{`Z}
\DeclareMathSymbol{\mathscr@char@a}{\mathord}{mathscrLC}{`a}
\DeclareMathSymbol{\mathscr@char@b}{\mathord}{mathscrLC}{`b}
\DeclareMathSymbol{\mathscr@char@c}{\mathord}{mathscrLC}{`c}
\DeclareMathSymbol{\mathscr@char@d}{\mathord}{mathscrLC}{`d}
\DeclareMathSymbol{\mathscr@char@e}{\mathord}{mathscrLC}{`e}
\DeclareMathSymbol{\mathscr@char@f}{\mathord}{mathscrLC}{`f}
\DeclareMathSymbol{\mathscr@char@g}{\mathord}{mathscrLC}{`g}
\DeclareMathSymbol{\mathscr@char@h}{\mathord}{mathscrLC}{`h}
\DeclareMathSymbol{\mathscr@char@i}{\mathord}{mathscrLC}{`i}
\DeclareMathSymbol{\mathscr@char@j}{\mathord}{mathscrLC}{`j}
\DeclareMathSymbol{\mathscr@char@k}{\mathord}{mathscrLC}{`k}
\DeclareMathSymbol{\mathscr@char@l}{\mathord}{mathscrLC}{`l}
\DeclareMathSymbol{\mathscr@char@m}{\mathord}{mathscrLC}{`m}
\DeclareMathSymbol{\mathscr@char@n}{\mathord}{mathscrLC}{`n}
\DeclareMathSymbol{\mathscr@char@o}{\mathord}{mathscrLC}{`o}
\DeclareMathSymbol{\mathscr@char@p}{\mathord}{mathscrLC}{`p}
\DeclareMathSymbol{\mathscr@char@q}{\mathord}{mathscrLC}{`q}
\DeclareMathSymbol{\mathscr@char@r}{\mathord}{mathscrLC}{`r}
\DeclareMathSymbol{\mathscr@char@s}{\mathord}{mathscrLC}{`s}
\DeclareMathSymbol{\mathscr@char@t}{\mathord}{mathscrLC}{`t}
\DeclareMathSymbol{\mathscr@char@u}{\mathord}{mathscrLC}{`u}
\DeclareMathSymbol{\mathscr@char@v}{\mathord}{mathscrLC}{`v}
\DeclareMathSymbol{\mathscr@char@w}{\mathord}{mathscrLC}{`w}
\DeclareMathSymbol{\mathscr@char@x}{\mathord}{mathscrLC}{`x}
\DeclareMathSymbol{\mathscr@char@y}{\mathord}{mathscrLC}{`y}
\DeclareMathSymbol{\mathscr@char@z}{\mathord}{mathscrLC}{`z}

\makeatother

\relax

\title{Collectivity from interference}

\author[a]{Boris Blok,}
\author[b]{Christian D. J\"akel,}
\author[c]{Mark Strikman,}
\author[d]{Urs Achim Wiedemann}

\affiliation[a]{Department of Physics, Technion - Israel Institute of Technology, Haifa, Israel}
\affiliation[b]{Department of Applied Mathematics, University of S\~ao Paulo (USP), Brazil}
\affiliation[c]{Physics Department, Penn State University, University Park, PA, USA}
\affiliation[d]{Theoretical Physics Department, CERN, CH-1211 Gen\`eve 23, Switzerland}

\emailAdd{blok@ph.technion.ac.il}
\emailAdd{jaekel@ime.usp.br}
\emailAdd{urs.wiedemann@cern.ch}
\emailAdd{mxs43@psu.edu}

\abstract{In hadronic collisions, interference between different production channels affects
momentum distributions of multi-particle final states. As this QCD interference does not
depend on the strong coupling constant $\alpha_s$, it is part of the no-interaction baseline 
that needs to be controlled prior to searching for other manifestations of collective dynamics,
\emph{e.g.}, in the analysis of azimuthal anisostropy coefficients $v_n$ at the LHC.  
Here, we introduce a model that is based on the QCD theory of multi-parton interactions and
that allows one to study interference effects in the production of $m$ 
particles in hadronic collisions with $N$ parton-parton interactions (``sources''). In an expansion in powers
of $1/(N_c^2-1)$ and to leading order in the number of sources $N$, we calculate interference effects in the
$m$-particle spectra and we determine from them the second and fourth order cumulant momentum 
anisotropies $v_n\lbrace 2\rbrace$ and $v_n\lbrace 4 \rbrace$. Without invoking any azimuthal asymmetry 
and any density dependent non-linear dynamics in the incoming state, and without invoking any interaction 
in the final state, we find that QCD interference alone can give rise to values for $v_n\lbrace 2\rbrace$ and 
$v_n\lbrace 4\rbrace$, $n$ even, that persist unattenuated for increasing number of sources, that may increase with
increasing multiplicity and that  agree with measurements in proton-proton (pp) collisions in terms of the order of 
magnitude of the signal and the approximate shape of the transverse momentum dependence. We further find 
that the non-abelian features of QCD interference can give rise to odd harmonic anisotropies. These findings
indicate that the no-interaction baseline including QCD interference effects can make a sizeable if not dominant
contribution to the measured $v_n$ coefficients in pp collisions. Prospects for analyzing QCD interference
contributions further and their possible relevance for proton-nucleus and nucleus-nucleus collisions are
discussed shortly. }

\begin{document}

\maketitle

\section{Introduction} \label{intro}

Multi-particle production in proton-proton (pp) collisions is typically modeled in terms of multiple parton-parton interactions 
without invoking explicitly density-dependent dynamics in the incoming wave 
functions or final state rescattering of the outgoing partons. In particular, multi-purpose event generators provide a 
reasonable modeling of many characteristics of the underlying event in proton-proton 
collisions~\cite{Sjostrand:2017cdm,Gieseke:2007ad,Gieseke:2017yfk,Schulz:2016vml}, but the simulation of effects that relate 
different parton-parton interactions is largely limited to ensuring consequences of global conservation laws (energy, momentum, color). The standard picture of multi-particle production in ultra-relativistic nucleus-nucleus (AA) collisions is 
radically different. Here, jet quenching provides unambiguous evidence for significant final state rescattering 
effects~\cite{Chatrchyan:2012nia,Aad:2014bxa,Abelev:2013kqa}. Rescattering is a precursor of fluid dynamics. 
Partonic systems in which rescattering is operational can be described by an effective kinetic theory that is known to 
hydrodynamize rapidly~\cite{Kurkela:2016vts}. Indeed, fluid dynamical modeling has been 
demonstrated to provide a phenomenologically valid basis for the simulation of soft 
multi-particle production in heavy ion collisions~\cite{Heinz:2013th}. 

The different dynamical pictures of multi-particle production in pp, pA and AA may be mutually compatible. 
The transverse size of the systems produced in pp collisions may be sufficiently small for rescattering
effects to be negligible, while pA and AA collisions may be sufficiently large and dense to be dominated
by multiple rescattering in the final state. However, the recent observation of heavy-ion like behavior in pp
(and pA) collisions at the LHC challenges this simple interpretation. On the one hand, the observation of 
a strong multiplicity-dependence of (multi-)strange hadron production in pp collisions~\cite{ALICE:2017jyt}
and of momentum anisotropies in pp and pA collisions ~\cite{Khachatryan:2015waa,Khachatryan:2016txc,Aaboud:2017acw,Aad:2015gqa}
seems incompatible with modeling such collisions as an essentially 
incoherent superposition of multiple partonic interactions supplemented by global constraints
(see, \emph{e.g.}, Refs.~\cite{Fischer:2016zzs,Bozek:2011if,Bzdak:2013zma,He:2015hfa} for attempts to model these phenomena).
On the other hand, the apparent absence of rescattering effects in inclusive jet and hadron production (above $p_T \sim O(1\, {\rm GeV})$) in pp 
and pA~\cite{Adam:2015hoa,ATLAS:2014cpa,Khachatryan:2016xdg} raises the question whether final state 
rescattering is sufficiently effective in the smaller collision systems to give rise to measurable signs of collectivity.  

This prompts us to ask whether physical phenomena could be at work that contribute to the recent observations of heavy-ion like 
behavior in pp collisions without invoking final state rescattering or density dependent dynamics in the incoming 
state.  Our focus will be on QCD interference effects, as these do not depend on the coupling constant $\alpha_s$ or 
on interaction probability while they are known to affect multi-particle distributions in the final state. We are mainly interested in 
understanding their contribution to the anisotropy coefficients
$v_m\lbrace 2n\rbrace$ that are measured in pp, pPb and PbPb collisions from connected $(2n)$-particle correlation functions 
via the so-called cumulant technique~\cite{Borghini:2000sa,Borghini:2001vi,Bilandzic:2010jr}.
QCD interference is known to lead
to momentum anisotropies in the 2-particle cumulant $v_2\lbrace 2\rbrace$ (for a rederivation, see section~\ref{sec3} below). 
However, also the measured higher order cumulants $v_m\lbrace 2n\rbrace$ show the same $p_T$-, rapidity- and multiplicity-dependent sizeable values (that are typically $~20\%$ smaller than $v_m\lbrace 2\rbrace$). This is commonly refered to as a signature of collectivity~\cite{Khachatryan:2015waa,Khachatryan:2016txc}, since it is consistent with a correlation amongst 
all particles in the event. The technical question that 
we shall address with explicit calculations in this manuscript is whether QCD interference can give rise to non-vanishing 
higher order cumulants $v_m\lbrace 2n\rbrace$ and how these are expected to scale with system size. 

Under the assumption that hadronic wave functions at ultra-relativistic energies carry saturated gluon distributions, 
multi-particle correlations have been calculated in the so-called CGC-formalism~\cite{Altinoluk:2015uaa,Altinoluk:2015eka,Lappi:2015vta,Dumitru:2014yza,Kovner:2016jfp,Dumitru:2010iy,Levin:2011fb,Kovner:2010xk,Kovner:2011pe,Kovchegov:2012nd,Dumitru:2014dra,Dusling:2012iga,Gotsman:2016fee,
Gotsman:2016owk,McLerran:2015sva}.
This formalism combines effects that are non-negotiably at work (\emph{i.e.}, QCD interference) with effects of a saturated gluon distribution that are searched for as signatures of QCD in a novel high-density regime. 
In contrast, we work in a simplified model that treats QCD interference exactly but that
does not invoke parton saturation effects. This may ultimately help to disentangle both classes of effects.  
 In section~\ref{sec2}, we define a QCD-inspired model of multi-particle production that is sufficiently simple to 
allow for explicit calculations of higher order cumulants. In sections~\ref{sec3} and
\ref{sec4}, we calculate $v_2$ from 2-, 4- and 6-particle cumulants, before summarizing our preliminary analysis of 
odd harmonics in section~\ref{sec5}. Section~\ref{sec6} discusses how the model defined in section~\ref{sec2} is
related to the theory of multi-parton interactions. This allows us to estimate the value of the only model parameter,
which we use in section~\ref{sec7} to obtain some numerical results. We conclude by summarizing the main
conclusions as well as important open questions.

\section{A model of multi-particle production}
\label{sec2}
\subsection{Defining the model}
\label{sec2a}
\begin{figure}[t]
\centering
\includegraphics[width=0.9\textwidth]{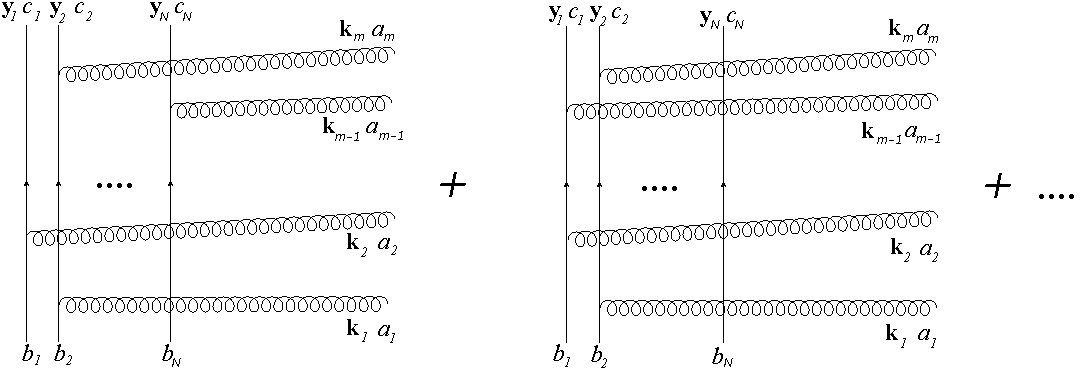} 
\caption{Two of the $N^m$ diagrammatic contribution to the $m$-gluon emission amplitude from $N$ sources. 
The $N$ partonic line sources in adjoint representation start with initial colors $b_j$ and, after emitting gluons,
end at final color $c_j$. Emitted gluons are regarded as being ordered in rapidity, see text for details. }
\label{fig1}
\end{figure}

The model for multi-particle production introduced here views a hadronic collision as an event consisting of $N$ parton-parton interactions occurring at positions ${\bf y}_i$, $i \in \left[1,N\right]$, in
the transverse plane. To each of the transverse positions  ${\bf y}_i$, the model associates a partonic line source which may be thought of as starting with initial color $b_i$ at the rapidity of the first 
colliding hadron, emitting gluons in the intermediate rapidity window and ending at the rapidity of the second hadron with final color $c_i$. 
Each multi-particle production amplitude is therefore of the type given in Fig.~\ref{fig1}. 
The model can be summarized as follows:  
\begin{enumerate}
\item 
Each hadron collision is characterized by a set $\lbrace {\bf y}_i, b_i\rbrace$, $i \in \left[1, N\right]$, of $N$
particle emitting sources distributed at transverse positions ${\bf y}_i$ with initial colors $b_i$ in the adjoint representation.
%
%
 %
\item Gluon emission from a source at position ${\bf y}_j$ and color $b_j$ is described by an eikonal vertex, 
\begin{equation}
\includegraphics[width=0.1\textwidth]{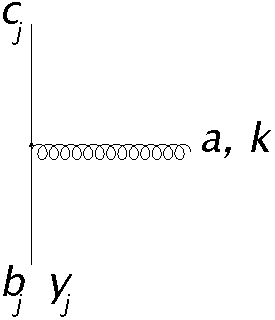} \quad
	= T_{b_j c_j}^a\,  \int d{\bf x}\, \vec{f}({\bf x}-{\bf y})\, e^{i\,{\bf k}.{\bf x}}
	\equiv T_{b_j c_j}^a\,  \vec{f}({\bf k})\, \exp\left[i\, {\bf y}.{\bf k}  \right]\, ,
	 \label{eq2.2}
\end{equation}
where the integration variable ${\bf x}$ is two-dimensional transverse, and the color structure of the vertex is defined by the adjoint
generators $T^a$ of $SU(N_c)$. The vertex function $\vec{f}$ is a two-dimensional vector in the transverse plane, that in cross sections
will appear dotted into another vertex function. For instance, for gluons in the non-abelian Coulomb field of an incoming source,  one may
write $\vec{f}({\bf k}) \propto g {\bf k} / {\bf k}^2$. In the following calculations, however, we do not assume a specific
functional shape of $\vec{f}({\bf k})$. The vector $\vec{f}({\bf k})$  parametrizes then the ${\bf k}$-dependent microscopic dynamics that gives 
rise to gluon emission. 
\item
When calculating cross sections of event samples,  the initial data are weighted with a {\it classical} probability distribution $\rho\left( \lbrace {\bf y}_i \rbrace \right)$. 
Denoting coordinates in the complex conjugate amplitude with primes, this means that initial data 
$\lbrace {\bf y}_i, b_i\rbrace$ and $\lbrace {\bf y'}_i, {b'}_i\rbrace$ are averaged with the weight
$\rho\left( \lbrace {\bf y}_i\rbrace \right)$ $\delta^{(2)}\left({\bf y}_i-{\bf y'}_i\right)\, \delta_{b_i,{b'}_i}$.
Also, final colors are summed over with the constraint $\delta_{c_i,{c'}_i}$.
\end{enumerate}
%

According to the model defined above, the spectrum for emission of $m$ particles of transverse momenta ${\bf k}_1$, ..., ${\bf k}_m$ from $N$ 
sources takes the form
\beq
\frac{d\Sigma}{d{\bf k}_1...d{\bf k}_m}=  \int \left(\prod_{i=1}^N d{\bf y}_i\right)   \rho(\lbrace y_i\rbrace )\,     
{\hat \sigma}\left(\lbrace {\bf k}_j\rbrace,\lbrace y_i\rbrace\right)\, .
\label{eq2.3}
\eeq
Here, ${\hat \sigma}(\lbrace {\bf k}_j\rbrace,\lbrace {\bf y_i}\rbrace)$  denotes the spectrum for the production of $m$ particles
of momentum $\lbrace {\bf k}_j\rbrace$, $j \in \left[1,m\right]$ from $N$ sources at specific positions $\lbrace {\bf y_i}\rbrace)$, 
$i \in \left[1,N\right]$. Phenomenologically relevant values for $m$ and $N$ may be fixed by noting that 
high-multiplicity proton-proton collisions at the LHC can contain $m \sim O(100)$  particles, and events of this multiplicity are
modeled in Monte Carlo event generators typically with $N \sim O(10)$ parton-parton interactions. However, the main focus of the present 
work is not on this phenomenologically relevant parameter range but on the qualitative question of 
whether QCD interference can give rise to momentum anisotropies $v_n$ that persist in higher order cumulants. 
To this end, the main aim of this manuscript is to calculate ${\hat \sigma}\left(\lbrace {\bf k}_j\rbrace,\lbrace y_i\rbrace\right)$  for arbitrary values of 
$m$ and~$N$, and to analyze in particular the limit of large $N$ in which possible asymmetries due to fluctuations in the number of sources are absent. We do this
with the following simplifications:
\begin{enumerate}
	\item {\it Neglecting longitudinal phase factors}\\
	Only transverse momenta and transverse coordinates are considered explicitly in the model. The rationale for this simplification is the following:\\
	One could supplement the model with longitudinal phase factors in the definition of the vertex function (\ref{eq2.2})
	by replacing $\vec{f}({\bf x}-{\bf y})\, e^{i\, {\bf k}.{\bf x}} \longrightarrow$ $\vec{f}(x- y)\, e^{i\, {\bf k}.{\bf x}} e^{i\,  k^+ x^- +i\,  k^- x^+ }  $, 
	where the indices $\pm$ denote components of light-cone coordinates and momenta. For high collision energy, however, when both 
	the emitting sources and the emitted gluons
	propagate close to the light cone, one has $k^- \approx 0$. This 
	implies that  $ e^{i\,  k^- x^+ } \approx 1$. \\
	If the remaining phase  $e^{i\,  k^+ x^- }$ were included in 
	the following calculations of gluon production cross sections, it would result in an additional multiplicative factor $e^{i\, k^+(y_i-y_j)^- }$ in those terms in which a 
	factor $e^{i\, {\bf k}.({\bf y}_i-{\bf y}_j)}$ occurs.
	Here, $y_i$, $y_j$ denote generic positions of sources from which the gluon of momentum $k$ is emitted in the amplitude and absorbed in the complex conjugate amplitude, respectively.
	However, identifying the particle emitting source with an energetic parton of light cone momentum fraction $p_i^+$, it follows from the uncertainty relation that 
	$y_i^- \sim 1/p_i^+$. For soft emitted gluons ($k^+ \ll p_i^+$), this phase is hence negligible, too,  
	$k^+(y_i-y_j)^- \sim \textstyle\frac{k^+}{ p_i^+} - \textstyle\frac{k^+}{ p_j^+} \ll 1$. \\
	We therefore conclude that longitudinal dynamics can be neglected when discussing phase interference. 
	After a first illustrative calculation, we shall explain at the end of section~\ref{sec3a} why gluons are correlated in transverse momentum even if they are separated by
	a significant rapidity interval. 
	\item {\it Emitted gluons do not cross.}\\
         In several simpler examples, it was demonstrated explicitly that contributions to the multi-gluon cross section of maximal power in $\ln (1/x)$ arise 
         in light cone gauge from ladder diagrams in which emissions are strongly ordered in rapidity, see e.g.~\cite{Salam:1999ft}. It is not known how these arguments extend 
         to the more complex problem of radiation of many soft gluons from multiple sources discussed here. However,  
	our aim is to devise a model that retains relevant features of QCD but that is simple enough to allow for the explicit calculation of soft multi-gluon interference for large
	$m$ and $N$. Motivated by the above-mentioned results for multi-gluon cross sections in simpler systems and by the need for computational simplicity, we therefore
	assume that multi-gluon radiation is dominated by ladder-type diagrams in which gluon lines do not cross, and we think of the emitted gluons as ordered in rapidity.  
	\item {\it $m$-particle emission cross sections will be symmetrized amongst the $m$ emittees.}\\
	We shall find that interference contributions to multi-particle emission cross sections are not always 
	symmetric under interchange between final state momenta ${\bf k}_i$. This is so, since the color constraints on gluon emission of the first ($i=1,2 ...$) and last ($i = ..,m-1,m$) gluons
	in the emission amplitude are different from those in between, see appendices ~\ref{appa} and ~\ref{appb} for technical details. As these differences are small and unimportant for
	our discussion, but since they lead to much longer expressions for higher order cumulants, we shall often randomize final results by averaging over all
         permutations $s$ of the $m$ outgoing momenta,
\begin{equation}
	\frac{d\Sigma}{d{\bf k}_1...d{\bf k}_m} \longrightarrow \frac{1}{m!} \sum_{s} \frac{d\Sigma}{d{\bf k}_{s(1)}...d{\bf k}_{s(m)}}\, .
	\label{eq2.4}
\end{equation}
	\item {\it No modelling of hadronization}\\
	Throughout this work, we calculate {\it partonic} spectra and momentum correlations. If hadronization would satisfy local parton-hadron duality (LPHD), then our result could
	be compared to measured hadron spectra and correlations. However, the simple LPHD prescription may not be phenomenologically viable for multi-particle correlations at soft transverse
	momentum. We regard it as a limitation of this work that we do not address uncertainties arising from the hadronization stage. We emphasize that our main focus is on  
	addressing the {\it qualitative} question of whether QCD interference can give rise to momentum
	anisotropies $v_n$ that persist in higher order cumulants. Since any valid hadronization prescription conserves momentum flow in azimuth, the qualitative answer to
	this question should not depend on details of the hadronization model, and realistic hadronization models are expected to preserve the order of magnitude of the azimuthal
	asymmetries found on the partonic level. However, hadronic particle spectra and correlations are generally softened and smeared compared to their partonic parents. This is
	in particular a caveat for the interpretation of the transverse momentum dependencies of azimuthal anisotropy coefficient $v_n$ discussed in section~\ref{sec7}. 
\end{enumerate}

%
\subsection{Azimuthal multi-particle correlations}
\label{sec2b}
To define two-particle correlation functions, we average the $m$-particle emission spectrum in (\ref{eq2.3})
with a phase factor $e^{in(\phi_1 - \phi_2)}$,
\begin{equation}
	T_n(k_1,k_2) = \binom{m}{2} \int_\rho \int_0^{2\pi} d\phi_1\, d\phi_2\, \exp\left[in(\phi_1 - \phi_2) \right]
	\left(\int \prod_{b=3}^m k_b\, dk_b\, d\phi_b \right)\, {\hat \sigma}\, ,
	\label{eq2.5}
\end{equation}
where $k_i$, $\phi_i$ denote the radial and azimuthal components of the two-dimensional transverse momenta ${\bf k}_i$. 
We also construct the corresponding norm 
\begin{equation}
	\overline{T}(k_1,k_2) =  \binom{m}{2}  \int_\rho \int_0^{2\pi} d\phi_1\, d\phi_2\, 
	\left(\int \prod_{b=3}^m k_b\, dk_b\, d\phi_b \right)\, {\hat \sigma}\, ,
	\label{eq2.6}
\end{equation}
where the Binomial coefficient $\binom{m}{2}$ counts the number of particle pairs in an event.
The integration $ \int_\rho .. \equiv  \int \left(\prod_{i=1}^N d{\bf y}_i\right)   \rho(\lbrace y_i\rbrace ) .. $ amounts to 
an average for a specific event sample that is defined by the source distribution $\rho$. 
The  angular two-particle correlation function 
$ \langle\langle e^{in(\phi_1 - \phi_2)} \rangle\rangle$ is then defined as~\cite{Borghini:2001vi,Bilandzic:2010jr}
 \begin{equation}
  \langle\langle e^{in(\phi_1 - \phi_2)} \rangle\rangle(k_1,k_2) \equiv  \frac{T_n(k_1,k_2)}{\overline{T}(k_1,k_2)}\, .
  \label{eq2.7}
 \end{equation}
  The experimentally measured (second-order cumulant) anisotropy coefficients $ v_n^2\lbrace{2}\rbrace$ can be identified
with (\ref{eq2.7}), 
\begin{equation}
v_n^2\lbrace{2}\rbrace(k_1,k_2) \equiv \langle\langle e^{in(\phi_1 - \phi_2)} \rangle\rangle(k_1,k_2)\, .
\label{eq2.8}
\end{equation}
We follow experimental practice by defining
\begin{equation}
	v_n\lbrace{2}\rbrace(k) \equiv \sqrt{ \langle\langle e^{in(\phi_1 - \phi_2)} \rangle\rangle(k,k) }\, .
	\label{eq2.9}
\end{equation}
However, (\ref{eq2.8}) cannot be expected to factorize, and in general $\langle\langle e^{in(\phi_1 - \phi_2)} \rangle\rangle(k_1,k_2)
\not= v_n\lbrace{2}\rbrace(k_1)\, v_n\lbrace{2}\rbrace(k_2)$ (see sections~\ref{sec4a} and~\ref{sec7}). 

Correlation functions for more than 2 particles can be defined analogously~\cite{Borghini:2001vi,Bilandzic:2010jr}. 
In particular, we shall calculate the
normalized azimuthal 4-particle correlation functions  $\langle\langle e^{in(\phi_1 +\phi_2 - \phi_3 - \phi_4)} \rangle\rangle$ from
\begin{equation}
	S_n(k_1,k_2,k_3,k_4) =  \binom{m}{4} \int_\rho \int_0^{2\pi} d\phi_1\, d\phi_2\, d\phi_3\, d\phi_4\, 
	e^{in(\phi_1 + \phi_2 - \phi_3 - \phi_4)}
	\left(\int \prod_{b=5}^m k_b\, dk_b\, d\phi_b \right)\, {\hat \sigma}\, ,
	\label{eq2.10}
\end{equation}
and from the corresponding normalization $\overline{S}$ obtained by evaluating (\ref{eq2.10}) without phase factors. 
The fourth order cumulants are then defined in the standard way, 
\begin{align}
	 \langle\langle e^{in(\phi_1+\phi_2 - \phi_3 - \phi_4)} \rangle\rangle_c  & = 
	 \langle\langle e^{in(\phi_1 +\phi_2 - \phi_3 - \phi_4)} \rangle\rangle
	 \nonumber \\
	& 	\qquad  
	 -  \langle\langle e^{in(\phi_1 - \phi_3)} \rangle\rangle  \langle\langle e^{in(\phi_2 - \phi_4)} \rangle\rangle
	 -  \langle\langle e^{in(\phi_1 - \phi_4)} \rangle\rangle  \langle\langle e^{in(\phi_2 - \phi_3)} \rangle\rangle\, ,
	 \label{eq2.11}
\end{align}
which defines the fourth order cumulant anisotropy coefficient 
\begin{equation}
v_n\lbrace{4}\rbrace \equiv  \sqrt[4]{-\langle\langle e^{in(\phi_1+\phi_2 - \phi_3 - \phi_4)} \rangle\rangle_c} \, .
\label{eq2.12}
\end{equation}
%

\section{The dipole interference term}
\label{sec3}
\subsection{Explicit calculation of a simple example: $N=2$, $m=2$}
\label{sec3a}
\begin{figure}[t]
\centering
\includegraphics[width=0.9\textwidth]{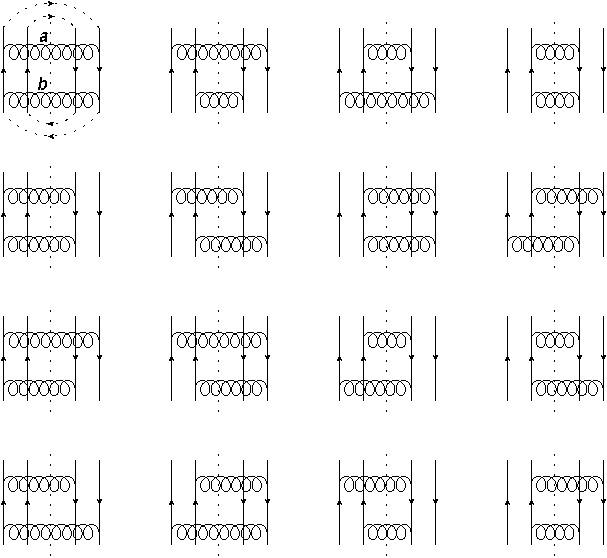} 
\caption{The 16 diagrammatic contributions to the $2$-gluon emission spectrum from $2$ sources. In each of the 16 diagrams, contributions 
from the amplitude (complex conjugate amplitude) are on the left (right) hand side of the dotted line. Averaging (summing) over the initial (final) 
state closes the color flow as illustrated by the dashed arrows in the top left hand diagram. This color flow implies that the eight diagrams in the 
third and fourth row vanish. 
}
\label{fig2}
\end{figure}

To illustrate the calculation of the spectrum ${\hat \sigma}$ in (\ref{eq2.3}), we discuss now the case of emitting $m=2$ gluons from $N=2$ sources. Fig.~\ref{fig2} shows the 16 diagrammatic contributions.
The first row of Fig.~\ref{fig2} shows {\it diagonal} contributions in which all gluons are emitted from the same source in the amplitude and in the complex conjugate amplitude. These contributions are free of interference effects. Denoting by $a$ and $b$ the colors of the two emitted gluons,  we find for the top left and top right diagrams the color factor (we work in the adjoint representation)
\begin{equation}
	{\rm Tr}\left[T^{a} T^{b} T^{b} T^{a}  \right] {\rm Tr}\left[  \mathbb{1} \right] = N_c^2\, \left( N_c^2 -1\right)^2\, . 
	\label{eq3.1}
\end{equation}
The second and third diagrams on the top row of Fig.~\ref{fig2} have the color factor ${\rm Tr}\left[T^{a} T^{a}\right]$ 
${\rm Tr}\left[ T^{b} T^{b}  \right] = N_c^2\, \left( N_c^2 -1\right)^2$, 
so that all diagrams on the top row of Fig.~\ref{fig2} have the same color factor (\ref{eq3.1}). 

The second row of Fig.~\ref{fig2} shows the four diagrammatic contributions for which both emitted gluons are {\it off-diagonal}, \emph{i.e.}, they are emitted 
from one source in the amplitude and they are absorbed by the other source in the complex conjugate amplitude. The corresponding color factor reads
\begin{equation}
	{\rm Tr}\left[T^{a} T^{b}  \right] {\rm Tr}\left[  T^{b} T^{a}  \right] = N_c^2\, \left( N_c^2 -1\right)\, .
	\label{eq3.2}
\end{equation}
Compared to the contribution (\ref{eq3.1}) from diagonal gluon exchanges, they are $O\left(1/(N_c^2 - 1)\right)$-suppressed.

The third and fourth row of Fig.~\ref{fig2} shows contributions with one diagonal and one off-diagonal gluon exchange. In contrast to a QED emission, these vanish in QCD since the color trace of one of the two source lines is $\propto {\rm Tr}\left[  T^{a}  \right]$ or $\propto {\rm Tr}\left[  T^{b}  \right]$. 

The emission vertices (\ref{eq2.2}) carry positive (negative) phases $\propto e^{i {\bf k}.{\bf y}_i}$
($\propto e^{-i {\bf k}.{\bf y}_i}$)  in the amplitude (complex conjugate amplitude). The squared amplitude can then be written easily, 
\bea
&&{\hat \sigma}\left(\lbrace {\bf k_1}, {\bf k}_2\rbrace,\lbrace {\bf y}_1, {\bf y}_2\rbrace\right) \propto  N_c^2\, \left( N_c^2 -1\right)^2  
\left|\vec{f}({\bf k}_1)\right|^2\, \left|\vec{f}({\bf k}_2)\right|^2   
 \nonumber \\
 && \qquad   \qquad \times  
  \Big\lbrace 4 + \frac{1}{(N_c^2-1)} \left(  
     e^{i({\bf k}_1+{\bf k}_2).({\bf y}_1-{\bf y}_2)} + e^{i({\bf k}_1-{\bf k}_2).({\bf y}_1-{\bf y}_2)}  \right.
   \nonumber \\
&& \qquad   \qquad \qquad   \qquad \qquad   \qquad
     \left. + e^{i({\bf k}_1+{\bf k}_2).({\bf y}_2-{\bf y}_1)} + e^{i({\bf k}_1-{\bf k}_2).({\bf y}_2-{\bf y}_1)} \right)   \Big\rbrace\, .
  \label{eq3.3}
\eea
Here, the leading factor 4 counts the four diagrams in the first row of Fig.~\ref{fig2}, for which all gluon emissions are
diagonal and thus all phases cancel. The four $O\left( 1/(N_c^2-1) \right)$-suppressed phases in (\ref{eq3.3}) correspond to the four diagrams in the second row of Fig.~\ref{fig2} (They are written in the same order in which they arise in the figure.) 
Here and in the following, we do not specify the normalization since it drops out of the correlation functions 
that we are interested in.

The two-gluon emission spectrum (\ref{eq2.3}) can then be calculated for any given probability distribution 
$\rho\left(\lbrace {\bf y}_i \rbrace \right)$ of sources. In particular, for a Gaussian ansatz
\begin{equation}
	\rho\left({\bf y}_1,{\bf y}_2 \right) = \frac{1}{(2\pi B)^2} \exp\left[ - \frac{{\bf y}_1^2}{2B} -\frac{{\bf y}_2^2}{2B} 
	\right]\, ,
	\label{eq3.4}
\end{equation}
that may be regarded as characterizing a collision at vanishing impact parameter for which the localization of
sources does not have a statistically preferred azimuthal orientation, 
one finds after averaging over the relative distance $\Delta {\bf y} \equiv {\bf y}_1-{\bf y}_2$,
\beq
\frac{d\Sigma}{d{\bf k}_1d{\bf k}_2} \propto  \left|\vec{f}({\bf k}_1)\right|^2\, \left|\vec{f}({\bf k}_2)\right|^2  
  \left[ 1 + \frac{\left( e^{-B({\bf k}_1+{\bf k}_2)^2} + e^{-B({\bf k}_1-{\bf k}_2)^2} \right)}{\left(N_c^2 -1\right)}   \right]
 .\label{eq3.5}
\eeq
This is consistent with results obtained in ~\cite{Lappi:2015vta,Altinoluk:2015uaa,Altinoluk:2015eka}. 
The simple model defined in section~\ref{sec2a} thus shares important commonalities with other approaches.\footnote{In eq.~(37) of 
Ref.~\cite{Lappi:2015vta} and in eq.~(18) of Ref.~\cite{Altinoluk:2015uaa}, the two-gluon spectrum was calculated from
so-called glasma graphs. These calculations used a Gaussian average similar to (\ref{eq3.4}),
and they took a formal limit $2\pi B  e^{-B({\bf k})^2/2} \to (2\pi)^2 \delta^{(2)}({\bf k})$ while associating the factor $1/B$ with the inverse of a
large but finite transverse surface. Expression (\ref{eq3.5}) is consistent with these limiting cases and it matches the results given 
explicitly in Ref.~\cite{Altinoluk:2015eka}. For a more thorough related derivation of the two-gluon spectrum, see also Ref.~\cite{Kovchegov:2012nd}. }

  Eq.~(\ref{eq3.5}) describes QCD dipole radiation. As the average distance 
  $\langle \Delta {\bf y} \rangle\propto \sqrt{B}$ between the legs of the dipole increases, interference effects decrease
  and the second term in (\ref{eq3.5}) becomes less important.  It is a characteristic feature of the two-gluon emission spectrum (\ref{eq3.5})
  that interference effects enhance the emission equally  for gluon pairs that are close
  in momentum space, ${\bf k}_1 \approx {\bf k}_2$ and for those that are recoiling against each other, ${\bf k}_1 \approx - {\bf k}_2$, 
  \cite{Altinoluk:2015eka}. Also, it has been noted repeatedly that spectra like (\ref{eq3.5}) are symmetric with respect to ${\bf k}_i \to -{\bf k}_i$,  so that they cannot give rise to odd harmonics~\cite{Lappi:2015vta}. The area $B$ may be interpreted in terms of the
inverse saturation scale $1/Q_s^2$ of a saturated parton density~\cite{Lappi:2015vta,Altinoluk:2015uaa,Altinoluk:2015eka}.
As we discuss in section~\ref{sec6}, the ansatz (\ref{eq3.4}) and a parameter range for $B$ can also be motivated within the theory of multi-parton interactions.

We comment at this point on the physical interpretation of the two-gluon correlation in (\ref{eq3.5}). This correlation arises from QCD interference of different production 
amplitudes but it should not be regarded as being the consequence of interference {\it between} the two gluons. Against the latter interpretation speaks the finding 
that an enhancement due to QCD interference is observed not only when the gluons sit close together in transverse momentum space (the term $e^{-B({\bf k}_1-{\bf k}_2)^2} $
in (\ref{eq3.5})), but also when they are recoiling against each other (the term $e^{-B({\bf k}_1+{\bf k}_2)^2} $ in (\ref{eq3.5}) ). Rather, the interference pattern in (\ref{eq3.5})
is consistent with the picture that an azimuthal asymmetry in gluon emission arises for each gluon individually from the interference of the production amplitudes in which
the gluon is linked to the first and the second source, respectively. What correlates the orientation of the two emitted gluons in azimuth is not their mutual interference but 
the fact that both are emitted from the {\it same} source pair. Since emission from a source dipole is symmetric with respect to the plane 
orthogonal to the dipole orientation, each gluon has the same propensity for ending up on the left or right hand side of that plane, and the probability of both gluons ending up in the
same hemisphere (the term $e^{-B({\bf k}_1-{\bf k}_2)^2} $ in (\ref{eq3.5})) or in opposite ones (the term $e^{-B({\bf k}_1-{\bf k}_2)^2} $ in (\ref{eq3.5})) is therefore equal. 

The above argument has noteworthy consequences beyond the simple example discussed in this subsection. First, 
for arbitrary gluon multiplicity $m$ and arbitrary number of sources $N$, the diagrams with exactly 2 off-diagonal and  $m-2$ diagonal gluons are of particular interest since they 
determine the full $O(1/(N_c^2-1))$ contribution to leading order in $N$ (see next subsection). It follows from the color traces involved that these diagrams are only non-vanishing
if both off-diagonal gluons connect to the same source pair. As a consequence, the above line of argument carries over to this more general case that we discuss
in the next subsection. Second, the above discussion shows explicitly that gluons need not be close to each other in transverse momentum space to be correlated, since they are 
correlated via common sources. The analogous argument applies to the rapidity dependence. As long as the two sources used for the calculation (\ref{eq3.5}) are eikonal and therefore
emit gluons from the same transverse positions in different rapidity windows, the two gluons will be correlated in transverse momentum due to the dipole orientation of their common 
source pair and irrespective of their rapidity difference. This line of argument extends to all emission patterns studied in the present paper. We therefore
expect that the omission of explicit rapidity dependencies in our model calculation does not change our conclusions qualitatively.
  
\subsection{Dipole interference term for arbitrary $m>2$ gluons from $N>2$ sources}
\label{sec3b}
The calculation of the full interference pattern for emission of a large number $m$ of 
gluons from a large number $N$ of sources is difficult. Here, we consider first the simpler problem 
of calculating only the dipole interference
terms that include $m-2$ diagonal and 2 off-diagonal gluons. This is 
the leading $O\left(1/(N_c^2-1)\right)$-correction to the emission spectrum
 \begin{eqnarray}
         &&{\hat \sigma}
  	\propto (N_c^2 -1)^N N_c^m \left(\prod_{i=1}^m \left|\vec{f}({\bf k}_i)\right|^2 \right)\,
  		\nonumber \\
  		&& \qquad \times \Big\lbrace N^m + F^{(2)}_{\rm corr}(N,m)
  	\frac{N^{m-2}}{(N_c^2-1)} \sum_{(ab)} \sum_{(ij)} 4 \, 
  	{\cos\left({\bf k}_a.\Delta {\bf y}_{ij} \right)}  { \cos\left( {\bf k}_b.\Delta {\bf y}_{ij}   \right)} 
  	\nonumber \\
  	&& \qquad \qquad + O\left(\frac{1}{N}\frac{1}{(N_c^2-1)}\right) 
  	+ O\left(\frac{1}{(N_c^2-1)^2}\right)   \Big\rbrace	\, .
  	\label{eq3.6}
 \end{eqnarray}
There are $N^m$ possibilities of emitting
incoherently $m$ diagonal gluons from $N$ sources. The color trace of an incoherent gluon emission
gives one factor $N_c$ for each of the $m$ diagonal gluons and one factor  $(N_c^2-1)$ for 
each of the $N$ sources.  This explains the prefactor  $(N_c^2 -1)^N N_c^m\, N^m$ of the leading
term in (\ref{eq3.6}). 

For the subleading term in (\ref{eq3.6}), the factor $N^{m-2}$ accounts for the number of choices of
connecting $m-2$ diagonal gluons to $N$ sources. The  sum $\sum_{(ij)}$ goes over the $N(N-1)/2$ pairs of 
sources; the second term in (\ref{eq3.6}) is therefore of the same order  $O\left( N^m\right)$ as the
first one. The dipole interference of two off-diagonal gluons suppresses this term by a 
factor $1/(N_c^2-1)$ compared to the leading one, as explained in section~\ref{sec3a}.  

\begin{figure}[t]
\centering
\includegraphics[width=0.5\textwidth]{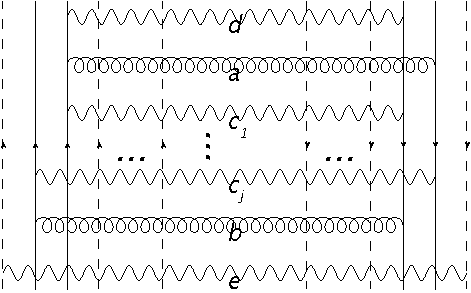} 
\caption{One specific contribution to the second term of the emission cross section (\ref{eq3.6}) in which 
two off-diagonal gluons (curly lines) are supplemented by some diagonal gluons (zagged lines). The
color trace associated to this particular contribution is ${\rm Tr}\left[T^eT^e \right]\, 
{\rm Tr}\left[ T^{c_j} T^b T^{c_j} T^a\right]
{\rm Tr}\left[T^{c_1} T^a T^d T^d T^{c_1} T^b\right] {\rm Tr}\left[\mathbb{1} \right]^2$. As argued in the text, 
each diagonal gluon that is sandwiched between the two off-diagonal ones and that links to the same sources
as the off-diagonal ones (here, the gluons $c_1$ and $c_j$) results in a reduction of the color factor by 
$1/2$. Therefore, diagonal gluons are not superimposed incoherently
to the dipole interference pattern. 
}
\label{fig3}
\end{figure}
%
\subsubsection{The color correction factor $F^{(2)}_{\rm corr}(N,m)$} 
\label{sec3b1}
To understand the factor $F^{(2)}_{\rm corr}(N,m)$ in (\ref{eq3.6}), consider the color trace for a gluon emission
diagram with one pair of off-diagonal gluons (colors $a$, $b$ in Fig.~\ref{fig3})
and with an arbitrary number of diagonal gluons. The following can be checked to be generally true: If a 
diagonal gluon is not connected to a source to which an off-diagonal gluon connects
(color $e$ in Fig.~\ref{fig3}), or if it is connected to such a source but is not sandwiched between the
two off-diagonal gluons (color $d$ in Fig.~\ref{fig3}), then the generators associated to this gluon
emission stand always next to each other in some color trace and they simplify thus according to the color identity
\begin{equation}
	T^dT^d = N_c\, \mathbb{1}\, .
	\label{eq3.7}
\end{equation}
In contrast, for those diagonal gluons that are sandwiched between the off-diagonal ones
 and that are connected to the same sources as the off-diagonal ones
(colors  $c_1$, $c_j$ in Fig.~\ref{fig3}), 
the generators in one of the color traces can always be brought into a form where they 
sandwich one of the generators of an off-diagonal gluon, 
\begin{equation}
	T^{c_j} T^a T^{c_j} = \frac{1}{2} N_c\, T^a\, .
	\label{eq3.8}
\end{equation}

For an ordered list of $m$ gluons with one off-diagonal pair, there are $(m-1-j)$ possibilities of sandwiching
$j = 0, \ldots , m-2$ diagonal gluons between the two off-diagonal ones. For each configuration with $j$ sandwiched
diagonal gluons, there are $\binom{j}{l} 2^{l} (N-2)^{j-l}$ possibilities of
linking $l$ of the sandwiched diagonal gluons to the sources to which the off-diagonal gluons are connected.
Each such contribution is then suppressed by a correction factor $1/2^l$. 

If one would ignore this correction factor $1/2^l$, one would assume that all $m-2$ diagonal gluons
are incoherently superimposed to the interference pattern of the two off-diagonal gluons.
The number of such incoherent superpositions is
\begin{equation}
  {\cal N}_{\rm incoh} =  \sum_{j=0}^{m-2} N^{m-2-j}  (m-1-j) \left( \sum_{l=0}^j 
 	 \binom{j}{l} 2^{l} (N-2)^{j-l}  \right) = \frac{m(m-1)}{2}\, N^{m-2}\, .
 	\label{eq3.9}
\end{equation}
Taking into account that the sum $\sum_{(ab)}$ in (\ref{eq3.6}) goes over $m(m-1)/2$ choices of 
selecting a pair of off-diagonal gluons from the ordered list, the factor $ {\cal N}_{\rm incoh}$ accounts 
for the $N$- and $m$-dependence of the prefactor of the $O\left( 1/(N_c^2-1)\right)$-suppressed term in (\ref{eq3.6})
if $F^{(2)}_{\rm corr}(N,m)$ equals unity.

However, each of the $l$ diagonal gluons that are sandwiched between the off-diagonal ones
comes with an extra factor 1/2 that corrects $ {\cal N}_{\rm incoh}$, and therefore
 \begin{eqnarray}
 	F^{(2)}_{\rm corr}(N,m) &=& \frac{1}{{\cal N}_{\rm incoh}}
 	\sum_{j=0}^{m-2} N^{m-2-j}  (m-1-j) \left( \sum_{l=0}^j \binom{j}{l}
 	 2^{l} (N-2)^{j-l} \frac{1}{2^{l}} \right) 
 	\nonumber \\
 	&=& \frac{2}{m(m-1)} N^{1-m} \left(N(N-1)^m + m N^m -N^{1+m}\right)\, .
 	\label{eq3.10}
 \end{eqnarray}
 One finds $F^{(2)}_{\rm corr}(N=2,m=2) = 1$, but in general, $F^{(2)}_{\rm corr}(N,m)  \leq 1$. For instance,
 $F^{(2)}_{\rm corr}(3,3) = 8/9$ and $F^{(2)}_{\rm corr}(4,4) = 27/32$ are consistent with cases explicitly calculated in
 the appendices ~\ref{appa} and ~\ref{appb}. We note the following limiting cases:
 \begin{enumerate} 
 \item {\it The limit $m = {\rm const.}$, $N\to \infty$.} \\
 Increasing the number of sources at fixed multiplicity $m$ favors incoherent particle production and hence
 \begin{equation}
 	\lim_{N\to\infty} F^{(2)}_{\rm corr}(N,m)\Big\vert_{m={\rm const.}} = 1\, . 
 	\label{eq3.11}
 \end{equation}
 The same limiting value is reached for other color correction factors $F^{(*)}_{\rm corr}(N,m)$ that we encounter
 in the next section.
 \item {\it The limit $m \to \infty$ for fixed average multiplicity per source $\overline{m} = m/N$.}\\
 This limit is consistent with analyses of LHC pp data which indicate  that the multiplicity of hard processes is proportional to the soft multiplicity~\cite{Azarkin:2014cja}.
 The color correction factors $F^{(*)}_{\rm corr}(N,m)$ are generally finite in this limit. In particular,
  \begin{equation}
  	\lim_{m\to\infty} F^{(2)}_{\rm corr}(m/\overline{m},m) = \frac{2\overline{m}+2 e^{-\overline{m}} - 2}{\overline{m}^2}\, .
  	\label{eq3.12}
  \end{equation}
 For $\overline{m} = 1$, the correction factor (\ref{eq3.12}) is $2/e \approx 0.73$,
 but for $\overline{m} = 3$, it is 0.46 and for $\overline{m} = 5$, it equals 0.32. So, higher event multiplicity per 
 source leads to decorrelation that reduces the interference term in (\ref{eq3.6}). 
 \item {\it The high-multiplicity limit $m\to \infty$ for fixed $N$.}\\
 For fixed number of sources, the color correction factor behaves asymptotically like
 \begin{equation}
 	F^{(2)}_{\rm corr}(N,m) \Big\vert_{N = {\rm const.}} \sim \frac{2\, N}{m} + O\left( \frac{N^2}{m^2}\right)\, .
 	\label{eq3.13}
 \end{equation}
 Therefore, increasing multiplicity for a fixed number of sources leads to decorrelation.
  \end{enumerate}

 %
 \subsubsection{The second order cumulant $v_2^2\lbrace{2}\rbrace$ to leading $O(1/N_c^2-1)$.}
 \label{sec3b2}
  From the emission cross section (\ref{eq3.6}) to order $O(1/N_c^2-1)$, one obtains  
 from (\ref{eq2.8}) the anisotropy coefficient 
\begin{eqnarray}
	v_2^2\lbrace{2}\rbrace(k_1,k_2)
	&\equiv& \langle\langle e^{i2(\phi_1 - \phi_2)} \rangle\rangle(k_1,k_2)
	\nonumber\\
	&\equiv& 
	\frac{F^{(2)}_{\rm corr}(N,m)  \int_\rho
	 \frac{1}{N^2}\sum_{(ij)} 2^2 J_2\left(k_1 \Delta y_{ij}\right)\, J_2\left(k_2 \Delta y_{ij}\right)}{
	(N_c^2 -1) + F^{(2)}_{\rm corr}(N,m)  \int_\rho \frac{1}{N^2}\sum_{(ij)} 2^2 J_0\left(k_1 \Delta y_{ij}\right)\, J_0\left(k_2 \Delta y_{ij}\right)
	}
	\nonumber\\
	&=& F^{(2)}_{\rm corr}(N,m)
	 \frac{2^2 }{ (N_c^2 -1)}  \int_\rho \frac{1}{N^2}\sum_{(ij)} J_2\left(k_1 \Delta y_{ij}\right)\, J_2\left(k_2 \Delta y_{ij}\right)
	 \nonumber \\
	&&  \quad+ O\left(\frac{1}{ (N_c^2 -1)^2}  \right) \, .
	\label{eq3.14}
\end{eqnarray}
Here, the Bessel functions $J_2$ arise from the $\phi$-integration $\int_0^{2\pi} d{\phi}_a\, e^{i2\phi_a}
\cos {\bf k}_a.({\bf y}_i-{\bf y}_j)$ in (\ref{eq2.6}) and we use $\Delta y_{ij} \equiv |{\bf y}_i - {\bf y}_j|$. The
notational shorthand $ \int_\rho$ stands for the averaging over source distributions as defined in (\ref{eq2.3}),
$ \int_\rho ... \equiv \int \left(\prod_{i=1}^N d{\bf y}_i\right)   \rho(\lbrace y_i\rbrace) ... $.
The sum $\sum_{(ab)}$ over the $m(m-1)/2$ possible pairs of off-diagonal gluon momenta drops out
in calculating the average $\langle\langle e^{in(\phi_1 - \phi_2)} \rangle\rangle$ in (\ref{eq2.8}).
We highlight three observations:
\begin{enumerate}
 \item {\it For any multiplicity $m$, $v_2^2\lbrace{2}\rbrace$ is finite in the limit $N\to \infty$ of a large number of sources.}\\
 Since the sum $\sum_{(ij)}$ goes over $N(N-1)/2$ source pairs in (\ref{eq3.14}),
 the two-particle cumulant $v_2^2\lbrace{2}\rbrace(k_1,k_2)$ approaches a finite value for $N\to \infty$.
 Physically, this is so since the observable $v_2^2\lbrace{2}\rbrace(k_1,k_2)$ sums over all source pairs,
 and since each source pair contributes with a $1/(N_c^2-1)$-suppressed contribution to two-particle interference terms. 
 The signal strength $v_2^2\lbrace{2}\rbrace(k_1,k_2)$ does not decrease with the number of dipoles (or the multiplicity
 in the event) although each dipole is oriented in a statistically independent direction. 
 \item {\it $v_2^2\lbrace{2}\rbrace(k_1,k_2)$ does not factorize except for small transverse momentum. }\\
 In general, due to the source average $ \int_\rho$, one has $v_2^2\lbrace{2}\rbrace(k_1,k_2)  \not= 
 v_2\lbrace{2}\rbrace(k_1)\, v_2\lbrace{2}\rbrace(k_2)$. For a Gaussian source distribution (\ref{eq3.4}), however,
 factorization holds for soft transverse momenta to leading order in $Bk_1^2$ and $B k_2^2$, 
 \begin{equation}
     v_2^2\lbrace{2}\rbrace(k_1,k_2)  = F^{(2)}_{\rm corr}(N,m) \frac{1}{(N_c^2-1)}\, (Bk_1^2)\, (Bk_2^2) + O\left((Bk^2)^3\right)\, .
     \label{eq3.15}
 \end{equation}
 \item {\it For any finite number of sources $N$, $v_2^2\lbrace{2}\rbrace(k_1,k_2)$ vanishes in the high-multiplicity limit.}\\
 This is a direct consequence of (\ref{eq3.13}).  
 Based on intuition from QED, one may have expected that maximal azimuthal correlation arises if all gluons are emitted from the
 same color dipole (\emph{i.e.}, $N=2$). This is not the case. Emitting a large number $m$ of gluons from a large number of sources $N$
 can yield a larger signal $v_2^2\lbrace{2}\rbrace(k_1,k_2)$ than emission from a small number of sources, since the color between
 off-diagonal gluons is less decorrelated. 
\end{enumerate}

 \newpage
 \section{Beyond 2nd order cumulants: results for leading $O(N)$ and up to subleading $O\left(1/(N_c^2-1)^3\right)$ }
 \label{sec4}
 
 We extend now the calculations of section~\ref{sec3} to the fourth order cumulant $v_2\lbrace 4\rbrace$. To this end, we 
 have calculated the $m$-gluon emission cross section from $N$ sources up to $O\left(1/(N_c^2-1)^3\right)$. The result 
 is\footnote{In eq.~(\ref{eq4.1})  $\sum_{(abc)}$ sums over the $m(m-1)(m-2)/3!$ unordered triplets of outgoing momenta, \emph{i.e.},
each index $a$, $b$, $c$ runs from 1 to $m$ and the combinations $(abc)$, $(bac)$ and other permutations are counted like one
element in the sum. Similarly, $\sum_{(lm)(mn)(nl)}$ sums over the $N(N-1)(N-2)$ unordered triplets of three source pairs made
of three sources. In contrast, the comma in the sum $\sum_{(lm),(no)}$ indicates that this sum is over the $N(N-1)(N-2)(N-3)/2^2$
elements in the ordered set of doublets of source pairs, \emph{i.e.}, the entries $(lm),(no)$ and $(no),(lm)$ are counted separately.}
(see appendix~\ref{appb} for details)
  \begin{eqnarray}
         &&\hspace{-.5cm}{\hat \sigma}
  	\propto  N_c^m \left( N_c^2 - 1\right)^N  \left( \prod_{i=1}^m \left|\vec{f}({\bf k}_i)\right|^2 \right) N^{m-4}
  		\nonumber \\
  		&& \times \Big\lbrace  N^4 +  F_{\rm corr}^{(2)}(N,m) \frac{N^2}{(N_c^2-1)} \sum_{(ab)} \sum_{(lm)}  
  		2^2\,  {\cos\left({\bf k}_a.\Delta {\bf y}_{lm} \right)}  { \cos\left( {\bf k}_b.\Delta {\bf y}_{lm}  \right)}
  	\nonumber \\
  	&& \qquad \qquad + F_{\rm corr}^{(3i)}(N,m) \frac{N}{(N_c^2-1)^2} 
  		 \sum_{(abc)} \sum_{(lm)(mn)(nl)}  2^3 {\cos\left({\bf k}_a.\Delta {\bf y}_{lm} \right)}  
  	\nonumber \\
  	&& \qquad \qquad  \qquad \qquad  \qquad \qquad \qquad  \qquad \qquad \times	 
  	{ \cos\left( {\bf k}_b.\Delta {\bf y}_{mn}  \right)}
  	{ \cos\left( {\bf k}_c.\Delta {\bf y}_{nl}  \right)}
  	\nonumber \\
  	&& \qquad \qquad + F_{\rm corr}^{(4i)}(N,m) \frac{1}{(N_c^2-1)^2} \sum_{(lm),(no)}  \sum_{(ab)(cd)}  2^4\, 
  	{\cos\left({\bf k}_a.\Delta {\bf y}_{lm} \right)}  { \cos\left( {\bf k}_b.\Delta {\bf y}_{lm}  \right)}
  	\nonumber \\
  	&& \qquad \qquad  \qquad \qquad  \qquad \qquad \qquad  \qquad \qquad \times
  	{ \cos\left( {\bf k}_c.\Delta {\bf y}_{no}  \right)}\, { \cos\left( {\bf k}_d.\Delta {\bf y}_{no}  \right)}
  	\nonumber\\
  	&& \qquad \qquad + F_{\rm corr}^{(4ii)}(N,m) \frac{1}{(N_c^2-1)^3}  \sum_{(lm)(mn)(no)(ol)} 
  		\sum_{(abcd)}  2^4\, 
  	{\cos\left({\bf k}_a.\Delta {\bf y}_{lm} \right)}  { \cos\left( {\bf k}_b.\Delta {\bf y}_{mn}  \right)}
  	\nonumber \\
 		&&  \qquad \qquad  \qquad \qquad  \qquad \qquad \qquad  \qquad \qquad  \times
 	{ \cos\left( {\bf k}_c.\Delta {\bf y}_{no}  \right)}\, { \cos\left( {\bf k}_d.\Delta {\bf y}_{ol}  \right)} 
 	\nonumber \\
 	&& \qquad \qquad + F_{\rm corr}^{(5)}(N,m) \frac{N^{-1}}{(N_c^2-1)^3}  \sum_{[(lm)(mn)(nl)](op)} 
  		\sum_{(abc)(de)}  2^2\, 
  	{\cos\left({\bf k}_d.\Delta {\bf y}_{op} \right)}  { \cos\left( {\bf k}_e.\Delta {\bf y}_{op}  \right)}
  	\nonumber \\
 		&&  \qquad \qquad  \qquad \qquad  \qquad \qquad \qquad  \times 2^3
 	{ \cos\left( {\bf k}_a.\Delta {\bf y}_{lm}  \right)}\, { \cos\left( {\bf k}_b.\Delta {\bf y}_{mn}  \right)} 
 	\, { \cos\left( {\bf k}_c.\Delta {\bf y}_{nl}  \right)} 
 	\nonumber \\
 	&& \qquad \qquad + F_{\rm corr}^{(6)}(N,m) \frac{N^{-2}}{(N_c^2-1)^3}  \sum_{(lm)(no)(pq)} 
  		\sum_{(ab)(cd)(ef)}  2^2\, 
  	{\cos\left({\bf k}_a.\Delta {\bf y}_{lm} \right)}  { \cos\left( {\bf k}_b.\Delta {\bf y}_{lm}  \right)}
  	\nonumber \\
 		&&  \qquad \qquad  \qquad \qquad  \times 2^2\, 
 	{ \cos\left( {\bf k}_c.\Delta {\bf y}_{no}  \right)}\, { \cos\left( {\bf k}_d.\Delta {\bf y}_{no}  \right)} 
 	\, 2^2\, 
 	{ \cos\left( {\bf k}_e.\Delta {\bf y}_{pq}  \right)}\, { \cos\left( {\bf k}_f.\Delta {\bf y}_{pq}  \right)} 
 	\nonumber \\
  	&& \qquad \qquad 
  	+ O\left( \frac{1}{N}\right) + O\left(\frac{1}{(N_c^2-1)^4}\right)   
  	 \Big\rbrace \, ,
  	\label{eq4.1}
 \end{eqnarray}
 This expression contains color correction factors $F_{\rm corr}^{(*)}(N,m)$ that we determine in appendix~\ref{appc}
 in close analogy to the derivation given in section~\ref{sec3b1} for $F_{\rm corr}^{(2)}(N,m)$. 
 
  In equation (\ref{eq4.1}), only those contributions are written that are $O(N^m)$ and thus leading in the number 
  of sources. To subleading order in the number of sources, there is a large number of additional 
  interference diagrams. For instance, one can have four off-diagonal gluons emitted from one single source pair,
  and this contribution is $O(N^{-2})$ suppressed compared to the terms given in (\ref{eq4.1}). 
   For the cases $N=m=3$ and $N=m=4$, we have determined all these contributions
  explicitly in appendices~\ref{appa} and ~\ref{appb}. These appendices provide combinatorical and calculational
  details on how to determine (\ref{eq4.1}). However, we have neither calculated nor classified completely the 
  subleading contributions $O(N^{-1})$ for an arbitrary number of sources $N$. For technical reasons, we therefore
  limit the present discussion to leading $O(N)$, which amounts to calculating $v_2\lbrace 4\rbrace$ to $O(N^0)$. 
 
 We are particularly interested in the question to what extent interference effects could give rise to asymmetries in the final state 
momentum distributions even if there are no asymmetries in the initial source distributions. We therefore specialize to a factorized ansatz 
of the $N$-source distribution in terms of a product of azimuthally symmetric single-source probabilities,
\begin{equation}
	\rho\left( \lbrace {\bf y}_j\rbrace \right) = \prod_{j=1}^N \rho({\bf y}_j)\, . 
	\label{eq4.2}
\end{equation}	
This ansatz includes factorizing Gaussian source models at vanishing impact parameter ${\bf b}$, 
 $\rho\left( \lbrace {\bf y}_j\rbrace \right) =\prod_{j=1}^N \left( \frac{1}{(2\pi B)}\exp(-{\bf y}_j^2/(2B))  \right)$,
 that will be motivated further in section~\ref{sec6}.

\subsection{4-particle cumulant to $O(N^0)$ and $O\left(1/(N_c^2 - 1)^2\right)$ }
\label{sec4a}
Restricting our discussion to the second harmonics, we define the normalized 4-point correlation function 
in terms of equations (\ref{eq2.10}) and (\ref{eq4.1}), 
\begin{eqnarray}
  \langle\langle e^{i2(\phi_1 + \phi_2 - \phi_3 - \phi_4)} \rangle\rangle(k_1,k_2,k_3,k_4) 
  =  \frac{S_2(k_1,k_2,k_3,k_4)}{\overline{S}(k_1,k_2,k_3,k_4)}\, .
  \label{eq4.3}
\end{eqnarray}
Here, we discuss this expression first to leading order $O\left(1/(N_c^2 - 1)^2\right)$, when only one term of (\ref{eq4.1})
contributes, 
\begin{eqnarray}
  && \langle\langle e^{i2(\phi_1 + \phi_2 - \phi_3 - \phi_4)} \rangle\rangle(k_1,k_2,k_3,k_4) 
  =  \frac{1}{(N_c^2-1)^2} \frac{2^4}{N^4} F_{\rm corr}^{(4i)}(N,k)
  \nonumber \\  
  && \qquad \qquad \int_{\rho} \sum_{(lm),(no)} 
        \Big\lbrace  J_2\left(k_1 \Delta y_{lm}\right)\, 	 J_2\left(k_2 \Delta y_{lm}\right)\, 
	 J_2\left(k_3 \Delta y_{no}\right)\, J_2\left(k_4 \Delta y_{no}\right)\, e^{i4(\phi_{lm} - \phi_{no})}
	 \nonumber \\
	 && \qquad \qquad \qquad  \qquad \qquad 
	 + J_2\left(k_1 \Delta y_{lm}\right)\, 	 J_2\left(k_3 \Delta y_{lm}\right)\, 
	 J_2\left(k_2 \Delta y_{no}\right)\, J_2\left(k_4 \Delta y_{no}\right)
	 \nonumber \\
	 && \qquad \qquad \qquad  \qquad \qquad 
	 + J_2\left(k_1 \Delta y_{lm}\right)\, 	 J_2\left(k_4 \Delta y_{lm}\right)\, 
	 J_2\left(k_2 \Delta y_{no}\right)\, J_2\left(k_3 \Delta y_{no}\right) \Big\rbrace
	 \nonumber \\
	  && \qquad \qquad \qquad  \qquad \qquad + O\left(\frac{1}{N}\frac{1}{(N_c^2-1)^2} \right) 	
	  + O\left(\frac{1}{(N_c^2-1)^3} \right) \, .
  \label{eq4.4}
\end{eqnarray}
Here, the angles $\phi_{lm}$, $\phi_{no}$ denote the azimuthal orientations of the dipoles $(lm)$ and $(no)$.
For factorizing distributions of the type (\ref{eq4.2}), different dipoles
are not correlated in angular orientation, and the source average over the phase $e^{i4(\phi_{lm} - \phi_{no})}$ in the first
term of (\ref{eq4.4}) vanishes. With the help of the two-point function (\ref{eq3.12}), 
the  fourth order cumulant (\ref{eq2.11}) can then be written in the following compact form
\begin{eqnarray}
  && \langle\langle e^{i2(\phi_1 + \phi_2 - \phi_3 - \phi_4)} \rangle\rangle_{\rm c}(k_1,k_2,k_3,k_4) 
  \nonumber\\
  && =  \int_{\rho}  \frac{F_{\rm corr}^{(4i)}(N,m)}{N^4}    \Big\lbrace  \left( \sum_{(lm),(no)}  \frac{2^2}{(N_c^2-1)}
	  J_2\left(k_1 \Delta y_{lm}\right) J_2\left(k_3 \Delta y_{lm}\right) \right)
	  \nonumber \\
	  && \qquad \qquad \times \left(  \frac{2^2}{(N_c^2-1)}
	 J_2\left(k_2 \Delta y_{no}\right)\, J_2\left(k_4 \Delta y_{no}\right) \right) + \left( k_3 \longleftrightarrow k_4\right)  \Big\rbrace
	 \nonumber \\
	 &&   -  \int_{\rho}   \left( \frac{F_{\rm corr}^{(2)}(N,m)}{N^2} \right)^2
        \Big\lbrace  \left( \sum_{(lm)}  \frac{2^2}{(N_c^2-1)}
	  J_2\left(k_1 \Delta y_{lm}\right)\, 	 J_2\left(k_3 \Delta y_{lm}\right)\right)
	  \nonumber\\
	&& \qquad \qquad \times  \left(  \frac{2^2}{(N_c^2-1)} \sum_{(no)} 
	 J_2\left(k_2 \Delta y_{no}\right)\, J_2\left(k_4 \Delta y_{no}\right) \right)
	 + \left( k_3 \longleftrightarrow k_4\right)  \Big\rbrace 
	 \nonumber \\
	 && \qquad + O\left(\frac{1}{N}\right) + O\left(\frac{1}{(N_c^2-1)^2}\right)\, .
  \label{eq4.5}
\end{eqnarray}
%
%
For source distributions (\ref{eq4.2}), all dipole configurations $(lm)$, $(no)$ make identical contributions.
It is then sufficient to count the number of these dipoles. The sum $\sum_{(lm),(no)}$ in (\ref{eq4.5})
goes over $N(N-1)(N-2)(N-3)/2^2$ doublets of source pairs and thus includes $N^4/4 + O(N^3)$ terms.
Each of the sums $ \sum_{(lm)}$ and $\sum_{(no)}$ in (\ref{eq4.5}) go over $N(N-1)/2$ possibilities so that 
their product includes $N^4/4 + O(N^3)$ terms, too. To leading order in $1/N$, we can therefore replace in (\ref{eq4.5})
\begin{eqnarray}
 &&\left( \frac{F_{\rm corr}^{(4i)}(N,m)}{N^4} \left( \sum_{(lm),(no)} 1\right)  -  \frac{F_{\rm corr}^{(2)}(N,m)}{N^2} \left( \sum_{(lm)} 1\right)
  \frac{F_{\rm corr}^{(2)}(N,m)}{N^2} \left( \sum_{(no)} 1 \right) \right) \dots
  \nonumber \\
  &&\longrightarrow \frac{1}{4} \left(  F_{\rm corr}^{(4i)}(N,m) -  \left(F_{\rm corr}^{(2)}(N,m)\right)^2  \right) 
 + O\left( N^{-1} \right)\, .
   \label{eq4.6}
\end{eqnarray}
To order $O\left(N^0\right)$, this expression vanishes. In more detail:
 \begin{enumerate} 
 \item {\it In the limit $m = {\rm const.}$, $N\to \infty$, all color correction factors in (\ref{eq4.1}) become trivial,} \\
 \begin{equation}
 	\lim_{N\to\infty} F^{(*)}_{\rm corr}(N,m)\Big\vert_{m={\rm const.}} = 1\, . 
 	\label{eq4.7}
 \end{equation}
 \item {\it In the limit $m \to \infty$, for fixed average multiplicity per source $\overline{m} = m/N$, }\\
  \begin{equation}
  	\lim_{N\to\infty} F^{(2)}_{\rm corr}(N, N\overline{m})^2 = \lim_{N\to\infty}F^{(4i)}_{\rm corr}(N,N\overline{m})  
  	= \left(\frac{2\overline{m}+2 e^{-\overline{m}} - 2}{\overline{m}^2}\right)^2\, .
  	\label{eq4.8}
  \end{equation}
 \end{enumerate}
 As a consequence, (\ref{eq4.6}) vanishes in both limits and we find for the 4-particle cumulant to $O(N^0)$ and $O\left(1/(N_c^2 - 1)^2\right)$
 \begin{equation}
 v_2^4\lbrace{4}\rbrace(k_1,k_2,k_3,k_4) = 0 + O\left(1/N\right) + O\left(1/(N_c^2 - 1)^3\right)\, .
 \label{eq4.9}
 \end{equation}
For the third limit discussed in section~\ref{sec3b1} ($m\to\infty$ for constant $N$), subleading terms in $N$ would have to
be kept. The present calculation therefore does not give access to this limit.  
 
\subsection{4-particle cumulant to $O(N^0)$ and $O\left(1/(N_c^2 - 1)^3\right)$ }
\label{sec4b}

We extend now the calculation of the 4-particle cumulant (\ref{eq4.3}) to order $O\left(1/(N_c^2 -1)^3\right)$. All
terms in (\ref{eq4.1}) contribute to this order. In principle, the azimuthal integrations in (\ref{eq4.3}) can be done analytically, 
and the result can be expressed in terms of source averages over Bessel functions. 
The resulting expression are straightforward to obtain but they are lengthy. They simplify significantly if one assumes Gaussian source distributions
\begin{equation}
	\rho\left( \lbrace {\bf y}_i \rbrace \right) =\prod_{j=1}^N \left( \frac{1}{(2\pi B)}
	\exp\left[ -\frac{{\bf y}_j^2}{2B}  \right] \, \right)\, ,
	\label{eq4.10}
\end{equation}
and if one limits the analysis to small transverse momenta $B\, k_i^2 \ll 1$. For part of the following discussion, we
resort to this approximation in which the discussion of qualitative properties becomes more transparent. 

To calculate $\langle\langle e^{i2(\phi_1 + \phi_2 - \phi_3 - \phi_4)} \rangle\rangle_{\rm c}$ to $O\left(1/(N_c^2 - 1)^3\right)$ 
in (\ref{eq2.11}), we need to calculate the two-point correlation function to $O\left(1/(N_c^2 - 1)^2\right)$, and for this we need
to calculate the norm $\overline{T}(k_1,k_2)$ to $O\left(1/(N_c^2 - 1)^2\right)$. For the Gaussian source distribution (\ref{eq4.10}) and to 
lowest order in transverse momenta, we find 
\begin{equation}
   \overline{T}(k_1,k_2) = \# \frac{m(m-1)}{2} N^4 \Big\lbrace 1 + \frac{2}{N_c^2-1} F_{\rm corr}^{(2)}   \Big\rbrace
   + O\left(1/(N_c^2 -1)^2 \right)
   \, .\label{eq4.11}
\end{equation}
Here, the hash $\#$ denotes prefactors that are common to 
$\overline{T}$ and $T$ and that will therefore drop out in the calculation
of $\langle\langle e^{i2(\phi_1 - \phi_2)} \rangle\rangle_{\rm c}$. 
(Essentially, the hash stands for the factors written in the first line of (\ref{eq4.1}).)
The numerator of (\ref{eq2.7}) takes the form
\begin{eqnarray}
   T_2(k_1,k_2) &=& \# \frac{m(m-1)}{2} N^4 \, (Bk_1^2)\, (Bk_2^2) \, 
   \Big\lbrace \frac{F_{\rm corr}^{(2)} }{N_c^2-1}  
   	\nonumber \\
	&& \qquad \qquad   + \frac{F_{\rm corr}^{(3)}}{(N_c^2-1)^2}  (m-2)
   +  \frac{F_{\rm corr}^{(4i)}}{(N_c^2-1)^2}  (m-2)(m-3)  \Big\rbrace\, .
   \label{eq4.12}
\end{eqnarray}
To leading order $O\left(1/(N_c^2 - 1)\right)$, the normalized azimuthal two-particle correlation function reduces then to (\ref{eq3.15}),
but there are higher order corrections
 \begin{eqnarray}
  \langle\langle e^{i2(\phi_1 - \phi_2)} \rangle\rangle(k_1,k_2) &\equiv&  \frac{T(k_1,k_2)}{\overline{T}(k_1,k_2)}
     \nonumber \\
     &=& (Bk_1^2)\, (Bk_2^2)\, \Big\lbrace  \frac{1}{(N_c^2-1)}\, F^{(2)}_{\rm corr}
     \nonumber \\ 
     && \quad + \frac{1}{(N_c^2-1)^2} \left( F^{(4i)}_{\rm corr}(m-2)(m-3) 
	+ F^{(3)}_{\rm corr}(m-2) - 2 (F^{(2)}_{\rm corr})^2 \right) \Big\rbrace 
     \nonumber \\
     &&+ O\left( N^{-1} \right) + O\left( 1/(N_c^2 -1)^3 \right)\, .
     \label{eq4.13}
 \end{eqnarray}
The $m$-gluon emission cross section (\ref{eq4.1}) includes contributions that involve interference between 3 or
4 off-diagonal gluons and that enter (\ref{eq4.12}). For instance, the term proportional to $F^{(3)}_{\rm corr}$ in (\ref{eq4.12})
includes a sum $\sum_{(abc)}$ over $m!/ (3! (m-3)!)$ triplets of off-diagonal gluons. In the contribution of this
term to $\langle\langle e^{i2(\phi_1 - \phi_2)} \rangle\rangle(k_1,k_2)$, only those terms in $\sum_{(abc)}$ survive for 
which two of the three gluons match the phases. As a consequence, the sum reduces to $\sum_{(12c)}$, which is a sum
over $(m-2)$ terms. This is the reason for the factor $F^{(3)}_{\rm corr}\, (m-2)$ in (\ref{eq4.13}). The factor
$(m-2)(m-3)$ multiplying $F^{(4i)}_{\rm corr}$ in (\ref{eq4.13}) can be understood analogously. The factor 
$(F_{\rm corr}^{(2)})^2$ comes from expanding the normalization $1/\bar T$ to $O\left(1/(N_c^2 - 1)\right)$. 

Analogously, one obtains the phase factor (\ref{eq2.10}) which determines the numerator of the 
4-particle correlation function $ \langle\langle e^{i2(\phi_1 +\phi_2 - \phi_3 - \phi_4)} \rangle\rangle$,
\begin{eqnarray}
	S_2(k_1,k_2,k_3,k_4) &=& \# \left( \begin{array}{c} m \\ 4 \end{array}\right)\,  N^4\, (Bk_1^2)\, (Bk_2^2)\, (Bk_3^2)\, (Bk_4^2)
	\nonumber\\
		&& \times \Big\lbrace  \frac{1}{(N_c^2-1)^2}\, 2\, F^{(4i)}_{\rm corr}
		 \nonumber \\
		&& \quad + \frac{1}{(N_c^2-1)^3} \left(2F^{(4ii)}_{\rm corr} + 4F^{(5)}_{\rm corr} (m-4) + 2F^{(6)}_{\rm corr} (m-5)(m-6)\right)
		\Big\rbrace 
		\nonumber \\
     &&+ O\left( N^{-1} \right) + O\left( 1/(N_c^2 -1)^4 \right)\, .
     \label{eq4.14}		
\end{eqnarray}
To write down $ \langle\langle e^{i2(\phi_1 +\phi_2 - \phi_3 - \phi_4)} \rangle\rangle$ to $O\left(1/(N_c^2 - 1)^3\right)$, one needs the
normalization $\overline{S}$ to $O\left(1/(N_c^2 - 1)\right)$ since (\ref{eq4.14}) starts 
at $O\left(1/(N_c^2 - 1)^2\right)$. One finds, in close analogy to (\ref{eq4.11}),
\begin{equation}
   \overline{S}(k_1,k_2,k_3,k_4)  = \#  \left( \begin{array}{c} m \\ 4 \end{array}\right)\,  N^4 \Big\lbrace 1 + \frac{2}{N_c^2-1} F_{\rm corr}^{(2)}   \Big\rbrace
   + O\left(1/(N_c^2 -1)^2 \right)
   \, .\label{eq4.15}	
\end{equation}
These expressions define $ \langle\langle e^{i2(\phi_1 +\phi_2 - \phi_3 - \phi_4)} \rangle\rangle$ according to (\ref{eq4.3}). The
resulting connected 4-particle correlation function (\ref{eq2.11}) reads
\begin{eqnarray}
	\langle\langle e^{i2(\phi_1 +\phi_2 - \phi_3 - \phi_4)} \rangle\rangle_c &=& (Bk_1^2)\, (Bk_2^2)\, (Bk_3^2)\, (Bk_4^2)
	\nonumber \\
		&& \Big\lbrace  \frac{1}{(N_c^2-1)^2}\,   \left(2\, F^{(4i)}_{\rm corr} - 2 F^{(2)}_{\rm corr} F^{(2)}_{\rm corr} 
		+ O\left( N^{-1} \right) \right)
		\nonumber \\
		&& \quad + \frac{1}{(N_c^2-1)^3} \left( 2\, F^{(6)}_{\rm corr}(m-4)(m-5) - 4\, F^{(2)}_{\rm corr} \, F^{(4i)}_{\rm corr} (m-2)(m-3) \right.
		\nonumber \\
		&& \quad \qquad + 4\, F^{(5)}_{\rm corr}(m-4) - 4\, F^{(2)}_{\rm corr} F^{(3)}_{\rm corr}(m-2)
		\nonumber \\
		 && \quad \qquad \left. + 2\, F^{(4ii)}_{\rm corr} + 8 \left( F^{(2)}_{\rm corr}\right)^3 - 4\, F^{(4i)}_{\rm corr} F^{(2)}_{\rm corr}
		 \right) + O\left( N^{-1} \right) \Big\rbrace
		 \nonumber \\
		 && + O\left( 1/(N_c^2 -1)^4 \right)\, .
		 \label{eq4.16}
\end{eqnarray}
We discuss now limiting cases of this expression. 
%
\subsubsection{The limit $N\to \infty$ for constant multiplicity $m$ to order $O(1/(N_c^2-1)^3)$}
\label{sec4b1}
Following eq.~(\ref{eq4.7}), all color correction factors reduce to unity in this limit, and the connected 4-point correlation function 
(\ref{eq4.16}) reads
\begin{eqnarray}
	\langle\langle e^{i2(\phi_1 +\phi_2 - \phi_3 - \phi_4)} \rangle\rangle_c &=& (Bk_1^2)\, (Bk_2^2)\, (Bk_3^2)\, (Bk_4^2)
		\,  \frac{2}{(N_c^2-1)^3} \left(7+m-m^2  \right)
		\nonumber \\
		&& \quad 
		 + O\left( 1/(N_c^2 -1)^4 \right)\, .
		 \label{eq4.17}
\end{eqnarray}
According to equation (\ref{eq2.12}), the 4-th order cumulant defines a real-valued 4-th order anisotropy coefficient $v_2\lbrace 4\rbrace$
only if it is negative. Remarkably, this condition is satisfied for sufficiently large multiplicity $m$, since 
(\ref{eq4.17}) turns negative for $m> 3.7$. The $m^2$-term in (\ref{eq4.17}) will
dominate for relatively small multiplicities already (say for $m> 5$), and the 4-th order cumulant reads then
\begin{equation}
 	v_2\lbrace 4\rbrace(k) \simeq  \frac{1}{(N_c^2-1)^{3/4}}  2^{1/4} \sqrt{m}\, B\, k^2\, .
 	\label{eq4.18}
\end{equation}
We conclude that without any azimuthal asymmetry in the initial state [see eq.~(\ref{eq4.10})] and without any (coupling-constant 
dependent) interaction in the final state, 
QCD interference can give rise to non-vanishing negative fourth-order cumulants 
that have an interpretation in terms of azimuthal harmonics $v_2\lbrace 4\rbrace(k)$. In this sense, our calculation provides 
a proof of principle that 
the baseline of vanishing interaction in the final state and of vanishing azimuthal correlation 
in the initial state does not correspond to a vanishing value $v_2\lbrace 4\rbrace(k)$.

Equation (\ref{eq4.18}) provides a proof of principle for the arguments above, but its range of validity is limited as we discuss now.
We have calculated $v_2\lbrace 2\rbrace(k)$ and $v_2\lbrace 4\rbrace(k)$ to leading order in the number of sources and in the 
large-$N_c$ limit. In particular, the second order cumulant (\ref{eq4.13}) contains a leading term $\propto F_{\rm corr}^{(2)}/(N_c^2-1)$
and a subleading term $\propto F_{\rm corr}^{(4i)} m^2/(N_c^2-1)^2$. A similar observation can be made for the 4-particle correlator 
(\ref{eq4.16}) where the term suppressed by one power $1/(N_c^2 - 1)$ is enhanced by $m^2$. This seems to indicate
that the expansion in powers of  $1/(N_c^2 - 1)$ converges only as long as $m^2 < (N_c^2 - 1)$. 

The radius of convergence may be larger than the above estimate for the following reason:
In equations~(\ref{eq4.13}) and (\ref{eq4.16}), terms proportional
to $m$ ($m^2$) arise from integrating out the transverse momentum $q$ of one (two) of the off-diagonal gluons involved in
an interference term. In the simplest case\footnote{For instance, the contribution to (\ref{eq4.12}) obtained from 
integrating the term proportional to $F_{\rm corr}^{(3i)}$ in (\ref{eq4.1}) over
the third transverse momentum ${\bf k}_c$ yields $(m-2)$ choices of the transverse momentum ${\bf k}_c$. Accordingly, the
contribution to the two-particle correlator (\ref{eq4.13}) is $\propto F_{\rm corr}^{(3i)}\, a\, (m-2)$.}, 
this $q$-integration leads to contributions of the type
\begin{equation}
  a \equiv \frac{\int d{\bf q}\int d{\bf z}\, J_0(qz)\, \rho({\bf z})\, f({\bf q})^2}{\int d{\bf q}\, \vert \vec{f}({\bf q})\vert^2}\, .
  \label{eq4.19}
\end{equation}
For technical simplicity, we have worked in this subsection
to lowest order in small transverse momentum. This amounts
to the assumption that the emission vertex $f({\bf q})$  is dominated by transverse momenta that are much smaller than the
inverse transverse size of the source.  In this limit, $J_0(qz) \approx 1$ and thus $a = 1$. In general, however, $a \leq 1$.
Indeed, depending on the emission vertices $f({\bf q})$ and on the source density $\rho({\bf z})$, 
$a$ may be significantly smaller than unity, and e.g. the $F_{\rm corr}^{(3i)}$-term in (\ref{eq4.13})
should yield a contribution $\propto F_{\rm corr}^{(3i)}\, (m-2)\, a$.

Integrating out transverse gluon momenta from other contributions in (\ref{eq4.1}) 
can yield more complicated expressions than (\ref{eq4.19}), but the maximal value is always obtained in the limit $B\, q^2 \ll 1$
studied here, and the value starts decreasing  when the integral over the transverse momentum extends
to values that start resolving the transverse distance between sources. Therefore, rather than being limited to $m^2 < (N_c^2 - 1)$, 
the region of validity of the calculations in section~\ref{sec4} is expected to extend up to higher multiplicities 
\begin{equation}
 m_{\rm max} \sim \sqrt{N_c^2 - 1}/a\, ,
 \label{eq4.20}
 \end{equation}
where $a \ll 1$. Physically, $a$ is a penalty factor for integrating out off-diagonal gluons. It depends on the source size
and the emission vertex $f$. 

%
\subsubsection{$N\to \infty$ for constant average multiplicity $\overline{m}$ to order $O(1/(N_c^2-1)^3)$}
\label{sec4b2}
In the limit $N\to \infty$ for fixed average multiplicity $\overline{m} \equiv m/N$, the color correction factors 
satisfy several interesting identities. Defining $ F^{(*)}_{\rm corr} (\overline{m}) \equiv \lim_{N\to \infty} F^{(*)}_{\rm corr} (N,N\overline{m}) $,
one finds 
\begin{eqnarray}
	F^{(4i)}_{\rm corr} (\overline{m}) &=& F^{(2)}_{\rm corr} (\overline{m})\, F^{(2)}_{\rm corr} (\overline{m}) \, ,
	\label{eq4.21} \\
	F^{(5)}_{\rm corr} (\overline{m}) &=& F^{(3)}_{\rm corr} (\overline{m})\, F^{(2)}_{\rm corr} (\overline{m}) \, ,
	\label{eq4.22} \\
	F^{(6)}_{\rm corr} (\overline{m}) &=& F^{(2)}_{\rm corr} (\overline{m})\, F^{(2)}_{\rm corr} (\overline{m})\, F^{(2)}_{\rm corr} (\overline{m})\, .
	\label{eq4.23}
\end{eqnarray}
One may use these relations to simplify the connected 4-particle correlation function (\ref{eq4.16}),
\begin{eqnarray}
	\langle\langle e^{i2(\phi_1 +\phi_2 - \phi_3 - \phi_4)} \rangle\rangle_c &=& 
	 \frac{1}{(N_c^2-1)^3} \, (Bk_1^2)\, (Bk_2^2)\, (Bk_3^2)\, (Bk_4^2)
	\nonumber \\
		&&\times \Big\lbrace  
		 \left( F^{(2)}_{\rm corr} \right)^3 (-2m^2+2m+20)
		 - 8\, F^{(2)}_{\rm corr} F^{(3)}_{\rm corr}
		+  2\, F^{(4ii)}_{\rm corr} + O\left( N^{-1} \right) 
		 \Big\rbrace
		 \nonumber \\
		 &&+ O\left( 1/(N_c^2 -1)^4 \right)\, .
		 \label{eq4.24}
\end{eqnarray}
As in the limit of section~\ref{sec4b1}, this expression is negative for sufficiently large multiplicity $m\gtrsim 4$ and thus lends itself
to a collective interpretation of $v_2\lbrace 4\rbrace$.
However, since the limit $N\to \infty$ at constant $\overline{m}$ implies $m\to \infty$, one cannot parallel the argument of 
section~\ref{sec4b1} that the expansion is well-defined for sufficiently small multiplicity~$m$. To take this limit, one would
need information about subleading orders in $N$. [For instance, if there were terms of $O(N^{-1})$ in (\ref{eq4.24}) that
are enhanced by $\sim m^3$, then these would contribute to leading order in the limit $N\to \infty$ at constant $\overline{m}$.]

\subsection{6-particle cumulant to $O(N^0)$ and $O\left(1/(N_c^2 - 1)^3\right)$ }
\label{sec4c}
From equation (\ref{eq4.1}), we can also evaluate the expectation value 
$\langle\langle e^{i2(\phi_1 + \phi_2 + \phi_3 - \phi_4 - \phi_5 - \phi_6 )} \rangle\rangle$ to
order  $O\left(1/(N_c^2 - 1)^3\right)$. Since the numerator of this expression starts at $O\left(1/(N_c^2 - 1)^3\right)$, 
the normalization is trivial. One finds from explicit calculation in the limit $(B k_i^2)\ll 1$
\begin{eqnarray}
&& \langle\langle e^{i2(\phi_1 + \phi_2 + \phi_3 - \phi_4 - \phi_5 - \phi_6 )} \rangle\rangle(k_1,k_2,k_3,k_4,k_5,k_6)
\nonumber \\
&&\qquad = (Bk_1^2)\, (Bk_2^2)\, (Bk_3^2)\, (Bk_4^2)\, (Bk_5^2)\, (Bk_6^2) \frac{1}{(N_c^2-1)^3} \, 6\, F^{(6)}_{\rm corr}
 \nonumber \\
 && \qquad \quad + O\left( 1/(N_c^2 -1)^4 \right)\, .
 \label{eq4.25}
\end{eqnarray}
The 6-th particle cumulant is defined as $\langle\langle 6 \rangle\rangle_c \equiv \langle\langle 6 \rangle\rangle - 9 \langle\langle 2 \rangle\rangle \langle\langle 4  \rangle\rangle + 12 \langle\langle 2  \rangle\rangle^3$ (in this shorthand notation, the numbers in brackets denote the number of phases). One therefore finds
\begin{eqnarray}
&& \langle\langle e^{i2(\phi_1 + \phi_2 + \phi_3 - \phi_4 - \phi_5 - \phi_6 )} \rangle\rangle_c(k_1,k_2,k_3,k_4,k_5,k_6)
\nonumber \\
&&\qquad = (Bk_1^2)\, (Bk_2^2)\, (Bk_3^2)\, (Bk_4^2)\, (Bk_5^2)\, (Bk_6^2) \frac{1}{(N_c^2-1)^3} \, 
\nonumber \\
&& \qquad  \quad \times \Big\lbrace 6\, F^{(6)}_{\rm corr} - 18\, F^{(2)}_{\rm corr}\, F^{(4i)}_{\rm corr} + 12 \, \left(F^{(2)}_{\rm corr}\right)^3
\Big\rbrace  + O\left( 1/(N_c^2 -1)^4 \right)
 \nonumber \\
 &&\qquad   = 0 + O\left( 1/(N_c^2 -1)^4 \right)\, .
 \label{eq4.26}
\end{eqnarray}
Here, the fact that the contribution of $O\left( 1/(N_c^2 -1)^3 \right)$ vanishes to leading $O(1/N)$ is a consequence of 
\begin{equation}
	\lim_{N\to\infty} \Big\lbrace 6\, F^{(6)}_{\rm corr} - 18\, F^{(2)}_{\rm corr}\, F^{(4i)}_{\rm corr} + 12 \, \left(F^{(2)}_{\rm corr}\right)^3
\Big\rbrace = 0\, .
 \label{eq4.27}
\end{equation}
This limit vanishes, irrespective of whether it is taken for fixed multiplicity $m$ or for fixed average multiplicity $\overline m = m/N$. 
The latter statement is checked easily with the help of eqs.~(\ref{eq4.21}) -- (\ref{eq4.23}).
Consistent with our finding (\ref{eq4.9}) that the term 
proportional to $1/\sqrt{N_c^2 - 1}$ vanishes in $v_2\lbrace 4\rbrace$, we find therefore that $v_2\lbrace 6\rbrace$ vanishes to the 
same order. 

We did not check whether (\ref{eq4.26}) vanishes to order $O\left( 1/(N_c^2 -1)^4 \right)$. If it would not vanish (and if it would have
a positive sign), then $v_2\lbrace 6\rbrace \sim O(1/(N_c^2-1)^{2/3})$. This would be peculiar, as $v_2\lbrace 6\rbrace$ 
would then be parametrically larger than $v_2\lbrace 4\rbrace$. It is also conceivable that the first non-vanishing order
to $v_2\lbrace 6\rbrace $ is $O\left( 1/(N_c^2 -1)^5 \right)$. In this case, $v_2\lbrace 6\rbrace \sim O(1/(N_c^2-1)^{5/6})$ 
would be parametrically smaller than $v_2\lbrace 4\rbrace$. In this case, the second, fourth and sixth
cumulant would follow the systematics $v_2\lbrace 2k\rbrace \sim O\left( 1/(N_c^2 -1)^{(2k-1)/2k} \right)$ which would support
the idea that higher order cumulants take similar values. These are open questions that lie outside the scope of the present
manuscript. They would involve significant further calculations, but we believe that they can be addressed with the 
techniques used in this section. 

\section{Higher harmonics}
\label{sec5}
In sections~\ref{sec3} and ~\ref{sec4}, we have focussed on the calculation of the second harmonics 
$ \langle\langle e^{i2(\phi_1 - \phi_2)} \rangle\rangle$ and 
$ \langle\langle e^{i2(\phi_1+\phi_2 - \phi_3 - \phi_4)} \rangle\rangle_c$. Here, we discuss the calculation of other
even and odd harmonics. 
\subsection{Higher even harmonics} 
\label{sec5a}
When calculating $ \langle\langle e^{in(\phi_1 - \phi_2)} \rangle\rangle$
and $ \langle\langle e^{in(\phi_1+\phi_2 - \phi_3 - \phi_4)} \rangle\rangle_c$, one encounters 
elementary azimuthal integrals of the form 
\begin{equation}	
	\int d\phi\, e^{in\phi} \cos\left(q \Delta y \cos(\phi) \right) \, .
	\label{eq5.1}
\end{equation}
Here, $q$ is the modulus of a generic transverse momentum, $\Delta y$ is the modulus of a generic transverse
dipole separation, and $\phi$ is the relativ azimuthal angle between transverse momentum and dipole separation. 
All cosine-terms in (\ref{eq4.1}) can be written as $\cos\left(q \Delta y \cos(\phi) \right)$ for suitable choices of
$q$, $\Delta y$, and $\phi$. 

The integrals (\ref{eq5.1}) are non-vanishing for all even integers $n$. For $n=2$, one finds for instance
\begin{eqnarray}	
	\int d\phi\, e^{i2\phi} \cos\left(q \Delta y \cos(\phi) \right) &=& -2\pi J_2\left(q \Delta y \right) 
	\nonumber \\
	&=& \frac{-\pi}{4} \left(q \Delta y  \right)^2 + O\left(  (q \Delta y)^4  \right)\, .
	\label{eq5.2}
\end{eqnarray}
This leads to the Bessel functions $J_2$ in the calculation of the two-particle correlation (\ref{eq3.14}) and
the four particle correlation (\ref{eq4.4}). 
Explicit expressions can also be given for higher even harmonics.
For instance
\begin{eqnarray}	
	\int d\phi\, e^{i4\phi} \cos\left(q \Delta y \cos(\phi) \right) &=& 2\pi\, J_4\left(q \Delta y \right) 
		 \nonumber \\
		 &=& \frac{\pi}{192} \left(q \Delta y  \right)^4 + O\left(  (q \Delta y)^6  \right)\, .
	\label{eq5.3}
\end{eqnarray}
With the help of these expressions, the analysis of section~\ref{sec4} could be repeated for arbitrary even harmonics.
Since the small-$q$ limit of (\ref{eq5.1}) is $\propto q^n$, one concludes
immediately that the small-$k$-behavior of $v_n({\bf k})$ is $\propto k^n$. In particular, the parametric dependence of the
fourth harmonic at small $q$ is $\sim \left(q \Delta y  \right)^4$  
while that of the second harmonics is $\sim \left(q \Delta y  \right)^2$. Within the approximation of small transverse
momenta, $\sim (B\, q^2) \ll 1$ explored in section~\ref{sec4b}, this implies that $v_4 \propto v_2\, v_2$ for second 
and fourth order cumulants. We discuss this point further in section~\ref{sec7}. 
%
\subsection{Higher odd harmonics} 
\label{sec5b}
In contrast to the even harmonics, the integral (\ref{eq5.1}) vanishes for odd integers $n$. As a consequence,  
the odd harmonic flow coefficients $v_n$ vanish up to $O\left( 1/(N_c^2-1)^4\right)$ and $O\left( 1/N\right)$, 
since the spectrum (\ref{eq4.1}) shows only cosine-terms to this order. 

Based on the idea that interference patterns are momentum conjugates of spatial distributions, one may
naively expect that odd harmonics make some (possibly subleading) non-vanishing contribution whenever the spatial 
distribution shows odd harmonic eccentricities, \emph{i.e.}, for $N\geq 3$ sources. Motivated by this idea, we have 
calculated in appendix~\ref{appa} all terms contributing to $N=m=3$ in search of odd harmonics. However, 
the emission spectrum for  $N=m=3$ turned out to be free of odd harmonics. 

\subsubsection{Odd harmonics for the case $N=m=4$}
\label{sec5b1}
\begin{figure}[t]
\centering
\includegraphics[width=0.8\textwidth]{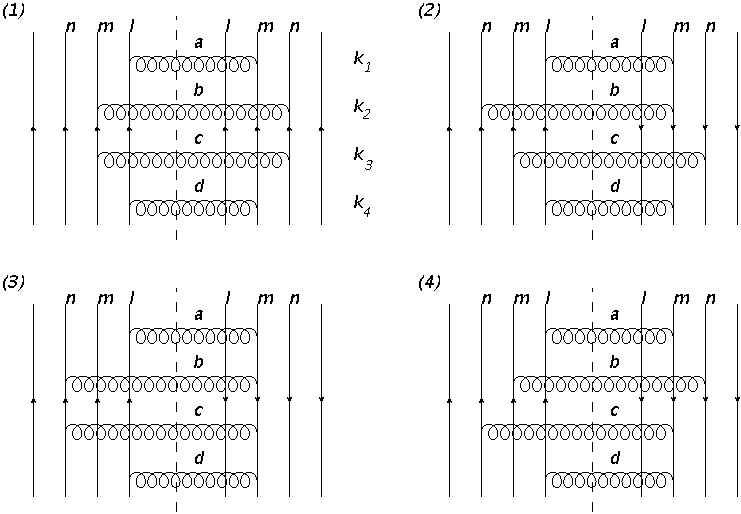} 
\caption{A set of diagrams that contribute to the $\sin$-terms in (\ref{eqb.13}), and thus to odd azimuthal harmonics.
}
\label{fig4}
\end{figure}

We have classified and calculated in appendix~\ref{appb} all contributions to the emission spectrum for $N=m=4$.
Two classes of diagrams were found to lead to odd harmonics, see eqs.~(\ref{eqb.13}) and (\ref{eqb.16}). 
Referring for technical details and explicit results to the appendix, we limit the discussion here in the main text
to provide qualitative insight into how properties of the $SU(N_c)$ color algebra 
can give rise to odd harmonics. To this end,  we consider in Fig.~\ref{fig4} a set of diagrams with four off-diagonal gluons that are emitted
from three sources $l$, $m$, $n$ that combine to two pairs of sources $(lm)$, $(mn)$ .

%
For this contribution, one checks easily that the color trace changes depending on
whether the sandwiched gluons of momentum ${\bf k}_2$ and ${\bf k}_3$ link from $m$ to $n$ or from 
$n$ to $m$. For the four diagrams in Fig.~\ref{fig4}, one finds

\begin{eqnarray}
	{\rm Tr}\left[ \mathbb{1} \right]\, {\rm Tr}\left[T^c\, T^b \right]\, {\rm Tr}\left[ T^b\, T^c\, T^d\, T^a\right]\, {\rm Tr}\left[ T^a T^d\right] 
	 = N_c^4\, \left( N_c^2 - 1\right)^2 \qquad \hbox{[for Fig~\ref{fig4}(1)]}
	 \nonumber \\
	{\rm Tr}\left[ \mathbb{1}\right]\, {\rm Tr}\left[ T^b\, T^c\right]\, {\rm Tr}\left[T^c\, T^d\, T^b\, T^a \right]\, {\rm Tr}\left[T^a\, T^d \right]\,  = \frac{1}{2}N_c^4\, \left( N_c^2 - 1\right)^2
	\qquad \hbox{[for Fig~\ref{fig4}(2)]}
	\nonumber \\
	{\rm Tr}\left[ \mathbb{1}\right]\, {\rm Tr}\left[ T^b\, T^c \right]\, {\rm Tr}\left[ T^d\, T^c\, T^b\, T^a\right]\, {\rm Tr}\left[ T^a\, T^d \right]\,  = N_c^4\, \left( N_c^2 - 1\right)^2
	\qquad \hbox{[for Fig~\ref{fig4}(3)]}
	\nonumber \\
	{\rm Tr}\left[ \mathbb{1}\right]\, {\rm Tr}\left[T^c\, T^b \right]\, {\rm Tr}\left[ T^b\, T^d\, T^c\, T^a\right]\, {\rm Tr}\left[ T^a\, T^d\right]\,  = \frac{1}{2} N_c^4\, \left( N_c^2 - 1\right)^2
	\qquad \hbox{[for Fig~\ref{fig4}(4)]}
	\nonumber
\end{eqnarray}
 Combining the factors $1/2$ of these color traces with the ${\bf k}_2$- and ${\bf k}_3$-dependent phase factors of the four diagrams in Fig.~\ref{fig4}, 
 we find
 \begin{eqnarray}
 	&&e^{i\, {\bf k}_2.{\bf \Delta y}_{mn}} \left(  e^{i\, {\bf k}_3.{\bf \Delta y}_{mn}} + \frac{1}{2} e^{-i\, {\bf k}_3.{\bf \Delta y}_{mn}}\right)
 	+ e^{-i\, {\bf k}_2.{\bf \Delta y}_{mn}} \left( \frac{1}{2}  e^{i\, {\bf k}_3.{\bf \Delta y}_{mn}} + e^{-i\, {\bf k}_3.{\bf \Delta y}_{mn}}\right)
 	\nonumber\\
 	&& = 3 \cos\left( {\bf k}_2.{\bf \Delta y}_{mn} \right)\, \cos\left( {\bf k}_3.{\bf \Delta y}_{mn} \right) 
 	- \sin\left( {\bf k}_2.{\bf \Delta y}_{mn} \right)\, \sin\left( {\bf k}_3.{\bf \Delta y}_{mn} \right) \, .
 	\label{eq5.6}
 \end{eqnarray}
 For each diagram that contributes with a phase $e^{-i\, {\bf k}_j.{\bf \Delta y}_{mn}}$, there is a diagram in which the off-diagonal
 gluon with momentum ${\bf k}_j$ links from $m$ to $n$ rather than from $n$ to $m$ while all other gluons are linked
 in the same way. If the prefactors of both these diagrams were the same, then these phases would add to a term 
  $e^{-i\, {\bf k}_j.{\bf \Delta y}_{mn}} + e^{+i\, {\bf k}_j.{\bf \Delta y}_{mn}}$ that is proportional to a cosine, and odd
  harmonics in ${\bf k}_j$ would not occur. In the example above, it is only the non-abelianess of $SU(N_c)$ that leads
  to different prefactors of the phases $e^{-i\, {\bf k}_j.{\bf \Delta y}_{mn}}$ and $e^{+i\, {\bf k}_j.{\bf \Delta y}_{mn}}$,
  thus giving rise to terms that change sign under ${\bf k}_j \to - {\bf k}_j$, see eq.~(\ref{eq5.6}). This is true for all
  odd harmonics that we have found in our calculations.

 \subsubsection{General comments on odd harmonics} 
 \label{sec5b2}
  In calculations of multi-particle correlations in the framework of saturation physics~\cite{Dumitru:2010iy,Levin:2011fb,Kovner:2010xk,Kovchegov:2012nd,Dumitru:2014dra,Dusling:2012iga,Dumitru:2014yza},
it has been a persistent problem to find non-vanishing values of odd harmonic anisotropy coefficients such as $v_3$. 
Solutions to this problem include the proposal that the proton wavefunction consists of a few patches in the transverse plane in each 
of which color fields point in preferred direction~\cite{Kovner:2011pe}, as well as the recent observation that odd harmonics arise as a 
high parton density effect directly from the non-linear QCD evolution~\cite{Kovner:2016jfp}. The calculation in section~\ref{sec5b1}
points to a third possible origin of odd harmonic contributions: without any coupling-constant dependent interaction in the initial
or final state and without any asymmetry in the initial state, odd harmonic contributions to multi-particle production can arise from 
non-abelian properties of QCD interference.

 For the odd harmonic contributions analyzed in section~\ref{sec5b1}, it was important that the number of off-diagonal 
 gluons was larger than the number of sources to which they were connected. As a consequence, these contributions 
 seem to be  $O(1/N)$-suppressed. We note, however, that we have studied odd harmonic contributions only for
 $N=m=4$. We did not try to determine the color correction factor when further diagonal gluons are added to the diagrams
 studied here 
 (this would require  combinatorical arguments that go beyond those developped in appendix~\ref{appc}), and we 
 do not know the full $N$- and $m$-dependence of the leading odd harmonic contribution.  
  It remains an interesting open
 question to find a classification of all diagrams  of $O(1/N)$ and to establish how odd harmonics manifest themselves in 
 the limit of a large number of sources.

 \section{Relation to the theory of multi-parton interactions (MPIs)}
 \label{sec6}
 Starting from ideas in the mid-80s~\cite{PT,mufti}, the treatment of multiple parton interactions (MPIs) in perturbative QCD
 has been developed further in recent years~\cite{stirling,BDFS1,stirling1,BDFS2,Diehl2,BDFS4,Gauntnew,BO,mpi2015}. 
  Here, we give simple arguments that relate this theory to the model defined in section~\ref{sec2} and that motivate
 values for the source parameter $B$ in pp collisions. 
 
 Hadronic cross sections involving $N$ partonic interactions are customarily parametrized as products of $N$ independent
 parton-parton interactions $\sigma_i$
\begin{equation}
\sigma_{N\, {\rm MPI}}=\frac{\sigma_1....\sigma_N}{K_N}\, .
\label{eq6.1}
\end{equation}
The physics of MPIs enters here in the coefficient $K_N$ that parametrizes deviations from an incoherent superposition
of $N$ interactions. $K_N$ is dimensionfull $\propto ({\rm area})^{N-1}$. For the case of double parton interactions ($N=2$),
$K$ reduces to the effective cross section $\sigma_{\rm eff}$ that is constrained experimentally. Under certain mild assumptions (such as neglecting 
contributions to MPIs from $1\to 2$ splittings), $K_N$ can be expressed through $N$-particle Generalized Parton Distributions 
(GPDs) $G_N$~\cite{BDFS1}
\begin{eqnarray}
\frac{1}{K_N}=    \int \left( \prod_{i=1}^N \frac{d{\bf \Delta}_i}{(2\pi)^2} \right)
\frac{ G_N(\lbrace x_i\rbrace, \lbrace Q_i^2\rbrace ,\lbrace {\bf \Delta}_i\rbrace)\, 
G_N(\lbrace x_{i}'\rbrace ,\lbrace Q_i^2\rbrace , \lbrace {\bf \Delta}_i\rbrace)}{ \prod_{i=1}^N  (f(x_{i},Q_i^2)\, f(x_{i}',Q_i^2))}
\delta^{(2)}\left(\sum_{i=1}^N {\bf \Delta}_i\right)\, .
\label{eq6.2}
\end{eqnarray}
Here $f(x,Q^2)$ denotes the standard single parton distribution function. The $x_i$, ${\bf \Delta}_i$ denote the longitudinal
momentum fractions and the initial transverse momenta of the i-th parton in the incoming hadronic wave function. Neglecting the
weak dependence of this expression on $x_i$ and $Q_i$, $K_N$ can be expressed in terms of the function 
\beq
F_N({\bf \Delta}_1,\dots {\bf \Delta}_N)=\frac{G^2_N({\bf \Delta}_1,\dots {\bf \Delta}_N)}{\prod_{i=1}^N f(x_i,Q_i)\, f(x'_i,Q_i)}\, .
\label{eq6.3}
\eeq
For a process involving $N$ MPIs, an $m$-parton production cross section can then be written formally as
\begin{equation}
\frac{d\sigma_N( \lbrace {\bf k}_i\rbrace, \lbrace {\bf \Delta}_i \rbrace )}{d{\bf k}_1...d{\bf k}_m}\sim \vert
{\cal M}^2(\lbrace {\bf k}_i\rbrace; \lbrace {\bf \Delta}_i \rbrace)\vert\, 
F^2_N\left({\bf \Delta}_1,\dots ,{\bf \Delta}_N\right)\,
\delta^{(2)}\left(\sum_{j=1}^N {\bf \Delta}_j\right)\, \sigma_1\dots \sigma_N\, ,
 \label{eq6.4}
\end{equation}
where ${\cal M}^2$ is the squared amplitude for the production of m gluons from N partons in the nucleon wave function 
(``N sources'').
The corresponding $m$-particle spectrum is obtained by normalizing this expression with the cross section 
\begin{equation}
\sigma_N=\prod \int d{\bf \Delta}_i\,  F^2_N\left({\bf \Delta}_1,\dots ,{\bf \Delta}_N\right)\,  
\delta^{(2)}\left(\sum_{j=1}^N {\bf \Delta}_j\right)\, 
\sigma_1\dots \sigma_N\label{eq6.5}\, .
\end{equation}
To arrive at eqs.~(\ref{eq6.4}) and (\ref{eq6.5}), one assumes that $m$-parton production can be formulated in a pQCD
factorized formalism. Here, we do not try to quantify corrections to this assumption. Rather, we treat
a bold extrapolation of this perturbative approach to soft momenta as one way of getting insight into 
soft multi-particle production. 

To see the commonalities between the
model in section~\ref{sec2} and this formalism deduced from MPI theory, we introduce one further approximation 
by writing the generalized $N$-parton distribution functions $G_N$ in a mean-field approximation~\cite{BDFS1} 
as products of generalized
one-particle distribution functions $G_1(x,Q,\Delta)=f(x,Q)\, F_{2g}(\Delta)$~\cite{Frankfurt,Frankfurt1,Frankfurt2}
 with a two-gluon form factor that parametrizes the transverse momentum distribution
\beq
G_N(\lbrace x_i\rbrace, \lbrace Q_i^2\rbrace ,\lbrace {\bf \Delta}_i\rbrace)
=\prod_{i=1}^N G_1(x_i,Q_i,{\bf \Delta}_i)=\prod_{i=1}^N f(x_i,Q_i)F_{2g}({\bf \Delta}_i)\, .
\label{eq6.6}
\eeq
Choosing $F_{2g}^2(\Delta) = \exp(-B\Delta_i^2)$ for simplicity to be of Gaussian form (other functional dependencies could be explored),
one has $F_N(\Delta_1,\dots, \Delta_N)= \prod_{i=1}^N\exp(-B\Delta_i^2)$, and it is straightforward to switch from (\ref{eq6.5})
to coordinate space representation~\footnote{Here, the density $\rho\left(\lbrace {\bf y}_i\rbrace,{\bf b}\right)$ is a convolution of the normalized densities of colliding partons in the two incoming hadrons. For a Gaussian ansatz 
$$ \rho\left(\lbrace {\bf y}_i\rbrace,{\bf b}\right) = \prod_j \frac{1}{(4\pi B)^2} \exp\left[-\frac{{\bf y}_j^2}{4B} \right]\,
\exp\left[-\frac{\left({\bf y}_j-{\bf b}\right)^2}{4B} \right]\, . $$
In other sections, we have used this distribution for vanishing impact parameter ${\bf b}=0$ and normalized to unity, see eq.~(\ref{eq4.2}). }
\beq
\frac{d\sigma_N}{\sigma_N\, d{\bf k}_1...d{\bf k}_m}
=\frac{ \int \left(\prod_{i=1}^N d{\bf y}_i\right)\, 
\int d{\bf b}\, \vert {\cal M}^2({\bf k}_1,\cdots ,{\bf k}_m; {\bf y}_1,\cdots ,{\bf y}_N)\vert
\rho({\bf y}_1...{\bf y}_N,{\bf b})}{\int \left(\prod_{i=1}^N d{\bf y}_i \right)\, d{\bf b}\, \rho({\bf y}_1...{\bf y}_N, {\bf b})}. \label{eq6.7}
\eeq
This is the form of the $m$-particle emission spectrum (\ref{eq2.3}) with a source probability distribution $\rho$ of the form (\ref{eq4.2}).
In section~\ref{sec2}, we have supplemented the structure of (\ref{eq6.7}) with a particularly
simple model for calculating ${\hat \sigma}= \vert {\cal M}^2\vert $. Here, we see that this structure arises from MPI theory 
once one treats $N$-parton GPDs in mean field approximation. 

We note that this framework allows one to constrain the parameter $B$ in the source density by data. Namely, 
$1/K_N = \int d{\bf b}\, \prod d{\bf y}_i\, \rho(\lbrace {\bf y}_i\rbrace, {\bf b}) = 1/ N (4\pi B)^{N-1}$, and therefore
$\sigma_{\rm eff} = K_2 = 8 \pi B$, 
\begin{equation}
	B=  2\,  {\rm GeV}^{-2}
	\qquad \longleftrightarrow \qquad \sigma_{\rm eff} \approx 20\,  {\rm mb}
	\label{eq6.8}
\end{equation}
Experimentally favored values for $\sigma_{\rm eff}$ lie in the range of $15 \pm 5$ mb for pp collisions at the LHC~\cite{Kuechler:2016wxp,Aaboud:2016dea,Gunnellini:2016noc} and for $p\bar{p}$ collisions at 
Tevatron~\cite{Abe:1997bp}. Somewhat higher values $\sigma_{\rm eff}$ of order 35--40 mb 
have been obtained in a mean field approach that does not include
other mechanisms for MPI enhancement~\cite{BDFS2,BS2}. This will prompt us in the next section to
scan values in the range $1\,  {\rm GeV}^{-2} < B < 4\,  {\rm GeV}^{-2}$.

 \section{Numerical results}
 \label{sec7}
 Whenever there are multiple partons in the final state, 
 QCD interference contributes to  the azimuthal anisotropies $v_n$ for both even (see sections~\ref{sec3} and ~\ref{sec4}) 
 and odd (see section~\ref{sec5}) harmonics. This raises the question how contributions from QCD interference 
 compare in size and signature to those of other physically conceivable mechanisms. 
 What is the typical signal size with which QCD  interference can contribute to $v_n$? And what is the
 expected $p_T$-, rapidity- and multiplicity-dependence of the effects discussed here? The present study is not sufficient
 to provide complete answers to these questions, but we summarize in this section what can be said from 
 parametric considerations and from first numerical results.
 
 Parametrically,  leading contributions to $v_n$, $n$ even, are $O\left(1/(N_c^2-1)^{1/2}\right)$ for the
 second order cumulants and $O\left(1/(N_c^2-1)^{3/4}\right)$ for the fourth-order cumulants. This $N_c$-dependence
 is multiplied by functions that grow $\propto \vert {\bf k}\vert^n$ for very small transverse momenta but that reach magnitudes of 
 $O(1)$ for sufficiently large transverse momenta, see, \emph{e.g.}, eq.~(\ref{eq3.14}).  For $N_c = 3$, signal sizes $v_n \sim 0.1 - 0.3$ 
 seem therefore conceivable. The signal size is roughly independent of the number of sources $N$. While our analysis of the multiplicity  
 dependence of $v_n$ in section~\ref{sec4} does not allow us to draw conclusions in the limit of large multiplicity,
 it points to the possibility that $v_n$  rises with multiplicity.   Also, the model described in section~\ref{sec2} leads naturally
 to an approximately flat rapidity dependence. Albeit not established on the quantitative level needed for decisive tests,
 the above-mentioned features agree at least qualitatively with trends in the data.
 
\begin{figure}[t]
\centering
\includegraphics[width=.8\textwidth]{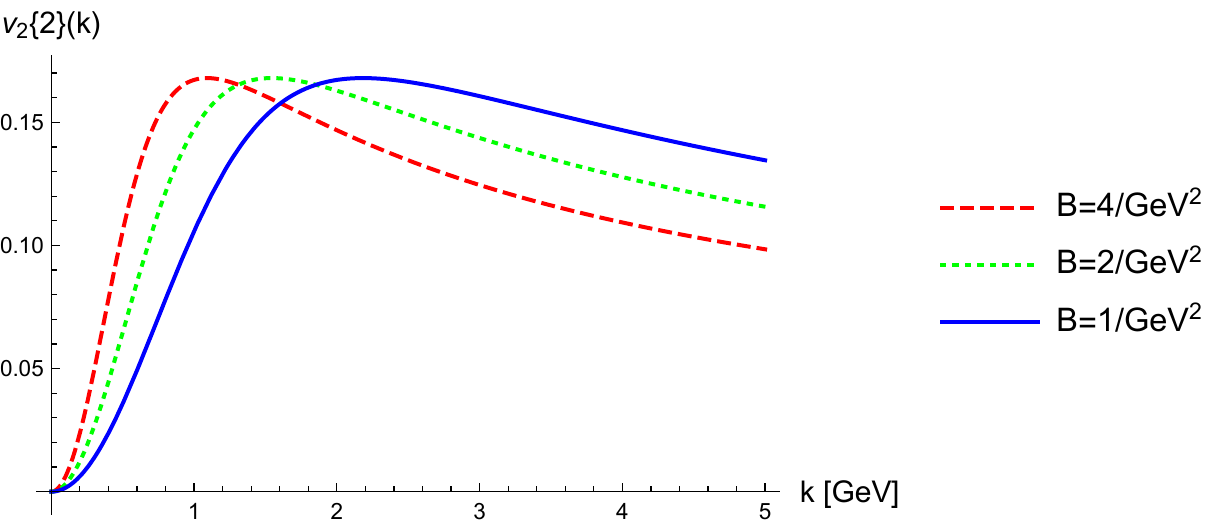} 
\caption{The 2nd order cumulant $v_2\lbrace 2\rbrace(k) = \sqrt{ v_2^2\lbrace 2\rbrace(k,k) }$ evaluated from 
(\ref{eq3.14})  for $F_{\rm corr}^{(2)} = 1$.
}
\label{fig5}
\end{figure}
 It is less clear whether also the transverse momentum dependence of the interference effects studied here can account 
 for the qualitative trends in the data. Amplitudes for the emission from different sources are expected to interfer only for 
 sufficiently small transverse momenta $k$ that do not resolve the separation $\Delta y$ between the emitters, $k < 1/\Delta y$. It is 
 therefore unclear whether QCD interference can contribute significantly to $v_n(k)$ in the multiple 
 GeV range where significant signal strength is observed. To address this question, we plot in Fig.~\ref{fig5} the 
 transverse momentum dependence of the second order cumulant for values of the source parameter $B$ favored
 by data on multi-parton interactions, see eq.~(\ref{eq6.8}). The main qualitative features of this second order cumulant
 are indeed consistent with main trends in the data, the signal strength increases up to transverse momenta of $~2$ GeV
 and a sizeable signal persists in the multi-GeV range. 
 
\begin{figure}[b]
\centering
\includegraphics[width=.8\textwidth]{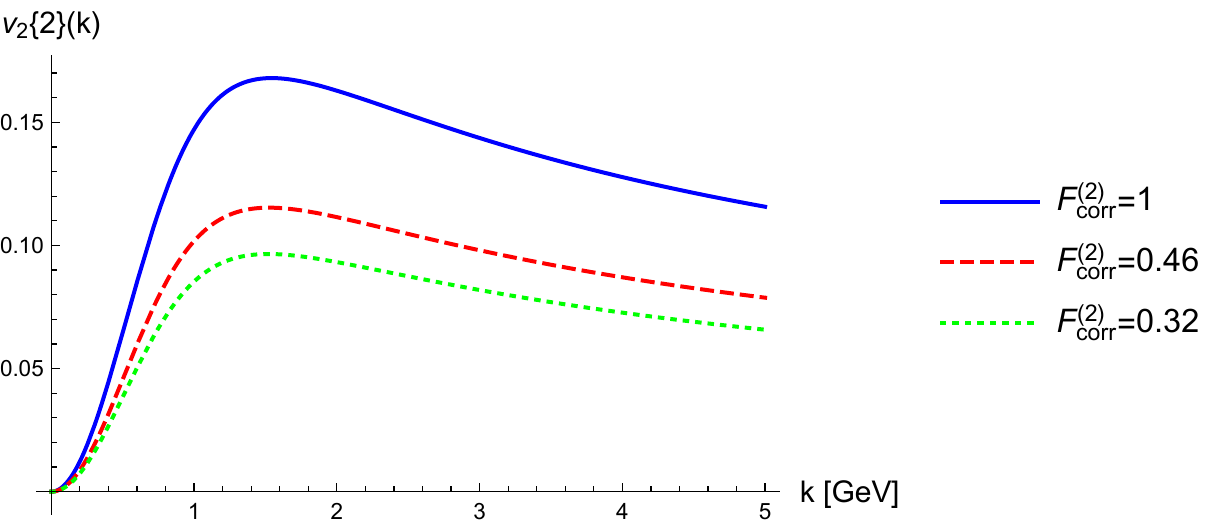} 
\caption{The 2nd order cumulant as in Fig.~\ref{fig5} for $B=2/{\rm GeV}^2$ but now compared to values of the
color correction factor $F_{\rm corr}^{(2)}$
= 0.45 (0.32) that correspond to an average of $\overline{m} =3$ (5) emitted particles per source. 
}
\label{fig6}
\end{figure}
As seen from Fig.~\ref{fig6}, the signal size obtained in the present formalism varies linearly with the value of the color
correction factor. 
The color correction factors used in Fig.~\ref{fig6} are for an average of $\bar m = 3$ or $5$ emitted particles per source, but a 
more extreme choice $\bar m = 20$ that may be at the upper end of what is phenomenologically viable would reduce the signal by
$F_{\rm corr}^{(2)} = 0.1$. 
In comparison to these uncertainties, the $O(1/(N_c^2-1))$- corrections in the denominator of (\ref{eq3.14})
turn out to be negligible.

\begin{figure}[t]
\centering
\includegraphics[width=.8\textwidth]{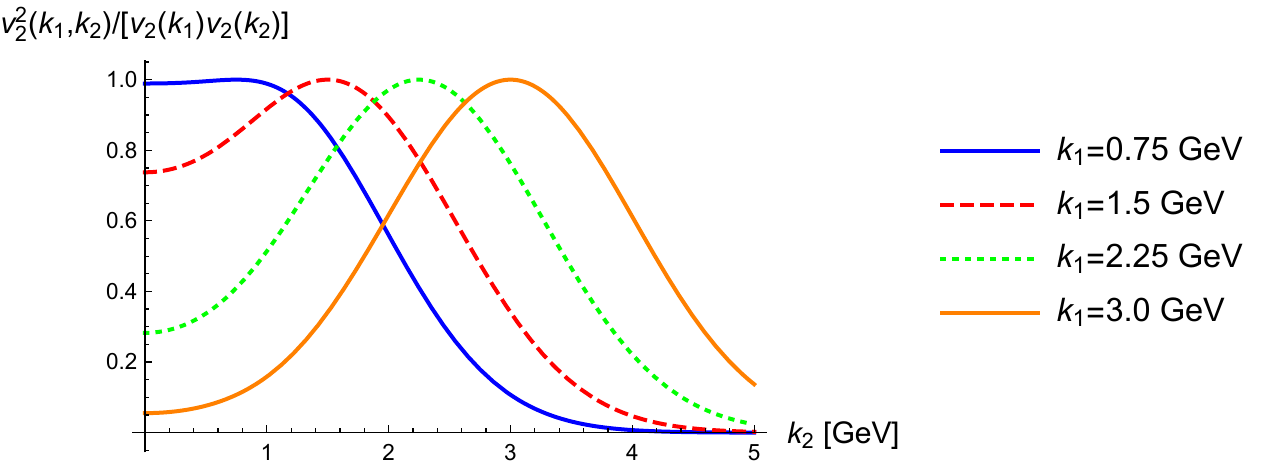} 
\includegraphics[width=.8\textwidth]{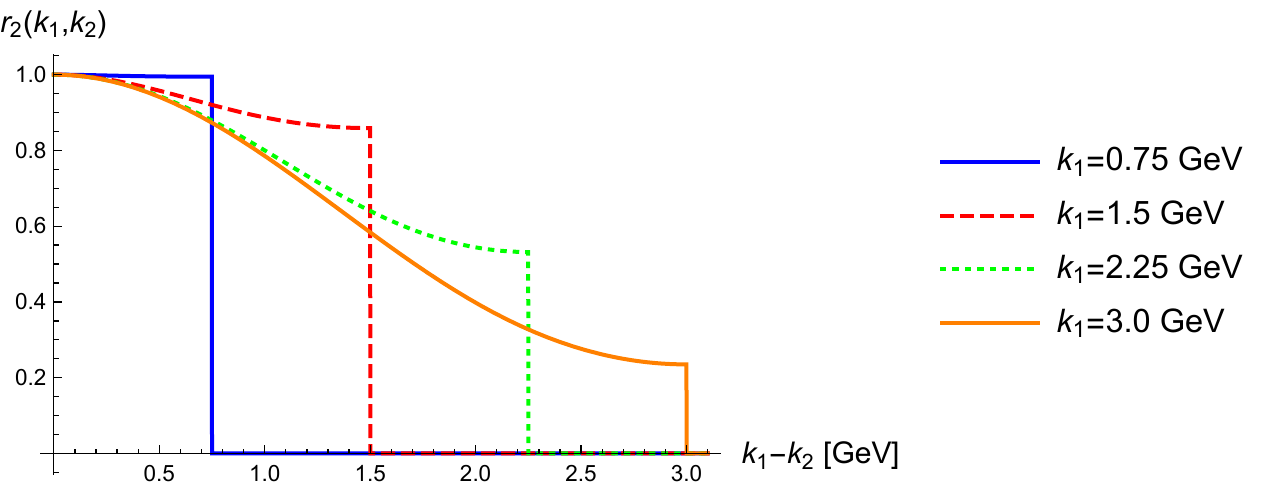} 
\caption{Upper panel: Illustration of factorization breaking $v_2^2\lbrace{2}\rbrace(k_1,k_2)  \not= v_2\lbrace{2}\rbrace(k_1)\, v_2\lbrace{2}\rbrace(k_2)$. Lower panel: The corresponding 
factorization ratio $r_2(k_1,k_2)$ as defined in equation (\ref{eq7.0}), plotted for selected `trigger' momenta $k_1$ against the relative momentum difference $k_1-k_2$.
}
\label{fig7}
\end{figure}
%
We next point to the fact that $v_2^2\lbrace{2}\rbrace(k_1,k_2)$ does not factorize. To this end,
we show in the upper panel of Fig.~\ref{fig7} the ratio  $v_2^2\lbrace{2}\rbrace(k_1,k_2)  / v_2\lbrace{2}\rbrace(k_1)\, v_2\lbrace{2}\rbrace(k_2)$. 
One sees that in the transverse momentum range up to $\sim 1.5$ GeV, deviations from factorization are at most $\sim 20\%$.
For higher transverse momenta, however, particle emission starts to decorrelate as soon as the difference between the
transverse momenta of the two emittees is larger than 1 - 2 GeV. For the experimentally measured hadronic momentum correlations, such deviations
from factorization are often characterized in terms of the factorization ratio
\begin{equation}
	r_2(k_1,k_2) = \frac{v_2\lbrace{2}\rbrace(k_1,k_2)}{\sqrt{v_2\lbrace{2}\rbrace(k_1)\, v_2\lbrace{2}\rbrace(k_2)}}\, ,
	\label{eq7.0}
\end{equation}
where one of the two momenta, say $k_1$, is regarded as trigger, and the other is the associate particle satisfying $k_2 < k_1$. The corresponding quantity is
plotted in the lower panel of Fig.~\ref{fig7} as a function of $k_1-k_2$. For pPb and PbPb collisions, the factorization ratio $r_2(k_1,k_2)$ has been measured
at LHC~\cite{Khachatryan:2015oea} and it is in agreement with fluid dynamic simulations~\cite{Shen:2015qta}, see also Ref.~\cite{Gardim:2012im} for earlier
discussions of the relation of $ r_2(k_1,k_2)$ to fluid dynamics. Fig.~\ref{fig7} shows features qualitatively similar
to those reported in Refs.~\cite{Khachatryan:2015oea,Shen:2015qta} in that $r_2$ decreases with increasing $k_1-k_2$ and that this decrease is more pronounced 
for increasing trigger $k_1$. Clearly, the present calculation is for pp collisions, and it shows a correlation on parton level. Hadronization may be expected to affect
the correlation $r_2$ significantly as it tends to smear and soften transverse momentum distributions. Given these substantial differences between the results of 
Refs.~\cite{Khachatryan:2015oea,Shen:2015qta} and the present calculation, we refrain from a quantitative comparison.

\begin{figure}[b]
\centering
\includegraphics[width=.55\textwidth]{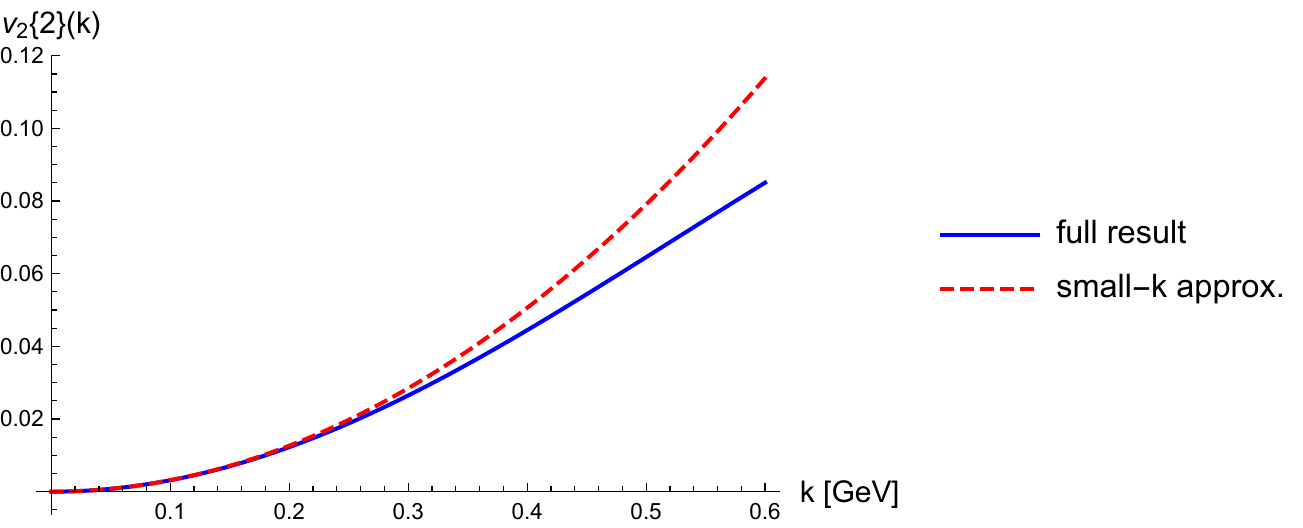} 
\includegraphics[width=.44\textwidth]{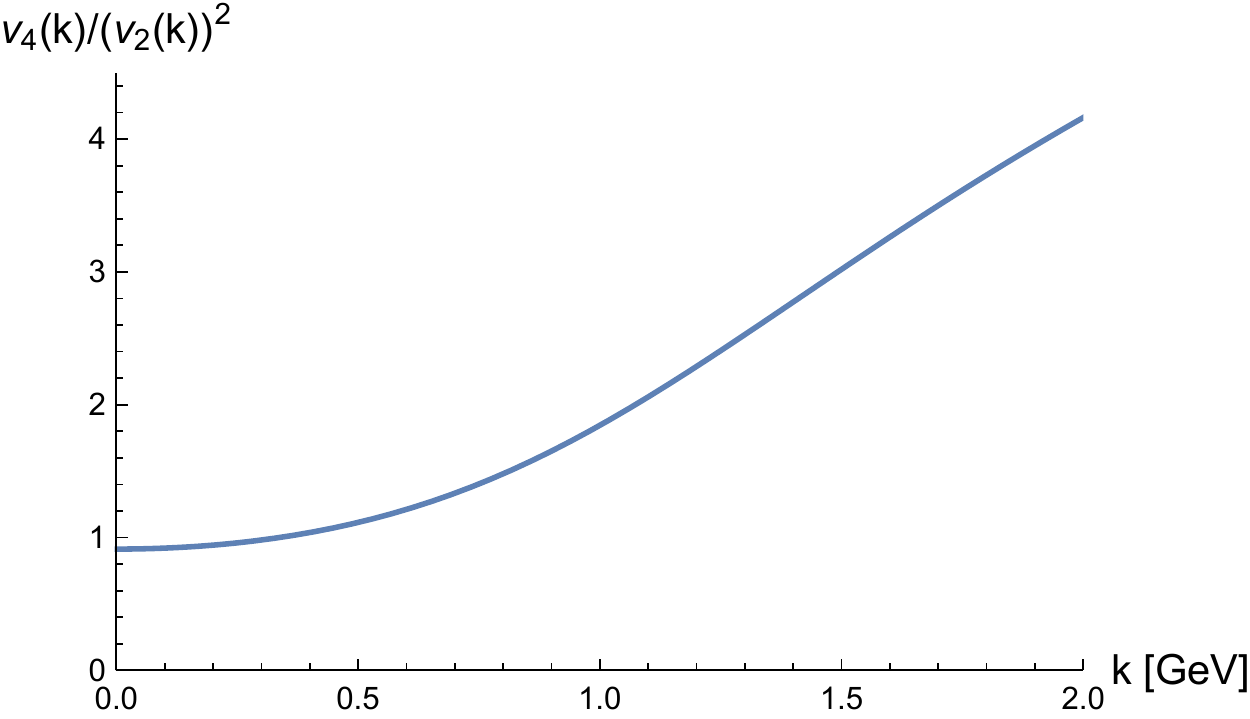} 
\caption{LHS: Comparison of the second cumulant $v_2\lbrace{2}\rbrace(k)$ calculated from (\ref{eq3.14}),
and the approximation (\ref{eq4.13}) for small transverse momentum. RHS: The ratio of $v_4\lbrace{2}\rbrace(k)$
and $(v_2\lbrace{2}\rbrace(k))^2$ that is known to be constant in the limit of small transverse momentum.
Both calculations are for $B = 2/GeV^2$ and $F_{\rm corr}^{(2)} = 1$. 
}
\label{fig8}
\end{figure}
%
 In section~\ref{sec4}, we have limited one part of our discussion to the small-$k$ approximation in which many
 expressions simplify. The left hand side of Fig.~\ref{fig8} indicates the range of validity of this approximation.
As discussed in section~\ref{sec5a}, it is a structural property of the small-$k$ approximation that 
 \begin{equation}
 	v_4\lbrace{2}\rbrace(k) = {\rm const.} \; v_2\lbrace{2}\rbrace(k)\, v_2\lbrace{2}\rbrace(k)\, .
 	\label{eq7.1}
 \end{equation}
 Similar relations between different harmonic coefficients are known to arise in fluid dynamic models as a consequence of
 mode-mode coupling~\cite{Teaney:2012ke,Floerchinger:2013tya,Yan:2015jma}. For the specific relation (\ref{eq7.1}),
 it was noted already in Ref.~\cite{Borghini:2005kd} that one expects for large transverse momentum a proportionality
 constant 1/2. It is therefore interesting to note that non-trivial relations between $v_4$ and $v_2$ may also arise
 from mechanisms that do not invoke interactions in the final state. The right hand side of Fig.~\ref{fig8} illustrates 
 this point further. For small $k$, $v_4$ is seen to be approximately proportional to $(v_2)^2$ with a proportionality
 factor of order unity. 
 
So far, the figures shown in this section were calculated to leading order in $1/(N_c^2-1)$ from the second order 
cumulant~(\ref{eq3.14}). To the next order in $1/(N_c^2-1)$, the second order cumulant $v_2\lbrace{2}\rbrace(k)$ receives an
additive correction, see  eq.~(\ref{eq4.13}).  However, as explained in section~\ref{sec4}, this is an expansion in powers 
of $m^2/(N_c^2-1)$ that does not converge for large multiplicities. An analogous statement 
applies to the 4-th order cumulant $v_2\lbrace{4}\rbrace(k)$ in equation~(\ref{eq4.16}) for which the 
contribution subleading in $1/(N_c^2-1)$ grows with $m^2$. Despite this caveat, we proceed here with a 
numerical exploration of $v_2\lbrace{4}\rbrace(k)$. To this end, we focus on the term $\propto m^2$ in (\ref{eq4.16}) 
that dominates for large multiplicity. Rather than going to the small-$k$-limit (\ref{eq4.18}), we keep the color correction 
factor in (\ref{eq4.16}), 
and we re-establish the large-$k$-behavior by replacing $B k^2$ with the integral over Bessel functions from which
it was obtained in a small-$k$-approximation, 
\begin{equation}
 	v_2^4\lbrace 4\rbrace(k) \approx  \frac{\left( F^{(2)}_{\rm corr}(N,m) \right)^3 }{(N_c^2-1)^{3}}  2\, d \, m^2\, 
 	\left(  \int_\rho
	 J_2\left(k_1 \Delta y \right)\, J_2\left(k_2 \Delta y \right)\, \right)^2\, .
	 \label{eq7.2}
\end{equation}
Here,  $d$ denotes a suppression factor that arises from integrating out one of the three dipoles in the term $\propto F_{\rm corr}^{(6)} =
\bigl( F^{(2)}_{\rm corr} \bigr)^3$ in (\ref{eq4.1}). The factor $d$ equals unity for $B\, k^2 \ll 1$ but it depends on the
vertex function and can be smaller than unity $d < 1$. For any given vertex function $f(k)$, it can be calculated in close
analogy to the suppression factor $a$ in (\ref{eq4.19}) and one has $d \sim O(a^2)$.

\begin{figure}[h]
\centering
\includegraphics[width=1.\textwidth]{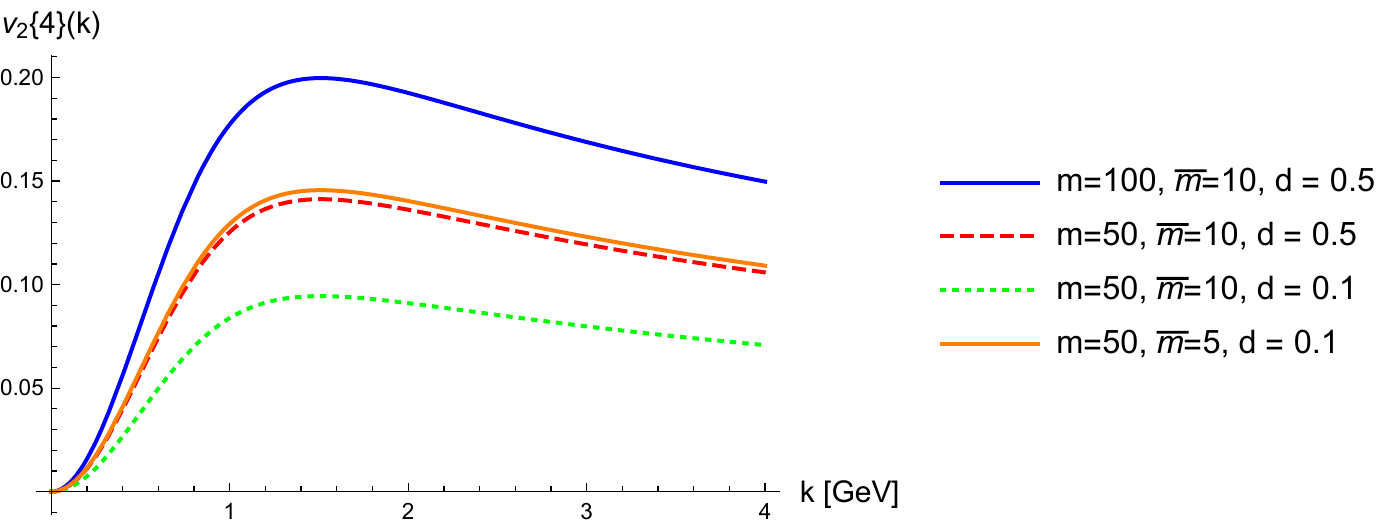} 
\caption{Approximation (\ref{eq7.2}) of the 4-th order cumulant $v_2^4\lbrace 4\rbrace(k)$ for 
different multiplicities $m$, average multiplicities $\overline{m}$ per source and values of the suppression factor $d$,
obtained from intergrating out one dipole radiation factor.
}
\label{fig9}
\end{figure}

Fig.~\ref{fig9} shows the fourth root of (\ref{eq7.2}) for multiplicities $m$ reached in high-multiplicity 
proton-proton collisions at the LHC, and for a number of sources $N$ comparable to the number of MPIs invoked in MC 
simulations of the underlying event of such pp collisions. We also vary the suppression factor $d$ 
over ranges that are easily obtained from (\ref{eq4.19}) and $d \simeq a^2$. While the results in Fig.~\ref{fig9}
fall short of a quantitative determination of $v_2^4\lbrace 4\rbrace(k)$, they support the qualitative statement that 
the size and shape of the 4-th order cumulant $v_2\lbrace 4\rbrace(k)$ resulting from QCD 
interference may be comparable to the size and shape of the second order cumulants $v_2\lbrace 2\rbrace(k)$
within the parameter range realized in high-multiplicity proton-proton collisions.
%

 \section{Conclusion}
Whenever multiple partons are produced in a hadronic collision, there are interference effects whose size varies with the 
momentum distribution and the colors of the outgoing partons. These QCD interference effects do not depend on 
$\alpha_s$. In this sense, they are part of a ``no-interaction'' baseline for $v_n$ that needs to be controlled prior 
to discussing non-linear dynamics in the incoming hadronic wavefunctions or rescattering and fluid dynamization 
in the final state.~\footnote{This $\alpha_s$-independence distinguishes the QCD interference effects discussed here
from other conceivable interference effects. For instance, in Ref.\cite{Gyulassy:2014cfa}, an initial state multiple scattering bremsstrahlung 
picture with recoil effects was explored as a source for finite harmonic flow coefficients in pA-collisions. This is based on
an LPM interference effect that is clearly unrelated to the interference effects studied in the present paper. The effects 
discussed in Ref.\cite{Gyulassy:2014cfa} are therefore expected to die out in the dilute limit when secondary scattering becomes unimportant 
while the effects included in our no-interaction baseline persist. }

Here we have used a simple model of multi-particle
production (see section~\ref{sec2}) to estimate how these QCD interference effects can contribute to the azimuthal
anisotropy coefficients $v_n$ measured with second and higher order cumulants in pp collisions. We have pointed 
out (section~\ref{sec6}) that this model can be realized in the mean field theory of multi-parton interactions (MPIs),
with MPIs representing the ``sources'' of individual parton-parton collisions and radiated gluons corresponding to 
radiation associated with the MPIs. 

Our calculations establish that the contribution of QCD interference to the anisotropy coefficients $v_n$, $n$ even, 
persists unattenuated for an increasing number of sources, that it can increase with increasing multiplicity,
and that it persists in higher order cumulants. We have further shown that odd harmonic anisotropy coefficients arise
due to the non-abelianness of the QCD interference pattern\footnote{The mechanism via which odd harmonics
arise is parametrically different: the contributions are $O(1/N)$, but they can be enhanced by powers of $m$
and work is needed to estimate them in the phenomenologically relevant parameter range.}.
In section~\ref{sec7}, we have supplemented these
structural results about how the QCD interference impacts anisotropy coefficients with a first numerical exploration. 
Both, the order of magnitude of the calculated $v_n$, as well as the shape of their transverse momentum dependence was found
to be of the order of magnitude and shape of the signals observed in $pp$ collisions at the LHC. Given the simple nature
of the model in section~\ref{sec2} (schematic treatment of particle production, absence of hadronization, etc.), these
qualitative commonalities between model and data must not be over-interpreted. However, they make it certainly 
conceivable that  the no-interaction baseline including QCD interference effects can make a sizeable if not dominant
contribution to the measured $v_n$ coefficients in pp collisions. 

This leads naturally to the question whether and to what extent QCD interference effects could contribute also to
the anisotropy coefficients measured in AA collisions. Here, however, marked differences need to be considered. First,
jet quenching provides unambiguous experimental evidence for significant final state rescattering which
is an $\alpha_s$-dependent  microscopic mechanism that underlies hydrodynamization and that
can translate spatial gradients into momentum anisotropies. Given the strength of jet quenching signals, 
it is thus inconceivable that QCD interference can account 
for the totality of the observed $v_n$ signals in AA. Second, final state scattering can destroy coherence
and the resulting interference. 
While QCD interference clearly shifts the no-interaction baseline for $v_n$, it is therefore questionable that effects of interactions 
contribute additively on top of this baseline. 
Rather than seeking a common explanation of $v_n$ across system size, it may therefore also be instructive to explore the 
opposite hypothesis, namely that   the physics mechanisms underlying the $v_n$-signals in pp and AA are qualitatively 
different, being dominated by QCD interference for pp while being dominated by fluid dynamics for AA. The  
similarities in the  experimental signal of pp, pA and AA may be viewed as disfavouring this hypothesis, but 
the qualitatively different evidence for final state rescattering in both systems suggests that different mechanisms are 
at work in pp and AA and these may therefore also be at the origin of the measured  $v_n$. The present work adds to this 
discussion only by illustrating that QCD interference may account for the order of magnitude and main qualitative features 
of $v_n$ signals in pp. As far as pA collisions are concerned, the absence of experimental evidence for significant final
state scattering supports the idea that the qualitative conclusions about QCD interference drawn here for pp
carry over to pA. 

We have aimed at controlling in this paper interference effects between an arbitrary number $m$ of gluons emitted
from an arbitrary number $N$ of sources. This was achieved within an expansion in powers of
$1/(N_c^2-1)$ and to leading order in $N$ by resumming the effects of an arbitrary number of diagonal
gluons in color correction factors. We hope that these results can be useful also for the discussion of other models,
such as models based on saturation physics\footnote{In the notation of section~\ref{sec2}, the initial
transverse density of sources is $\sim N/B$. Since the $v_2\lbrace 2\rbrace$ and 
$v_2\lbrace 4\rbrace$ calculated in section~\ref{sec4} are $N$-independent and finite to leading order in $1/N$, 
a high or saturated initial parton density is 
not a prerequisite for the effects discussed here while it is the basis of calculations in the so-called CGC-formalism.} that
encompass QCD interference effects and with which our calculation agrees to lowest order, see eq.~(\ref{eq3.5}). 
Within these models, the importance of QCD interference effects has been emphasized repeatedly (see e.g. Ref.~\cite{McLerran:2015sva}),
but they are not studied in isolation and are difficult to disentangle from effects of finite partonic density. Indeed, due to the greater complexity
of calculations in these models, correlation functions have not been analyzed to the same level of detail as in the present manuscript, and we
are not aware of similar parametric statements about the leading $N$-independence and $1/(N_c^2-1)$-dependence of higher order cumulants.

Future developments may also better relate the results reported here to recent efforts of improving MC simulations of the underlying event in TeV-scale
proton-proton collisions. On the one hand, while all modern multi-purpose MC  event generators model the underlying event in terms of multi-parton interactions, the QCD
interference effects discussed here are not included in these simulations. On the other hand, there are efforts to go beyond an essentially incoherent
superposition of MPIs supplemented with conservation laws, e.g. by modeling effects of overlapping strings and studying whether these could give rise
to signatures of collectivity~\cite{Bierlich:2014xba,Bierlich:2016vgw,Bierlich:2017vhg}. For earlier works, see e.g. Ref.~\cite{Braun:2012kn} and
approaches based on pomeron dynamics~\cite{Abramovsky:1980yc,Abramovsky:1988zh}. It would be interesting to understand how QCD interference 
can be included in MC simulations and how this compares e.g. to effects of overlapping strings or other models.

Finally, within the set-up of this manuscript, the expansion in $O\left(1/(N_c^2-1)\right)$ analyzed here provided first qualitative insights, 
but we also discussed its limitations. In particular, a more systematic 
control over $1/N$-suppressed terms may allow for more quantitative statements in the range of multiplicity and number of sources that are phenomenologically interesting,
and it would give access to odd harmonic anisotropy coefficients. Moreover, to gain better control in the phenomenologically interesting range of multiplicities, 
one would ideally like to resum all contributions that come with powers in $m^2/(N_c^2-1)$. 
We expect that such further advances are possible.

\appendix

\section{Multi-gluon emission for $N=m=3$}
\label{appa}
In this appendix, we calculate the cross section for producing $m=3$ gluons from $N=3$ sources. 
This will illustrate several statements that we have generalized to arbitrary $N$ and $m$ in the main text. 

For $N$ sources and $m$ emitted gluons, 
there are $N^{2m}$ different diagrams, since each of the $m$ gluons can be attached to any of the $N$ sources in the amplitude, 
and to any of the $N$ sources in the complex conjugate amplitude. For $N=m=3$, these $N^{2m} = 3^6$ diagrammatic contributions 
can be classified as follows:
  \begin{enumerate}[I]
  \item 3 diagonal gluons ($27$ diagrams)
   \item 2 diagonal gluons and 1 off-diagonal gluon ($6 \times 27$ diagrams)
  \item 1 diagonal gluon and 2 off-diagonal gluons   ($12 \times 27$ diagrams)
  \item 3 off-diagonal gluons ($8 \times 27$ diagrams)
  \end{enumerate}
 To illustrate how to enumerate these diagrams, consider, \emph{e.g.}, case II:
 there are 3 choices for the source to which each of the diagonal gluon can be attached. For the off-diagonal gluon, one has 3 choices for the
 source in the amplitude times 2 choices in the complex conjugate amplitude. As exactly one of the three gluons is off-diagonal, one has also 3 choices for selecting the off-diagonal gluon amongst all three gluons. Combining these
 factors leads to $3\times3\times (3\times 2) \times 3 = 6 \times 27$ different diagrams.
 %
\begin{figure}[t]
\centering
\includegraphics[width=1.0\textwidth]{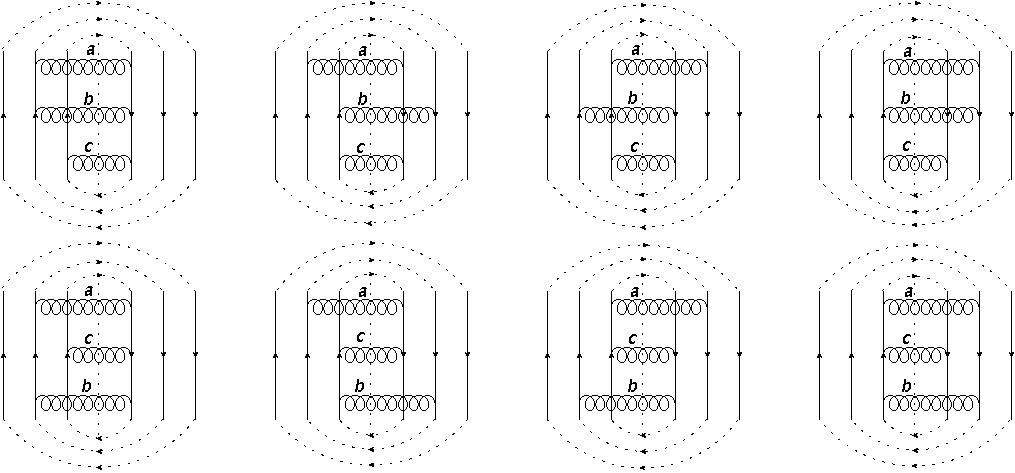} 
\caption{Some of the contributions with two off-diagonal and one diagonal gluon that contribute to the $k=3$-gluon radiation spectrum from $N=3$ sources. The color factor (\ref{eqa.2}) for diagrams in the top row is 
twice as large as the color factor (\ref{eqa.3}) for those in the bottom row. 
}
\label{fig5A}
\end{figure}

  We turn now to the calculation of the different contributing cases:
  For three diagonal gluons radiated off three sources (\underline{\bf case I}), all $27$ diagrams contributing to the 
 3-gluon radiation cross section have the same color factor $(N_c^2 -1)^3 N_c^3$. As phase factors 
 cancel for diagonal gluons, the contribution to ${\hat \sigma}$ is
  \begin{equation}
        {\hat \sigma}_{I}\left(\lbrace {\bf k_1}, {\bf k}_2, {\bf k}_3\rbrace,\lbrace {\bf y}_1, {\bf y}_2, {\bf y}_3\rbrace\right)
  	\propto (N_c^2 -1)^3 N_c^3 \left|\vec{f}({\bf k}_1)\right|^2\, \left|\vec{f}({\bf k}_2)\right|^2 \, \left|\vec{f}({\bf k}_3)\right|^2\, 27\, .
  	\label{eqa.1}
  \end{equation}
  One also checks easily that the color factors of all diagrams with only 1 off-diagonal gluon (\underline{\bf case II}) vanish,
  \begin{equation}
  	{\hat \sigma}_{II} = 0\, .
  	\label{eqa.2}
  \end{equation}
\underline{\bf Case III} includes $8 \times 27$ diagrams for which the two off-diagonal gluons connect to three sources; these have vanishing 
color factors. In addition, there are  $4\times 27$ diagrams for which the two off-diagonal gluons connect to two sources. 
  For those, we denote with $a$ and $b$ the colors of the off-diagonal gluons  and with $c$ the color of the diagonal one. 
  If the diagonal gluon links to the source that is not touched
  by the off-diagonal gluons (case not shown in Fig.~\ref{fig5A}), then the resulting color factor is 
  ${\rm Tr}\left[T^c T^c \right] {\rm Tr}\left[T^a T^b \right]
  {\rm Tr}\left[T^a T^b \right] = N_c^3 (N_c^2 - 1)^2$. If the diagonal gluon touches instead a source that is also touched
  by the off-diagonal gluons, but is not sandwiched between the off-diagonal gluons (see, \emph{e.g.}, top row of Fig.~\ref{fig5A}),  
   then
  \begin{equation}
  {\rm Tr}\left[1 \right] {\rm Tr}\left[T^c T^cT^a T^b \right]  {\rm Tr}\left[T^a T^b \right] = N_c^3 (N_c^2 - 1)^2\, .
  \label{eqa.3}
  \end{equation} 
  However, for those diagrams for which the diagonal gluon is sandwiched between the off-diagonal ones and for which the diagonal
  and off-diagonal ones have one source in common (see second row of Fig.~\ref{fig5A}), one finds  
  \begin{equation}
     {\rm Tr}\left[1 \right] {\rm Tr}\left[T^c T^a T^c T^b \right]  {\rm Tr}\left[T^a T^b \right] 
     = \frac{1}{2} N_c^3 (N_c^2 - 1)^2\, .
     \label{eqa.4}
  \end{equation}
  
The total contribution of the diagrams of case III to the 3-gluon
emission cross section takes then the form
 \begin{eqnarray}
         &&{\hat \sigma}^{(ord)}_{III}\left(\lbrace {\bf k_1}, {\bf k}_2, {\bf k}_3\rbrace,\lbrace {\bf y}_1, {\bf y}_2, {\bf y}_3\rbrace\right)
  	\propto (N_c^2 -1)^3 N_c^3 
  	\left|\vec{f}({\bf k}_1)\right|^2\, \left|\vec{f}({\bf k}_2)\right|^2 \, \left|\vec{f}({\bf k}_3)\right|^2
  		\nonumber \\
  		&& \qquad \times \Big\lbrace \frac{3}{(N_c^2-1)} \sum_{(ij)} 4 \,
  	{\cos\left({\bf k}_1.\Delta {\bf y}_{ij} \right)}  { \cos\left( {\bf k}_2.\Delta {\bf y}_{ij}  \right) }
  	        \nonumber \\
  	        && \qquad \qquad \qquad + \frac{2}{(N_c^2-1)} \sum_{(ij)} 4 \, 
  	{\cos\left({\bf k}_1.\Delta {\bf y}_{ij} \right)}  { \cos\left( {\bf k}_3.\Delta {\bf y}_{ij}  \right) }
  	        \nonumber \\
  	        && \qquad \qquad \qquad + \frac{3}{(N_c^2-1)} \sum_{(ij)} 4 \, 
  	{\cos\left({\bf k}_2.\Delta {\bf y}_{ij} \right)}  { \cos\left( {\bf k}_3.\Delta {\bf y}_{ij}  \right)}  \Big\rbrace\, .
  	\label{eqa.5}
 \end{eqnarray}
Here, we use the subscript {\it (ord)} to indicate that the momenta ${\bf k}_1$, ${\bf k}_2$ and ${\bf k}_3$
are ordered from top to bottom in the emission diagrams. The color trace (\ref{eqa.4}) appears only for 2 of 
the 3 diagrams for which the diagonal gluon carries ${\bf k}_2$, and this reduces the prefactor of the 
second term in (\ref{eqa.5}) to $2\times 1/2 + 1 = 2$.

Following (\ref{eq2.4}), we randomize the external momenta to obtain
 \begin{eqnarray}
         &&{\hat \sigma}_{III}\left(\lbrace {\bf k_1}, {\bf k}_2, {\bf k}_3\rbrace,\lbrace {\bf y}_1, {\bf y}_2, {\bf y}_3\rbrace\right)
  	\propto (N_c^2 -1)^3 N_c^3 
  	\left|\vec{f}({\bf k}_1)\right|^2\, \left|\vec{f}({\bf k}_2)\right|^2 \, \left|\vec{f}({\bf k}_3)\right|^2
  		\nonumber \\
  		&& \qquad \qquad \times
  	\frac{8}{3(N_c^2-1)} \frac{N}{3} \sum_{(ab)} \sum_{(ij)} 4 \, 
  	{\cos\left({\bf k}_a.\Delta {\bf y}_{ij} \right)}  { \cos\left( {\bf k}_b.\Delta {\bf y}_{ij}   \right)}
  	\, .
  	\label{eqa.6}
 \end{eqnarray}
 The prefactor $8/9$ in this expression is consistent with the factor $F^{(2)}_{\rm corr}(N,m)$, obtained for
 $N=m=3$ from eq.~(\ref{eq3.10}).
 
 
\begin{figure}[t]
\centering
\includegraphics[width=1.0\textwidth]{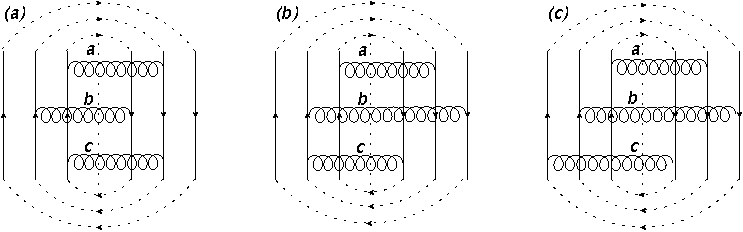} 
\caption{Examples for diagrammatic contributions with 3 off-diagonal gluons that link between one (a), two (b)
and three (c) different pairs of sources. Contributions of type (b) vanish while contributions of type (a) and (c)
have color traces that differ by a factor $(N_c^2-1)/4$, see text for more details.
}
\label{fig6A}
\end{figure}
 
  \underline{\bf Case IV} concerns diagrams with 3 off-diagonal gluons. 
 Examples for such contributions are depicted in Fig.~\ref{fig6A}. If all three gluons are linked to the same pair of 
 sources (see, \emph{e.g.}, Fig.~\ref{fig6A}(a)), we find for the symmetrized expression
 \begin{eqnarray}
         &&{\hat \sigma}_{IVa}\left(\lbrace {\bf k_1}, {\bf k}_2, {\bf k}_3\rbrace,\lbrace {\bf y}_1, {\bf y}_2, {\bf y}_3\rbrace\right)
  	\propto (N_c^2 -1)^3 N_c^3 
  	\left|\vec{f}({\bf k}_1)\right|^2\, \left|\vec{f}({\bf k}_2)\right|^2 \, \left|\vec{f}({\bf k}_3)\right|^2
  		\nonumber \\
  		&& \qquad \qquad \times
  	\frac{1}{4(N_c^2-1)} \sum_{(ij)} 8 \, 
  	{\cos\left({\bf k}_1.\Delta {\bf y}_{ij} \right)}  { \cos\left( {\bf k}_2.\Delta {\bf y}_{ij}   \right)}
  	{ \cos\left( {\bf k}_3.\Delta {\bf y}_{ij}   \right)}
  	\, .
  	\label{eqa.7}
 \end{eqnarray}
 If the 3 off-diagonal gluons are linked to exactly two pairs of sources (see, \emph{e.g.}, Fig.~\ref{fig6A}(b)), then one of 
 the color traces vanishes,
 \begin{equation}
 	{\hat \sigma}_{IVb} = 0\, .
 	\label{eqa.8}
 \end{equation}
 If the three off-diagonal gluons are linked to exactly three pairs of sources (see, \emph{e.g.}, Fig.~\ref{fig6A}(c)), one finds
 \begin{eqnarray}
         &&M_{IVc}\left(\lbrace {\bf k_1}, {\bf k}_2, {\bf k}_3\rbrace,\lbrace {\bf y}_1, {\bf y}_2, {\bf y}_3\rbrace\right)
  	\propto (N_c^2 -1)^3 N_c^3 
  	\left|\vec{f}({\bf k}_1)\right|^2\, \left|\vec{f}({\bf k}_2)\right|^2 \, \left|\vec{f}({\bf k}_3)\right|^2
  		\nonumber \\
  		&& \qquad \qquad \times
  	\frac{1}{(N_c^2-1)^2} \sum_{(a,b,c)} 8 \, 
  	{\cos\left({\bf k}_a.\Delta {\bf y}_{12} \right)}  { \cos\left( {\bf k}_b.\Delta {\bf y}_{23}   \right)}
  	{ \cos\left( {\bf k}_c.\Delta {\bf y}_{31}   \right)}
  	\, .
  	\label{eqa.9}
 \end{eqnarray}
 Parametrically, contributions to (\ref{eqa.9}) are suppressed by an extra factor $(N_c^2-1)$ compared to (\ref{eqa.7}).
 The prefactors of these expressions can be understood by noting that contributions to (\ref{eqa.7})
 (see Fig.~\ref{fig6A}(a)) have a color factor
 \begin{equation}
 	{\rm Tr}\left[ \mathbb{1} \right]\, {\rm Tr}\left[ T^a T^c T^b\right]\,  {\rm Tr}\left[  T^a T^b T^c \right] = \frac{1}{4} N_c^3 (N_c^2-1)^2\, ,
 	\label{eqa.10}
 \end{equation}
 while contributions to (\ref{eqa.9}) (see Fig.~\ref{fig6A}(c)) have a color trace 
 \begin{equation}
 	{\rm Tr}\left[ T^c T^b \right]\, {\rm Tr}\left[ T^b T^a \right]\,  {\rm Tr}\left[  T^a  T^c \right] =  N_c^3 (N_c^2-1)\, .
 	\label{eqa.11}
 \end{equation}
 %

\section{Multi-gluon emission for $N=m=4$}
\label{appb}
Here, we give details of the calculation of $m=4$ gluons emitted from $N=4$ sources. We consider a total
of $N^{2m} = 4^8$ diagrams, classified as
  \begin{enumerate}[I]
  \item 4 diagonal gluons ($4^4$ diagrams)
   \item 3 diagonal gluons and 1 off-diagonal gluon ($12 \times 4^4$ diagrams)
    \item 2 diagonal gluons and 2 off-diagonal gluon ($2 \times 3^3 \times 4^4$ diagrams)  
  \item 1 diagonal gluon and 3 off-diagonal gluons   ($4 \times 3^3 \times 4^4$ diagrams)
  \item 4 off-diagonal gluons ($3^4 \times 4^4$ diagrams)
  \end{enumerate}
 The \underline{\bf cases I and II} are trivial: diagrams with only diagonal gluons are counted with color factor 
 $N_c^4 (N_c^2-1)^4$ and diagrams with exactly one off-diagonal gluon have a vanishing color trace, 
  \begin{eqnarray}
        {\hat \sigma}_{I}
  	&\propto& (N_c^2 -1)^4 N_c^4 \left|\vec{f}({\bf k}_1)\right|^2\, \left|\vec{f}({\bf k}_2)\right|^2 \, 
  	\left|\vec{f}({\bf k}_3)\right|^2\, \left|\vec{f}({\bf k}_4)\right|^2\, 4^4\, ,
  	\label{eqb.1}\\
  	{\hat \sigma}_{II} &=& 0\, . 
  \end{eqnarray}
  For the three other cases, however, qualitatively novel features arise as we discuss now. 
\begin{figure}[t]
\centering
\includegraphics[width=1.0\textwidth]{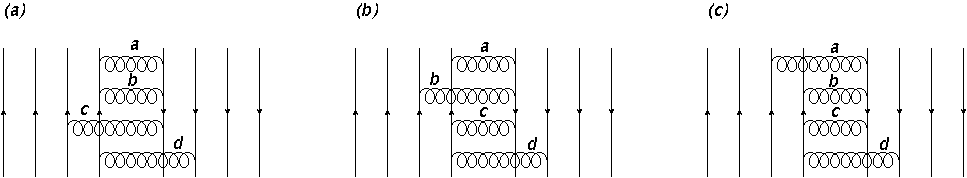} 
\caption{Examples of diagrammatic contributions with 2 off-diagonal and 2 diagonal gluons that are 
all connected to the same pair of sources. The three diagrams differ by the number of 0, 1 and 2 
diagonal gluons placed in between the off-diagonal gluons, and this changes the color trace, see text for
details.
}
\label{fig7A}
\end{figure}
  %
  \newline
  \underline{\bf Case III} includes diagrams with two off-diagonal gluons linked 
  to 2, 3 or 4 sources. Only the first of these possibilities gives non-vanishing contributions.
  It includes $9 \times 4^4$ diagrams.   Fig.~\ref{fig7A} depicts some of them. One checks easily 
  that these contributions have different color traces, namely
  \begin{equation}
  {\rm Tr}\left[ \mathbb{1} \right]\,  {\rm Tr}\left[ \mathbb{1} \right]\,  {\rm Tr}[T^a T^b T^c T^d T^b T^a]\,  {\rm Tr}[T^c T^d] = N_c^4 (N_c^2 -1)^3 \, , 
  \label{eqb.3}
  \end{equation}
  if in between the off-diagonal exchanges there is no diagonal gluon exchange (see, \emph{e.g.}, Fig.~\ref{fig7A}(a)),
   \begin{equation}
  {\rm Tr}\left[ \mathbb{1} \right]\,  {\rm Tr}\left[ \mathbb{1} \right]\,  {\rm Tr}[T^a T^b T^c T^d T^c T^a]\,  {\rm Tr}[T^b T^d] = \frac{1}{2} N_c^4 (N_c^2 -1)^3 \, ,
   \label{eqb.4}
  \end{equation}
  if one of the two diagonal gluons is exchanged in between the off-diagonal ones (see, \emph{e.g.}, Fig.~\ref{fig7A}(b)),
  and
   \begin{equation}
  {\rm Tr}\left[ \mathbb{1} \right]\,  {\rm Tr}\left[ \mathbb{1} \right]\,  {\rm Tr}[T^a T^b T^c T^d T^c T^b]\,  {\rm Tr}[T^a T^d] = \frac{1}{4} N_c^4 (N_c^2 -1)^3 \, ,
   \label{eqb.5}
  \end{equation}
  if both diagonal gluon exchanges occur in between the off-diagonal ones (see, \emph{e.g.}, Fig.~\ref{fig7A}(c)). This 
  illustrates the general statements about diagonal gluons sandwiched between off-diagonal ones that are made in section~\ref{sec3b1}.
  
  Evaluating these color traces for each of the $9 \times 4^4$ diagrams, we find $\Delta {\bf y}_{ij} \equiv ({\bf y}_i-{\bf y}_j)$
 \begin{eqnarray}
         &&{\hat \sigma}^{(ord)}_{III}
  	\propto (N_c^2 -1)^3 N_c^4 
  	\left|\vec{f}({\bf k}_1)\right|^2\, \left|\vec{f}({\bf k}_2)\right|^2 \, \left|\vec{f}({\bf k}_3)\right|^2
  	\,  \left|\vec{f}({\bf k}_4)\right|^2\, 4
  		\nonumber \\
  		&& \quad \times \lbrace  4^2 \sum_{(ij)}  
  		\left[
  	{\cos\left({\bf k}_1.\Delta {\bf y}_{ij} \right)}  { \cos\left( {\bf k}_2.\Delta {\bf y}_{ij}  \right)}
  	+
  	{\cos\left({\bf k}_3.\Delta {\bf y}_{ij} \right)}  { \cos\left( {\bf k}_4.\Delta {\bf y}_{ij}  \right)} \right.
  	\nonumber \\
  	&& \qquad \qquad \qquad \left.+ 
  	{\cos\left({\bf k}_2.\Delta {\bf y}_{ij} \right)}  { \cos\left( {\bf k}_3.\Delta {\bf y}_{ij}  \right) }\right]
  	        \nonumber \\
  	        && \qquad + 3\cdot4 \sum_{(ij)}  \left[
  	{\cos\left({\bf k}_1.\Delta {\bf y}_{ij} \right)}  { \cos\left( {\bf k}_3.\Delta {\bf y}_{ij}  \right) }
  	+ {\cos\left({\bf k}_2.\Delta {\bf y}_{ij} \right)}  { \cos\left( {\bf k}_4.\Delta {\bf y}_{ij}  \right) }\right]
  	        \nonumber \\
  	        && \qquad + 9  \sum_{(ij)} \, 
  	{\cos\left({\bf k}_2.\Delta {\bf y}_{ij} \right)}  { \cos\left( {\bf k}_3.\Delta {\bf y}_{ij}  \right)}  \rbrace\, .
  	\label{eqb.6}
 \end{eqnarray}
The prefactors in this expression follow from considering the 6 choices of distributing two diagonal gluons in an
  ordered list of four gluons. For three of these choices, the
  diagonal gluons are not sandwiched between the off-diagonal ones (such, \emph{e.g.},~Fig.~\ref{fig7A}(a)) and there are 
  $4^2$ possibilities to connect the diagonal gluons to sources. This fixes the prefactor of the first
  three terms in (\ref{eqb.6}). In addition, there are 2 choices, for which one of the two diagonal gluons is sandwiched
  between the off-diagonal one (see, \emph{e.g.},~ Fig.~\ref{fig7A}(b)). In this class of diagrams, there are $4\times 2$
  diagrams with color trace (\ref{eqb.3}) and there are $ 4\times 2$ diagrams with color trace (\ref{eqb.4}).
  This determines the prefactor in the second line of (\ref{eqb.6}), namely $4\times 2 + 4\times 2\times 1/2 = 4\times 3$. 
  Finally, one can sandwich both diagonal gluons in between the off-diagonal ones (see, \emph{e.g.},~Fig.~\ref{fig7A}(3)).
  Amongst the $4^2$ possibilities of connecting the diagonal gluons to sources, there are then 4 with trace (\ref{eqb.3}),
  there are 8 with trace (\ref{eqb.4}) and there are another 4 with trace (\ref{eqb.5}), giving the prefactor
  $4\times 1 + 8 \times 1/2 + 4 \times 1/4 = 9$  of the last term of (\ref{eqb.6}). 
  
  Randomizing according to (\ref{eq2.4}) over (\ref{eqb.6}), we obtain
 \begin{eqnarray}
         &&{\hat \sigma}_{III}
  	\propto (N_c^2 -1)^3 N_c^4 
  	\left|\vec{f}({\bf k}_1)\right|^2\, \left|\vec{f}({\bf k}_2)\right|^2 \, \left|\vec{f}({\bf k}_3)\right|^2
  	\,  \left|\vec{f}({\bf k}_4)\right|^2\, 
  		\nonumber \\
  		&& \qquad\qquad  \times  \frac{27}{32} 4^2 \sum_{(ab)} \sum_{(ij)}  
  		4  {\cos\left({\bf k}_a.\Delta {\bf y}_{ij} \right)}  { \cos\left( {\bf k}_b.\Delta {\bf y}_{ij}  \right)}\, .
  	\label{eqb.7}
 \end{eqnarray}
 Here, the sum $\sum_{(ab)}$ goes over the $m(m-1)/2 = 6$ pairs of emitted gluons. The correction factor
 $27/32$ is obtained by pulling the overall factor $4^2 m(m-1)/2$ out of each of the terms in (\ref{eqb.6}) and 
 summing up the remaining terms $   \left( 3 + 2\times 3/4 + 9/4^2\right)/6 = 27/32$. This coincides with
 $F^{(2)}_{\rm corr}(4,4)$, see eq.~(\ref{eq3.10}).

 \underline{\bf Case IV} contains diagrams in which three off-diagonal gluons 
 \begin{enumerate}[i)]
 \item are linked to exactly one pair of sources $(i,j)$.
 \item are linked to two pairs of sources $(ij)$, $(jl)$ made of three sources.
 \item are linked to two pairs of sources $(ij)$, $(lm)$ made of four sources.
 \item are linked to three pairs of sources $(ij)$, $(jl)$, $(li)$ made of three sources.
 \item are linked to 3 pairs made of four sources.
 \end{enumerate}
 These contributions have i) $3\times 4^4$, ii) $9\times 4^5$, iii) $9 \times 4^4$, iv) $3 \times 4^5$
 and v) $3\times 4^6$ diagramms respectively, that sum up to a total of $3^3 \times 4^5$ diagrams. 
The contributions ii), iii) and v) can be checked easily to have vanishing color traces. 
The contribution IVi) yields  (we present symmetrized cross sections according to (\ref{eq2.4}))
  \begin{eqnarray}
         &&M_{IVi}
  	\propto \frac{1}{4} N_c^4 \, (N_c^2 -1)^3 
  	\left|\vec{f}({\bf k}_1)\right|^2\, \left|\vec{f}({\bf k}_2)\right|^2 \, \left|\vec{f}({\bf k}_3)\right|^2
  	\,  \left|\vec{f}({\bf k}_4)\right|^2\,  \frac{7}{8} N
  		\nonumber \\
  		&& \qquad \qquad \times \sum_{(abc)} \sum_{(lm)}  8\,
  	{\cos\left({\bf k}_a.\Delta {\bf y}_{lm} \right)}  { \cos\left( {\bf k}_b.\Delta {\bf y}_{lm} \right)}
  	{ \cos\left( {\bf k}_c.\Delta {\bf y}_{lm}  \right)}\, ,
  	\label{eqb.8}
 \end{eqnarray}
 where the sum $\sum_{(abc)}$ runs over the four possibilities of selecting three of four particle momenta in the final state. 
 The color traces of diagrams contributing to (\ref{eqb.8}) are $\frac{1}{4} N_c^4 \, (N_c^2 -1)^3 $ if the diagonal gluon is not
 linked to source $l$ or $m$, but it is $\frac{1}{8} N_c^4 \, (N_c^2 -1)^3 $ if the diagonal gluon is
sandwiched between the off-diagonal ones and linked to $l$ or $m$. After symmetrization, this smaller color trace of some
diagrams leads to a color correction factor $7/8$. To keep track of the origin of different factors, we write in (\ref{eqb.8})
the factor $N$ for the number of possibilities to link the diagonal gluon to one of the $N$ sources. Compared to (\ref{eqb.1}),
the contribution (\ref{eqb.8}) is $O(N^{-1})$ suppressed.
  
Similarly, for diagrams contributing to case IViv, we find color traces $N_c^4 \, (N_c^2 -1)^2 $, as well as 
$\frac{1}{2} N_c^4 \, (N_c^2 -1)^3 $ for some diagrams. After symmetrization, the reduced color trace can be 
absorbed in a color correction factor $7/8$ and one finds
  \begin{eqnarray}
         &&M_{IViv}
  	\propto (N_c^2 -1)^2 N_c^4 
  	\left|\vec{f}({\bf k}_1)\right|^2\, \left|\vec{f}({\bf k}_2)\right|^2 \, \left|\vec{f}({\bf k}_3)\right|^2
  	\,  \left|\vec{f}({\bf k}_4)\right|^2\, \frac{7}{8}\, N 
  		\nonumber \\
  		&& \quad \times  \sum_{(abc)} \sum_{(lm)(mn)(nl)}  8\,
  	{\cos\left({\bf k}_a.\Delta {\bf y}_{lm} \right)}  { \cos\left( {\bf k}_b.\Delta {\bf y}_{mn}  \right)}
  	{ \cos\left( {\bf k}_c.\Delta {\bf y}_{nl}  \right)}\, .
  	\label{eqb.9}
 \end{eqnarray}
 For the prefactors of these expressions, there is the following consistency check: One recalls  
 that each diagram contributes phase factors with prefactor 1 times a color trace. Hence, if one ignores the color 
 correction factor and if one multiplies the 4 terms in the sum $ \sum_{(abc)} $
 times the $N(N-1)(N-2)$ terms in $\sum_{(lm)(mn)(nl)} $ times the explicit prefactor $N\, 8$, the result should
 match the number $3 \times 4^5$ of diagrams contributing to case IViv.

\underline{\bf Case V} includes diagrams with four off-diagonal gluons that are 
 \begin{enumerate}[i)]
 \item emitted from two sources, \emph{i.e.}, one pair of sources $(i,j)$.
 \item emitted from three sources that combine to two pairs of sources $(lm)$, $(mn)$.
 \item emitted from three sources that combine to three pairs of sources $(lm)$, $(mn)$, $(n,l)$.
 \item emitted from four sources that combine to two pairs of sources $(lm)$ and $(no)$.
 \item emitted from four sources that combine to three pairs of sources $(lm)$, $(mn)$, $(no)$.
 \item emitted from four sources that combine to four pairs of sources $(lm)$, $(mn)$, $(no)$, $(ol)$.
 \end{enumerate}
 
 These contributions include i) $6\times 4^2$  ii) $6\times 4^4$ + $2\times 3^2\times 4^3$  iii) $3^2\times 4^4$ iv) 
 $2^5\times 3^2$ + $6\times 4^3$ v) $3^3\times 4^4$ + $3^2\times 4^4$ vi) $2\times 3^2\times 4^3$ + $2\times 3^2 \times 4^4$
 diagrams, respectively, and they add up to $3^4\times 4^4$ diagrams.
 
 Case Vi) comprises diagrams with more complicated color trace
 \begin{equation}
 	{\rm Tr}\left[ \mathbb{1} \right]^2 {\rm Tr}\left[ T^a T^b T^c T^d\right]\, {\rm Tr}\left[ T^d T^c T^b T^a\right]
 	= \frac{1}{8} N_c^2 \left( N_c^4 + 11 N_c^2 - 12\right)\, \left(N_c^2-1\right)^2\, , 
 	\label{eqb.10}
 \end{equation}
that result in 
  \begin{eqnarray}
         &&M_{Vi}
  	\propto  \frac{1}{8} N_c^2 \left(N_c^2-1\right)^2 \left( N_c^4 + 11 N_c^2 - 12\right)
  	\left|\vec{f}({\bf k}_1)\right|^2\, \left|\vec{f}({\bf k}_2)\right|^2 \, \left|\vec{f}({\bf k}_3)\right|^2
  	\,  \left|\vec{f}({\bf k}_4)\right|^2\, 
  		\nonumber \\
  		&& \qquad \qquad \times  \sum_{(lm)}  2^4\, 
  	{\cos\left({\bf k}_1.\Delta {\bf y}_{lm} \right)}  { \cos\left( {\bf k}_2.\Delta {\bf y}_{lm}  \right)}
  	{ \cos\left( {\bf k}_3.\Delta {\bf y}_{lm}  \right)}\, { \cos\left( {\bf k}_4.\Delta {\bf y}_{lm}  \right)}\, .
  	\label{eqb.11}
 \end{eqnarray}
 It is a simple consistency check to test that the number of $N(N-1)/2$ terms in the sum times the prefactor $2^4$
 matches the number of diagrams in case Vi. 
 
Case Vii) comprises $6\times 4^4$ diagrams for which three off-diagonal gluons are attached to one pair of sources.
These diagrams vanish. It also comprises $2\times 3^2\times 4^3$ diagrams in which two off-diagonal gluons are attached
to each pairs of sources. For these latter diagrams, `most' color traces are $N_c^4 \left( N_c^2 - 1\right)^2$ but `some'
are $\frac{1}{2}N_c^4 \left( N_c^2 - 1\right)^2$. In the symmetrized expression, the modifications due to the smaller
color trace can be accounted for by a color correction factor $5/6$ and one obtains  
  \begin{eqnarray}
         &&{\hat \sigma}_{Vii1}
  	\propto  N_c^4 \left( N_c^2 - 1\right)^2
  	\left|\vec{f}({\bf k}_1)\right|^2\, \left|\vec{f}({\bf k}_2)\right|^2 \, \left|\vec{f}({\bf k}_3)\right|^2
  	\,  \left|\vec{f}({\bf k}_4)\right|^2\, \sum_{(lm)(mn)} 
  		\nonumber \\
  		&& \times \frac{5}{6} \sum_{(ab),(cd)}  2^4\, 
  	{\cos\left({\bf k}_a.\Delta {\bf y}_{lm} \right)}  { \cos\left( {\bf k}_b.\Delta {\bf y}_{lm}  \right)}
  	{ \cos\left( {\bf k}_c.\Delta {\bf y}_{mn}  \right)}\, { \cos\left( {\bf k}_d.\Delta {\bf y}_{mn}  \right)}\, .
  	\label{eqb.12}
 \end{eqnarray}
Here, the sum $ \sum_{(lm)(mn)}$ goes over the $N(N-1)(N-2)/2 = 4\times 3$ possibilities of picking two pairs of sources
that have one source in common. The sum $\sum_{(ab),(cd)} $ goes over the $m(m-1)/2$ $(m-2)(m-3)/2$ $= 6$ possibilities
of picking an ordered list of two pairs. As a consistency check, one can multiply these factors with the explicit prefactor 
$2^4$ in (\ref{eqb.12}) to recover the number of $2\times 3^2\times 4^3$ diagrams of this case. 

However, there is an additional contribution to this case in which one finds for the first time $\sin$-terms. This contribution breaks the 
 symmetry ${\bf k}_i \to  -{\bf k}_i$, and can therefore give rise to odd harmonics. It reads
 \begin{eqnarray}
         && {\hat \sigma}^{(ord)}_{Vii2}
  	\propto  N_c^4 \left( N_c^2 - 1\right)^2
  	\left|\vec{f}({\bf k}_1)\right|^2\, \left|\vec{f}({\bf k}_2)\right|^2 \, \left|\vec{f}({\bf k}_3)\right|^2
  	\,  \left|\vec{f}({\bf k}_4)\right|^2\, 
  		\nonumber \\
  		&& \times \sum_{(lm)(mn)} 
  		 \lbrace  - 4{\cos\left({\bf k}_1.\Delta {\bf y}_{lm} \right)}  { \sin\left( {\bf k}_2.\Delta {\bf y}_{mn}  \right)}
  	{ \sin\left( {\bf k}_3.\Delta {\bf y}_{mn}  \right)}\, { \cos\left( {\bf k}_4.\Delta {\bf y}_{lm}  \right)}	
  	\nonumber \\
  	&& \qquad \qquad - 4{\cos\left({\bf k}_1.\Delta {\bf y}_{mn} \right)}  { \sin\left( {\bf k}_2.\Delta {\bf y}_{lm}  \right)}
  	{ \sin\left( {\bf k}_3.\Delta {\bf y}_{lm}  \right)}\, { \cos\left( {\bf k}_4.\Delta {\bf y}_{mn}  \right)}
  	\nonumber \\
  	&& \qquad \qquad - 4 {\cos\left({\bf k}_1.\Delta {\bf y}_{lm} \right)}  { \sin\left( {\bf k}_2.\Delta {\bf y}_{mn}  \right)}
  	{ \sin\left( {\bf k}_3.\Delta {\bf y}_{lm}  \right)}\, { \cos\left( {\bf k}_4.\Delta {\bf y}_{mn}  \right)}
  	\nonumber \\
  	&& \qquad \qquad - 4{\cos\left({\bf k}_1.\Delta {\bf y}_{mn} \right)}  { \sin\left( {\bf k}_2.\Delta {\bf y}_{lm}  \right)}
  	{ \sin\left( {\bf k}_3.\Delta {\bf y}_{mn}  \right)}\, { \cos\left( {\bf k}_4.\Delta {\bf y}_{lm}  \right)}
  	\rbrace  \, .
  	\label{eqb.13}
 \end{eqnarray}
 We write ${\hat \sigma}^{(ord)}_{Vii2}$,
  without symmetrizing over gluon momenta, for  ${\bf k}_1$, ... ${\bf k}_4$ ordered from top to bottom
  in emission diagrams. The origin of these $\sin$-terms has been discussed in subsection~\ref{sec5b1}.

 In a similar way, one can inspect in detail all diagrams of the case Viii. The result can be expressed again in terms of 
 one contributions with $\cos$-terms only (that we write in symmetrized form) and one contribution that contains $\sin$-terms
 (and that we write first in un-symmetrized form)
  \begin{eqnarray}
         &&M_{Viii1}
  	\propto  N_c^4 \left( N_c^2 - 1\right)^2
  	\left|\vec{f}({\bf k}_1)\right|^2\, \left|\vec{f}({\bf k}_2)\right|^2 \, \left|\vec{f}({\bf k}_3)\right|^2
  	\,  \left|\vec{f}({\bf k}_4)\right|^2\, \sum_{\underline{(lm)}(mn)(nl)} 
  		\nonumber \\
  		&& \times \frac{1}{24} \sum_{(abcd)}  2^4\, 
  	{\cos\left({\bf k}_a.\Delta {\bf y}_{lm} \right)}  { \cos\left( {\bf k}_b.\Delta {\bf y}_{lm}  \right)}
  	{ \cos\left( {\bf k}_c.\Delta {\bf y}_{mn}  \right)}\, { \cos\left( {\bf k}_d.\Delta {\bf y}_{nl}  \right)}\, ,
  	\label{eqb.15}\\
  	&&M_{Viii2}^{(ord)}
  	\propto  N_c^4 \left( N_c^2 - 1\right)^2
  	\left|\vec{f}({\bf k}_1)\right|^2\, \left|\vec{f}({\bf k}_2)\right|^2 \, \left|\vec{f}({\bf k}_3)\right|^2
  	\,  \left|\vec{f}({\bf k}_4)\right|^2\, \sum_{\underline{(lm)}(mn)(nl)} 
  		\nonumber \\
  	&& \quad \times \Big\lbrace - 4{\cos\left({\bf k}_1.\Delta {\bf y}_{lm} \right)}  { \sin\left( {\bf k}_2.\Delta {\bf y}_{lm}  \right)}
  	{ \sin\left( {\bf k}_3.\Delta {\bf y}_{mn}  \right)}\, { \cos\left( {\bf k}_4.\Delta {\bf y}_{nl}  \right)}
  	\nonumber \\
  	&& \qquad - 4{\cos\left({\bf k}_1.\Delta {\bf y}_{lm} \right)}  { \sin\left( {\bf k}_2.\Delta {\bf y}_{lm}  \right)}
  	{ \sin\left( {\bf k}_3.\Delta {\bf y}_{nl}  \right)}\, { \cos\left( {\bf k}_4.\Delta {\bf y}_{mn}  \right)}
  	\nonumber \\
  	&& \qquad - 4 {\cos\left({\bf k}_1.\Delta {\bf y}_{lm} \right)}  { \sin\left( {\bf k}_2.\Delta {\bf y}_{mn}  \right)}
  	{ \sin\left( {\bf k}_3.\Delta {\bf y}_{nl}  \right)}\, { \cos\left( {\bf k}_4.\Delta {\bf y}_{lm}  \right)}
  	\nonumber \\
  	&& \qquad - 4{\cos\left({\bf k}_1.\Delta {\bf y}_{lm} \right)}  { \sin\left( {\bf k}_2.\Delta {\bf y}_{nl}  \right)}
  	{ \sin\left( {\bf k}_3.\Delta {\bf y}_{mn}  \right)}\, { \cos\left( {\bf k}_4.\Delta {\bf y}_{lm}  \right)}
  	\nonumber \\
  	&& \qquad - 4{\cos\left({\bf k}_1.\Delta {\bf y}_{mn} \right)}  { \sin\left( {\bf k}_2.\Delta {\bf y}_{lm}  \right)}
  	{ \sin\left( {\bf k}_3.\Delta {\bf y}_{lm}  \right)}\, { \cos\left( {\bf k}_4.\Delta {\bf y}_{nl}  \right)}
  	\nonumber \\
  	&& \qquad - 4{\cos\left({\bf k}_1.\Delta {\bf y}_{nl} \right)}  { \sin\left( {\bf k}_2.\Delta {\bf y}_{lm}  \right)}
  	{ \sin\left( {\bf k}_3.\Delta {\bf y}_{lm}  \right)}\, { \cos\left( {\bf k}_4.\Delta {\bf y}_{mn}  \right)}
  	\nonumber \\
  	&& \qquad - 4 {\cos\left({\bf k}_1.\Delta {\bf y}_{mn} \right)}  { \sin\left( {\bf k}_2.\Delta {\bf y}_{lm}  \right)}
  	{ \sin\left( {\bf k}_3.\Delta {\bf y}_{nl}  \right)}\, { \cos\left( {\bf k}_4.\Delta {\bf y}_{lm}  \right)}
  	\nonumber \\
  	&& \qquad - 4{\cos\left({\bf k}_1.\Delta {\bf y}_{nl} \right)}  { \sin\left( {\bf k}_2.\Delta {\bf y}_{lm}  \right)}
  	{ \sin\left( {\bf k}_3.\Delta {\bf y}_{mn}  \right)}\, { \cos\left( {\bf k}_4.\Delta {\bf y}_{lm}  \right)}
  	\Big\rbrace \, .
  	\label{eqb.16}
 \end{eqnarray}
 Here, the sum $\sum_{\underline{(lm)}(mn)(nl)} $ goes over triplets of source pairs, of which the source 
 pair $(lm)$ is connected to two off-diagonal gluons.
 
 For the case V iv), we obtain 
  \begin{eqnarray}
         &&M_{Viv1}
  	\propto  N_c^4 \left( N_c^2 - 1\right)^2
  	\left|\vec{f}({\bf k}_1)\right|^2\, \left|\vec{f}({\bf k}_2)\right|^2 \, \left|\vec{f}({\bf k}_3)\right|^2
  	\,  \left|\vec{f}({\bf k}_4)\right|^2\, \sum_{(lm)(no)} 
  		\nonumber \\
  		&& \times  \sum_{(ab),(cd)}  2^4\, 
  	{\cos\left({\bf k}_a.\Delta {\bf y}_{lm} \right)}  { \cos\left( {\bf k}_b.\Delta {\bf y}_{lm}  \right)}
  	{ \cos\left( {\bf k}_c.\Delta {\bf y}_{no}  \right)}\, { \cos\left( {\bf k}_d.\Delta {\bf y}_{no}  \right)} \, ,
  	\label{eqb.17}
 \end{eqnarray}
 where the sum $\sum_{(lm)(no)} $ goes over the $N(N-1)(N-2)(N-3)/8$ possibilities of picking two 
 independent pairs of sources. 
 
 Finally, there is a non-vanishing contribution for connecting four off-diagonal gluons to 4 different pairs of sources, namely 
  \begin{eqnarray}
         &&M_{Vvi}
  	\propto  N_c^4 \left( N_c^2 - 1\right)
  	\left|\vec{f}({\bf k}_1)\right|^2\, \left|\vec{f}({\bf k}_2)\right|^2 \, \left|\vec{f}({\bf k}_3)\right|^2
  	\,  \left|\vec{f}({\bf k}_4)\right|^2\, \sum_{(lm)(mn)(no)(ol)} 
  		\nonumber \\
  		&& \times  \sum_{(abcd))}  2^4\, 
  	{\cos\left({\bf k}_a.\Delta {\bf y}_{lm} \right)}  { \cos\left( {\bf k}_b.\Delta {\bf y}_{mn}  \right)}
  	{ \cos\left( {\bf k}_c.\Delta {\bf y}_{no}  \right)}\, { \cos\left( {\bf k}_d.\Delta {\bf y}_{ol}  \right)} \, .
  	\label{eqb.18}
 \end{eqnarray}
 Compared to (\ref{eqb.17}), this is suppressed by another power in $1/(N_c^2-1)$. All other cases vanish.
 
\section{Color correction factors}
In this appendix, we derive the color correction factors entering (\ref{eq4.1}). The factor $F^{(2)}_{\rm corr}$ was derived 
in section~\ref{sec3b1} already. 
\label{appc}
\subsection{$F_{\rm corr}^{(4i)}(N,m)$}
\label{appc1}
We consider the case of $m-4$ diagonal gluons and of four off-diagonal gluons that form two dipoles
$(lm)$, $(no)$ with four different sources. This is case V iv1 in appendix~\ref{appb}, which is
the leading $O\left(1/(N_c^2-1)^2\right)$ contribution to the 4-particle correlator. There are 
$\binom{m}{4}$
possibilities to assign the 4 off-diagonal gluons to an ordered list of $m$ gluons. 
For each assignment, there
are 
\begin{itemize}
\item $j_1$ diagonal gluons between the first and second off-diagonal gluon.
\item $j_2$ diagonal gluons between the second and third off-diagonal gluon.
\item $j_3$ diagonal gluons between the third and fourth off-diagonal gluon.
\item $m-4-j_1-j_2-j_3$ before the first or after the last off-diagonal gluon. 
\end{itemize}
The number of possibilities of distributing these $m-4$ gluons such that an arbitrary number of
$j_1+j_2+j_3$ gluons lies between the first and last off-diagonal gluon is
\begin{equation}
\sum_{j_3=0}^{m-4} \sum_{j_2=0}^{m-4-j_3} \sum_{j_1=0}^{m-4-j_2-j_3} \left(m-3-\left(j_1+j_2+j_3\right)\right)
= \binom{m}{4}\, .
 \label{eqc.1}
\end{equation}
This is a consistency check of our starting point that will be useful in the following. We also note that the total number of
different diagrams (for given sources $l$, $m$, $n$, $o$) is 
\begin{equation}
	{\cal N} = \binom{m}{4} \, N^{m-4}\, .
	\label{eqc.2}
\end{equation}
The four off-diagonal gluons can be paired into dipoles in three different ways. Denoting the off-diagonal gluons as
$1$, $\bar{1}$, $2$, $\bar{2}$ in a notation that makes their pairing clear, the three different paring are
\begin{enumerate}[A.]
\item Ordering of off-diagonal gluons: $1$, $\bar{1}$, $2$, $\bar{2}$
\item Ordering of off-diagonal gluons: $1$,  $2$, $\bar{1}$, $\bar{2}$
\item Ordering of off-diagonal gluons: $1$, $2$, $\bar{2}$, $\bar{1}$
\end{enumerate}
For \underline{case A.}, there are $j_1$ diagonal gluons sandwiched between  the first dipole pair $1$$\bar{1}$, and 
$j_3$ gluons sandwiched between $2$$\bar{2}$; $j_2$ gluons are not sandwiched between any dipole pair. 
 $l_1$ of the $j_1$ gluons will be linked to the same source pair $(lm)$ as $1$$\bar{1}$, and analogously for $l_3$
 out of the $j_3$ gluons sandwiched between $2$$\bar{2}$. As each diagonal gluon sandwiched between an off-diagonal
 pair reduces the color trace and thus the spectrum by a factor 2, we have
 \begin{eqnarray}
 && F_{\rm corr}^{(4i,A)}(N,m) = \frac{1}{{\cal N}} 
	\sum_{j_3=0}^{m-4} \sum_{j_2=0}^{m-4-j_3} \sum_{j_1=0}^{m-4-j_2-j_3} \left(m-3-\left(j_1+j_2+j_3\right)\right)
	N^{m-4-\left(j_1+j_2+j_3\right)}
	\nonumber \\
	&& \qquad \qquad \qquad \qquad
	\times \sum_{l_3=0}^{j_3}  \sum_{l_1=0}^{j_1} 
	\binom{j_1}{ l_1}\, \left( N-2\right)^{j_1-l_1}
 	N^{j_2}
 	\binom{j_3}{l_3}\,  \left( N-2\right)^{j_3-l_3}
	  \nonumber\\
	  && \qquad \qquad = \frac{(m-4)!}{m!} \left[-24 N^{4-m}(-3+m+3N)(N-1)^{m-1} \right.
	  \nonumber\\
	  &&  \qquad \qquad \qquad \qquad  \qquad \qquad \qquad \qquad \left. + 12N^2\left(m(m-1)-4mN+6N^2 \right) \right]\, .
	  \label{eqc.3}
 \end{eqnarray}
 We note that without the above-mentioned factors $1/2$, the terms in (\ref{eqc.3}) would be multiplied by factors $2^{l_1+l_3}$,
 and one would find $F_{\rm corr}^{(4i,A)}(N,m) = 1$.
 
 For \underline{case B.}, there are again $j_1$ diagonal gluons sandwiched between  the first dipole pair $1$$\bar{1}$, and 
$j_3$ gluons sandwiched between $2$$\bar{2}$. However, there are now $j_2$ gluons sandwiched between  $1$$\bar{1}$, and 
$2$$\bar{2}$. In this case, there will be $l_2$ out of $j_2$ gluons that are linked to the dipole $(lm)$ and there will be $\bar{l}_2$
of the $j_2$ gluons linked to $(no)$. The corresponding color factor reads
\begin{eqnarray}
	&& F_{\rm corr}^{(4i,B)}(N,m) = \frac{1}{{\cal N}(N,m)} 
	\sum_{j_3=0}^{m-4} \sum_{j_2=0}^{m-4-j_3} \sum_{j_1=0}^{m-4-j_2-j_3} \left(m-3-\left(j_1+j_2+j_3\right)\right)
	N^{m-4-\left(j_1+j_2+j_3\right)}
	\nonumber \\
	&& \qquad \qquad
	\times \sum_{l_3=0}^{j_3} \sum_{{\bar l}_2=0}^{j_2} \sum_{l_2=0}^{j_2-{\bar l}_2} \sum_{l_1=0}^{j_1} 
	\binom{j_1}{ l_1}\, \left( N-2\right)^{j_1-l_1}
  \frac{j_2!\,  \left( N-4\right)^{j_2-l_2-{\bar l}_2}}{l_2!\, {\bar l}_2!\, \left(j_2-l_2-{\bar l}_2 \right)!}
  \binom{j_3}{l_3}\,  \left( N-2\right)^{j_3-l_3}
  \nonumber \\
  && \qquad \quad = \frac{6\, (2m-5N)\, N^3}{m\, (m-1)\, (m-2)\, (m-3) } 
  \nonumber \\ 
  && \qquad \quad \quad + \frac{6\, N^{4-m}\, \left((N-2)^m + 4(N-1)^{m-1} (N+m-1)\right)}{m (m-1)(m-2)(m-3) }\, .
  \label{eqc.4}
\end{eqnarray}
 \underline{Case C.} gives the same result as case B. The color correction factor $F_{\rm corr}^{(4)}(N,m)$  in equation (\ref{eq4.1})
 therefore reads 
 \begin{eqnarray}
 	F_{\rm corr}^{(4i)}(N,m) &=& \frac{1}{3} F_{\rm corr}^{(4,A)}(N,m) + \frac{2}{3} F_{\rm corr}^{(4,B)}(N,m) 
 	\nonumber \\
 	&=& \frac{4(m-4)!}{m!} \left[N^2 \left(m(m-1) - 2mN  + N^2 \right) \right.
 	\nonumber \\
 	&& \qquad \qquad
 	\left.  + N^{4-m} \left((N-2)^m - 2 (N-1)^{m-1} (N-m-1) \right) \right]\, .
 	\label{eqc.5}
 \end{eqnarray}
 Fixing the average number of gluons emitted per source to $\overline{m} = m/N$, we arrive at a compact
expression in the high multiplicity limit, 
 \begin{equation}
 	\lim_{m\to\infty} F_{\rm corr}^{(4i)}(m/\overline{m},m) = \left(\frac{2e^{-\overline{m}}+2\overline{m} - 2}{\overline{m}^2}  \right)^2 \, .
 	\label{eqc.6}
 \end{equation}
 %
\subsection{$F_{\rm corr}^{(3i)}(N,m)$}
\label{appc2}
We consider the case of $m-3$ diagonal gluons and of three off-diagonal gluons that are linked between
the source pairs $(lm)$, $(mn)$ and $(nl)$. In close analogy to appendix~\ref{appc1}, we denote by $j_1$ ($j_2$)
the number of diagonal gluons between the first and second (the second and third) off-diagonal gluon. One checks
easily for each of the $j_1$ diagonal gluons: their color trace is $N_c$ if they link to a source other than $l$, $m$ or $n$.
Also for exactly one of the three sources $l$, $m$, $n$, the color trace is $N_c$, but it is $N_c/2$ for the other two sources.
Paralleling the logic that lead to (\ref{eqc.3}), we therefore find the norm
\begin{equation}
	{\cal N} = 
	\binom{m}{3}\, N^{m-3}\, ,
	\label{eqc.7}
\end{equation}
and 
 \begin{eqnarray}
 && F_{\rm corr}^{(3i)}(N,m) = \frac{1}{{\cal N}} 
	\sum_{j_2=0}^{m-3} \sum_{j_1=0}^{m-3-j_2} \left(m-2-\left(j_1+j_2\right)\right)
	N^{m-3-\left(j_1+j_2\right)}
	\nonumber \\
	&& \qquad \qquad \qquad \qquad
	\times \sum_{l_1=0}^{j_1}  \sum_{\bar{l}_1=0}^{j_1-l_1} 
	\frac{j_1!}{l_1!\, \bar{l}_1!\, \left(j_1 - l_1 -\bar{l}_1 \right)!} 1^{\bar{l}_1}\, 2^{l_1}\, \left( N-3\right)^{j_1-l_1-\bar{l}_1}
	  \nonumber\\
	&& \qquad \qquad \qquad \qquad
	\times \sum_{l_2=0}^{j_2}  \sum_{\bar{l}_2=0}^{j_2-l_2} 
	\frac{j_2!}{l_2!\, \bar{l}_2!\, \left(j_2 - l_2 -\bar{l}_2 \right)!} 1^{\bar{l}_2}\, 2^{l_2}\, \left( N-3\right)^{j_2-l_2-\bar{l}_2}
	  \nonumber\\
	  && \qquad \qquad = \frac{(m-3)!}{m!} \left[6 (m-2N) N^2 + 6 (N-1)^{m-1} N^{3-m} (-2+m+2N)  \right]\, .
	  \label{eqc.8}
 \end{eqnarray}
For $N=m=4$, we find $F_{\rm corr}^{(3i)}(4,4)= \frac{7}{8}$, which is consistent with the explicit calculation in appendix~\ref{appb}. 
The limiting value is
 \begin{equation}
 	\lim_{m\to\infty} F_{\rm corr}^{(3i)}(m/\overline{m},m) = \frac{6 e^{-\overline{m}} }{\overline{m}^3} \left(2+\overline{m}\right) 
 	+ \frac{6 }{\overline{m}^3} \left(\overline{m} - 2\right)  \, .
 	\label{eqc.9}
 \end{equation}
 %
\subsection{$F_{\rm corr}^{(4ii)}(N,m)$}
\label{appc3}
We consider the case of $m-4$ diagonal gluons and of four off-diagonal gluons that are linked between
the source pairs $(lm)$, $(mn)$, $(no)$ and $(ol)$. We denote by $j_1$, $j_2$ and $j_3$ the number of diagonal gluons 
that are sandwiched between the first and second, the second and third and the third and fourth off-diagonal gluon, respectively.
One checks easily that each of these diagonal gluons contributes to a color trace with factor $N_c$ if linked to any source other
than $l$, $m$, $n$ or $o$. Amongst the sources $l$, $m$, $n$ and $o$, there are always exactly two sources where the color
trace is $N_c$, while the color trace for the two other sources is $N_c/2$. Paralleling the logic of appendices ~\ref{appc1} and ~\ref{appc2},
we find
\begin{eqnarray}
	&& F_{\rm corr}^{(4ii)}(N,m) = \frac{1}{{\cal N}(N,m)} 
	\sum_{j_3=0}^{m-4} \sum_{j_2=0}^{m-4-j_3} \sum_{j_1=0}^{m-4-j_2-j_3} \left(m-3-\left(j_1+j_2+j_3\right)\right)
	N^{m-4-\left(j_1+j_2+j_3\right)}
	\nonumber \\
	&& \qquad \qquad
	\times \prod_{i=1}^{3} \left( \sum_{l_i=0}^{j_i} \sum_{{\bar l}_i=0}^{j_i-l_i} 
  \frac{j_i!}{l_i!\, {\bar l}_i!\, \left(j_i-l_i-{\bar l}_i \right)!}
	2^{l_i} 2^{\bar{l}_i} \left( N-2\right)^{j_i-l_i-\bar{l}_i} \frac{1}{2^{l_i}}\right)
  \nonumber \\
  && \qquad \quad = \frac{4!\, (m-4)!}{m!} \left[ (m-3N)N^3 \right.
  \nonumber \\
  && \qquad \qquad \qquad \qquad \left. + \frac{1}{2}(N-1)^{m-2} N^{4-m} \left(m^2 + 6(N-1)^2 + m(-5+4N) \right)  \right]\, .
  \label{eqc.10}
\end{eqnarray}
The limiting value is
 \begin{equation}
 	\lim_{m\to\infty} F_{\rm corr}^{(4ii)}(m/\overline{m},m) = \frac{e^{-2\overline{m}}  \left(9+12\overline{m}+6\overline{m}^2 +  e^{2\overline{m}}
 	\left(-9 + 6  \overline{m} \right)  \right)   }{2\overline{m}^4} 
 	  \, .
 	\label{eqc.11}
 \end{equation}
\subsection{$F_{\rm corr}^{(5)}(N,m)$, $F_{\rm corr}^{(6)}(N,m)$}
\label{appc4}
The complexity of the combinatorics increases if one considers diagonal gluons linked into diagrams with 5 or 6 off-diagonal gluons. 
For the terms $F_{\rm corr}^{(5)}(N,m)$ and $F_{\rm corr}^{(6)}(N,m)$ in (\ref{eq4.1}), however, results can be obtained in
close parallel to the derivations given in appendices~\ref{appc1}, ~\ref{appc2} and ~\ref{appc3}. We find
 \begin{equation}
 	\lim_{m\to\infty} F_{\rm corr}^{(5)}(m/\overline{m},m) = \frac{12\, e^{-2\overline{m}}  \left(2+\overline{m} +  e^{2\overline{m}}
 	\left(2 - 3\overline{m} + \overline{m}^2 \right) + e^{\overline{m}}\left(-4 + 2\overline{m} + \overline{m}^2 \right) \right)   }{\overline{m}^5} 
 	  \, ,
 	\label{eqc.12}
 \end{equation}
and
 \begin{equation}
 	\lim_{m\to\infty} F_{\rm corr}^{(6)}(m/\overline{m},m) = \frac{8\, e^{-3\overline{m}}  \left(1+e^{\overline{m}}
 	\left(-1 +\overline{m}  \right)  \right)^3   }{\overline{m}^6} 
 	  \, .
 	\label{eqc.13}
 \end{equation}
\subsubsection*{Acknowledgments}
B.B., C.J. and M.S. thank the CERN TH Department for hospitality and support during short stays in 2016 and 2017.
C.J. was supported by the Conselho Nacional de Desenvolvimento Cientifico e Tecnológico (CNPq).
M.S.'s  research  was  supported  by  the  US  Department  of  Energy  Office  of  Science,  Office  of
Nuclear Physics under Award No. DE-FG02-93ER40771.



\begin{thebibliography}{99}


\bibitem{Sjostrand:2017cdm}
  T.~Sjöstrand,
  arXiv:1706.02166 [hep-ph].

\bibitem{Gieseke:2007ad}
  S.~Gieseke, M.~H.~Seymour and A.~Siodmok,
  JHEP {\bf 0806} (2008) 001
  doi:10.1088/1126-6708/2008/06/001
  [arXiv:0712.1199 [hep-ph]].
  
\bibitem{Gieseke:2017yfk}
  S.~Gieseke, P.~Kirchgaeßer and F.~Loshaj,
  ``Soft Interactions in Herwig,''
  arXiv:1703.10808 [hep-ph].
  
\bibitem{Schulz:2016vml}
  H.~Schulz [SHERPA Collaboration],
  ``SHRiMPS – Status of soft interactions in SHERPA,'', in ~\cite{mpi2015}.
  
\bibitem{Chatrchyan:2012nia}
  S.~Chatrchyan {\it et al.} [CMS Collaboration],
  Phys.\ Lett.\ B {\bf 712} (2012) 176
  doi:10.1016/j.physletb.2012.04.058
  [arXiv:1202.5022 [nucl-ex]].
  
\bibitem{Aad:2014bxa}
  G.~Aad {\it et al.} [ATLAS Collaboration],
  Phys.\ Rev.\ Lett.\  {\bf 114} (2015) no.7,  072302
  doi:10.1103/PhysRevLett.114.072302
  [arXiv:1411.2357 [hep-ex]].
  
\bibitem{Abelev:2013kqa}
  B.~Abelev {\it et al.} [ALICE Collaboration],
  JHEP {\bf 1403} (2014) 013
  doi:10.1007/JHEP03(2014)013
  [arXiv:1311.0633 [nucl-ex]].
  
\bibitem{Kurkela:2016vts}
  A.~Kurkela,
  Nucl.\ Phys.\ A {\bf 956} (2016) 136
  doi:10.1016/j.nuclphysa.2016.01.069
  [arXiv:1601.03283 [hep-ph]].
  
\bibitem{Heinz:2013th}
  U.~Heinz and R.~Snellings,
  Ann.\ Rev.\ Nucl.\ Part.\ Sci.\  {\bf 63} (2013) 123
  doi:10.1146/annurev-nucl-102212-170540
  [arXiv:1301.2826 [nucl-th]].
  
\bibitem{ALICE:2017jyt}
  J.~Adam {\it et al.} [ALICE Collaboration],
  Nature Phys.\  {\bf 13} (2017) 535
  doi:10.1038/nphys4111
  [arXiv:1606.07424 [nucl-ex]].

\bibitem{Khachatryan:2015waa}
  V.~Khachatryan {\it et al.} [CMS Collaboration],
  Phys.\ Rev.\ Lett.\  {\bf 115} (2015) no.1,  012301
  doi:10.1103/PhysRevLett.115.012301
  [arXiv:1502.05382 [nucl-ex]].

\bibitem{Khachatryan:2016txc}
  V.~Khachatryan {\it et al.} [CMS Collaboration],
  Phys.\ Lett.\ B {\bf 765} (2017) 193
  doi:10.1016/j.physletb.2016.12.009
  [arXiv:1606.06198 [nucl-ex]].
  
\bibitem{Aaboud:2017acw}
  M.~Aaboud {\it et al.} [ATLAS Collaboration],
  Eur.\ Phys.\ J.\ C {\bf 77} (2017) no.6,  428
  doi:10.1140/epjc/s10052-017-4988-1
  [arXiv:1705.04176 [hep-ex]].
  
\bibitem{Aad:2015gqa}
  G.~Aad {\it et al.} [ATLAS Collaboration],
  Phys.\ Rev.\ Lett.\  {\bf 116} (2016) no.17,  172301
  doi:10.1103/PhysRevLett.116.172301
  [arXiv:1509.04776 [hep-ex]].

\bibitem{Fischer:2016zzs}
  N.~Fischer and T.~Sjöstrand,
  JHEP {\bf 1701} (2017) 140
  doi:10.1007/JHEP01(2017)140
  [arXiv:1610.09818 [hep-ph]].

\bibitem{Bozek:2011if}
  P.~Bozek,
  Phys.\ Rev.\ C {\bf 85} (2012) 014911
  doi:10.1103/PhysRevC.85.014911
  [arXiv:1112.0915 [hep-ph]].

\bibitem{Bzdak:2013zma}
  A.~Bzdak, B.~Schenke, P.~Tribedy and R.~Venugopalan,
  Phys.\ Rev.\ C {\bf 87} (2013) no.6,  064906
  doi:10.1103/PhysRevC.87.064906
  [arXiv:1304.3403 [nucl-th]].

\bibitem{He:2015hfa}
  L.~He, T.~Edmonds, Z.~W.~Lin, F.~Liu, D.~Molnar and F.~Wang,
  Phys.\ Lett.\ B {\bf 753} (2016) 506
  doi:10.1016/j.physletb.2015.12.051
  [arXiv:1502.05572 [nucl-th]].
  
\bibitem{Adam:2015hoa}
  J.~Adam {\it et al.} [ALICE Collaboration],
  Phys.\ Lett.\ B {\bf 749} (2015) 68
  doi:10.1016/j.physletb.2015.07.054
  [arXiv:1503.00681 [nucl-ex]].
  
\bibitem{ATLAS:2014cpa}
  G.~Aad {\it et al.} [ATLAS Collaboration],
  Phys.\ Lett.\ B {\bf 748} (2015) 392
  doi:10.1016/j.physletb.2015.07.023
  [arXiv:1412.4092 [hep-ex]].

\bibitem{Khachatryan:2016xdg}
  V.~Khachatryan {\it et al.} [CMS Collaboration],
  Eur.\ Phys.\ J.\ C {\bf 76} (2016) no.7,  372
  doi:10.1140/epjc/s10052-016-4205-7
  [arXiv:1601.02001 [nucl-ex]].

\bibitem{Borghini:2000sa}
  N.~Borghini, P.~M.~Dinh and J.~Y.~Ollitrault,
  Phys.\ Rev.\ C {\bf 63} (2001) 054906
  doi:10.1103/PhysRevC.63.054906
  [nucl-th/0007063].
  
\bibitem{Borghini:2001vi}
  N.~Borghini, P.~M.~Dinh and J.~Y.~Ollitrault,
  Phys.\ Rev.\ C {\bf 64} (2001) 054901
  doi:10.1103/PhysRevC.64.054901
  [nucl-th/0105040].
  
\bibitem{Bilandzic:2010jr}
 A.~Bilandzic, R.~Snellings and S.~Voloshin,
 Phys.\ Rev.\ C {\bf 83} (2011) 044913
 doi:10.1103/PhysRevC.83.044913
 [arXiv:1010.0233 [nucl-ex]].
  
  
\bibitem{Altinoluk:2015uaa}
  T.~Altinoluk, N.~Armesto, G.~Beuf, A.~Kovner and M.~Lublinsky,
  Phys.\ Lett.\ B {\bf 751} (2015) 448
  doi:10.1016/j.physletb.2015.10.072
  [arXiv:1503.07126 [hep-ph]].

\bibitem{Altinoluk:2015eka}
  T.~Altinoluk, N.~Armesto, G.~Beuf, A.~Kovner and M.~Lublinsky,
  Phys.\ Lett.\ B {\bf 752} (2016) 113
  doi:10.1016/j.physletb.2015.11.033
  [arXiv:1509.03223 [hep-ph]].

\bibitem{Lappi:2015vta}
  T.~Lappi, B.~Schenke, S.~Schlichting and R.~Venugopalan,
  JHEP {\bf 1601} (2016) 061
  doi:10.1007/JHEP01(2016)061
  [arXiv:1509.03499 [hep-ph]].
  
\bibitem{Dumitru:2014yza}
  A.~Dumitru, L.~McLerran and V.~Skokov,
  Phys.\ Lett.\ B {\bf 743} (2015) 134
  doi:10.1016/j.physletb.2015.02.046
  [arXiv:1410.4844 [hep-ph]].
  
\bibitem{Kovner:2016jfp}
  A.~Kovner, M.~Lublinsky and V.~Skokov,
  Phys.\ Rev.\ D {\bf 96} (2017) no.1,  016010
  doi:10.1103/PhysRevD.96.016010
  [arXiv:1612.07790 [hep-ph]].
 
\bibitem{Dumitru:2010iy}
  A.~Dumitru, K.~Dusling, F.~Gelis, J.~Jalilian-Marian, T.~Lappi and R.~Venugopalan,
  Phys.\ Lett.\ B {\bf 697} (2011) 21
  doi:10.1016/j.physletb.2011.01.024
  [arXiv:1009.5295 [hep-ph]].
 
\bibitem{Levin:2011fb}
  E.~Levin and A.~H.~Rezaeian,
  Phys.\ Rev.\ D {\bf 84} (2011) 034031
  doi:10.1103/PhysRevD.84.034031
  [arXiv:1105.3275 [hep-ph]].
  
\bibitem{Kovner:2010xk}
  A.~Kovner and M.~Lublinsky,
  Phys.\ Rev.\ D {\bf 83} (2011) 034017
  doi:10.1103/PhysRevD.83.034017
  [arXiv:1012.3398 [hep-ph]].
 
\bibitem{Kovner:2011pe}
  A.~Kovner and M.~Lublinsky,
  Phys.\ Rev.\ D {\bf 84} (2011) 094011
  doi:10.1103/PhysRevD.84.094011
  [arXiv:1109.0347 [hep-ph]].
 
\bibitem{Kovchegov:2012nd}
  Y.~V.~Kovchegov and D.~E.~Wertepny,
  Nucl.\ Phys.\ A {\bf 906} (2013) 50
  doi:10.1016/j.nuclphysa.2013.03.006
  [arXiv:1212.1195 [hep-ph]].
 
\bibitem{Dumitru:2014dra}
  A.~Dumitru and A.~V.~Giannini,
  Nucl.\ Phys.\ A {\bf 933} (2015) 212
  doi:10.1016/j.nuclphysa.2014.10.037
  [arXiv:1406.5781 [hep-ph]].
  
\bibitem{Dusling:2012iga}
  K.~Dusling and R.~Venugopalan,
  Phys.\ Rev.\ Lett.\  {\bf 108} (2012) 262001
  doi:10.1103/PhysRevLett.108.262001
  [arXiv:1201.2658 [hep-ph]].

\bibitem{Gotsman:2016fee}
  E.~Gotsman, E.~Levin and U.~Maor,
  Eur.\ Phys.\ J.\ C {\bf 76} (2016) no.11,  607
  doi:10.1140/epjc/s10052-016-4434-9
  [arXiv:1607.00594 [hep-ph]].
  
\bibitem{Gotsman:2016owk}
  E.~Gotsman, E.~Levin and U.~Maor,
  Phys.\ Rev.\ D {\bf 95} (2017) no.3,  034005
  doi:10.1103/PhysRevD.95.034005
  [arXiv:1604.04461 [hep-ph]].

\bibitem{McLerran:2015sva}
  L.~McLerran and V.~Skokov,
  Nucl.\ Phys.\ A {\bf 947} (2016) 142
  doi:10.1016/j.nuclphysa.2015.12.005
  [arXiv:1510.08072 [hep-ph]].
 

\bibitem{Salam:1999ft}
  G.~P.~Salam,
  JHEP {\bf 9903} (1999) 009
  doi:10.1088/1126-6708/1999/03/009
  [hep-ph/9902324].


\bibitem{Azarkin:2014cja}
  M.~Y.~Azarkin, I.~M.~Dremin and M.~Strikman,
  Phys.\ Lett.\ B {\bf 735} (2014) 244
  doi:10.1016/j.physletb.2014.06.040
  [arXiv:1401.1973 [hep-ph]].

\bibitem{PT}
  N.~Paver and D.~Treleani,
  Nuovo Cim.\ A {\bf 70} (1982) 215.
  doi:10.1007/BF02814035

\bibitem{mufti} M.\ Mekhfi, Phys. Rev. D{\bf 32}, 2371 (1985).
\bibitem{stirling} J.R.\ Gaunt and W.J.\ Stirling,
	  JHEP {\bf 1003}, 005 (2010)   [arXiv:0910.4347 [hep-ph]]; \\
	J.R.\ Gaunt, C.H.\ Kom, A.\ Kulesza and W.J.\ Stirling,
	  Eur.\ Phys.\ J.\  C {\bf 69}, 53 (2010)  [arXiv:1003.3953 [hep-ph]].
\bibitem{BDFS1}
  B.\ Blok, Yu.\ Dokshitzer, L.\ Frankfurt and M.\ Strikman,
  Phys.\ Rev.\  D {\bf 83}, 071501 (2011)
  [arXiv:1009.2714 [hep-ph]].
	\bibitem{stirling1} J.R.\ Gaunt and W.J.\ Stirling,
	  JHEP {\bf 1106},  048 (2011) [arXiv:1103.1888 [hep-ph]].
  \bibitem{BDFS2} B.\ Blok, Yu.\ Dokshitser, L.\ Frankfurt and M.\ Strikman,
  Eur.\ Phys.\ J.\ C {\bf72}, 1963  (2012)
  [arXiv:1106.5533 [hep-ph]].
\bibitem{Diehl2} M.\ Diehl, D.\ Ostermeier and A.\ Schafer,
	  JHEP {\bf 1203} (2012) 089
	  [arXiv:1111.0910 [hep-ph]].
 \bibitem{BDFS4}
 B.~Blok, Y.~Dokshitzer, L.~Frankfurt and M.~Strikman,
  Eur.\ Phys.\ J.\ C {\bf 74} (2014) 2926
  [arXiv:1306.3763 [hep-ph]].
  %
  \bibitem{Gauntnew}
  J.~R.~Gaunt, R.~Maciula and A.~Szczurek,
  Phys.\ Rev.\ D {\bf 90} (2014) no.5,  054017
  doi:10.1103/PhysRevD.90.054017
  [arXiv:1407.5821 [hep-ph]].
%
  \bibitem{BO}K.~Golec-Biernat and E.~Lewandowska,
  Phys.\ Rev.\ D {\bf 90} (2014) no.9,  094032
  [arXiv:1407.4038 [hep-ph]].

  \bibitem{mpi2015}`Proceedings of the Seventh International Workshop on Multiple Partonic Interactions at the Large Hadron Collider,'' 23-27 Nov 2015, Trieste, Italy. (http://indico.ictp.it/event/a14280/)



  \bibitem{Frankfurt2} L.~Frankfurt and M.~Strikman,
  Phys.\ Rev.\ D {\bf 66} (2002) 031502
  [hep-ph/0205223].
    \bibitem{Frankfurt}
	  L.\ Frankfurt, M.\ Strikman and C.\ Weiss,
	  Phys.\ Rev.\  D {\bf 69}, 114010 (2004)
	  
	   
 \bibitem{Frankfurt1}
  L.~Frankfurt, M.~Strikman and C.~Weiss,
  Phys.\ Rev.\ D {\bf 83} (2011) 054012
  [arXiv:1009.2559 [hep-ph]].
  
	  
  

\bibitem{Kuechler:2016wxp}
  J.~Kuechler [ALICE and ATLAS and CMS Collaborations],
  PoS LHCP {\bf 2016} (2016) 133.
  
\bibitem{Aaboud:2016dea}
  M.~Aaboud {\it et al.} [ATLAS Collaboration],
  JHEP {\bf 1611} (2016) 110
  doi:10.1007/JHEP11(2016)110
  [arXiv:1608.01857 [hep-ex]].
  
\bibitem{Gunnellini:2016noc}
  P.~Gunnellini [CMS Collaboration],
  ``Study of high $p_T$ particle production from double parton scatterings at the CMS experiment,''
  in~\cite{mpi2015}.
  
\bibitem{Abe:1997bp}
  F.~Abe {\it et al.} [CDF Collaboration],
  Phys.\ Rev.\ Lett.\  {\bf 79} (1997) 584.
  doi:10.1103/PhysRevLett.79.584
  
 \bibitem{BS2}
  B.~Blok and M.~Strikman,
  Phys.\ Lett.\ B {\bf 772} (2017) 219
  doi:10.1016/j.physletb.2017.06.049 
  [arXiv:1611.03649 [hep-ph]].
 
  
\bibitem{Khachatryan:2015oea}
  V.~Khachatryan {\it et al.} [CMS Collaboration],
  Phys.\ Rev.\ C {\bf 92} (2015) no.3,  034911
  doi:10.1103/PhysRevC.92.034911
  [arXiv:1503.01692 [nucl-ex]].

 \bibitem{Shen:2015qta}   C.~Shen, Z.~Qiu and U.~Heinz,   
  Phys.\ Rev.\ C {\bf 92} (2015) no.1,  014901
  doi:10.1103/PhysRevC.92.014901
  [arXiv:1502.04636 [nucl-th]].

\bibitem{Gardim:2012im}
  F.~G.~Gardim, F.~Grassi, M.~Luzum and J.~Y.~Ollitrault,
  Phys.\ Rev.\ C {\bf 87} (2013) no.3,  031901
  doi:10.1103/PhysRevC.87.031901
  [arXiv:1211.0989 [nucl-th]].
  
  
\bibitem{Yan:2015jma}
  L.~Yan and J.~Y.~Ollitrault,
  Phys.\ Lett.\ B {\bf 744} (2015) 82
  doi:10.1016/j.physletb.2015.03.040
  [arXiv:1502.02502 [nucl-th]].
 
\bibitem{Teaney:2012ke}
  D.~Teaney and L.~Yan,
  Phys.\ Rev.\ C {\bf 86} (2012) 044908
  doi:10.1103/PhysRevC.86.044908
  [arXiv:1206.1905 [nucl-th]].
  
\bibitem{Floerchinger:2013tya}
  S.~Floerchinger, U.~A.~Wiedemann, A.~Beraudo, L.~Del Zanna, G.~Inghirami and V.~Rolando,
  Phys.\ Lett.\ B {\bf 735} (2014) 305
  doi:10.1016/j.physletb.2014.06.049
  [arXiv:1312.5482 [hep-ph]].
  
\bibitem{Borghini:2005kd}
  N.~Borghini and J.~Y.~Ollitrault,
  Phys.\ Lett.\ B {\bf 642} (2006) 227
  doi:10.1016/j.physletb.2006.09.062
  [nucl-th/0506045].
 
\bibitem{Gyulassy:2014cfa}
  M.~Gyulassy, P.~Levai, I.~Vitev and T.~S.~Biro,
  Phys.\ Rev.\ D {\bf 90} (2014) no.5,  054025
  doi:10.1103/PhysRevD.90.054025
  [arXiv:1405.7825 [hep-ph]].
 
\bibitem{Bierlich:2014xba}
  C.~Bierlich, G.~Gustafson, L.~Lönnblad and A.~Tarasov,
  JHEP {\bf 1503} (2015) 148
  doi:10.1007/JHEP03(2015)148
  [arXiv:1412.6259 [hep-ph]].
  
\bibitem{Bierlich:2016vgw}
  C.~Bierlich, G.~Gustafson and L.~Lönnblad,
  arXiv:1612.05132 [hep-ph].
  
\bibitem{Bierlich:2017vhg}
  C.~Bierlich, G.~Gustafson and L.~Lönnblad,
  arXiv:1710.09725 [hep-ph].
  
\bibitem{Braun:2012kn}
  M.~A.~Braun, C.~Pajares and V.~V.~Vechernin,
  Nucl.\ Phys.\ A {\bf 906} (2013) 14
  doi:10.1016/j.nuclphysa.2013.02.200
  [arXiv:1204.5829 [hep-ph]].
  
\bibitem{Abramovsky:1980yc}
  V.~A.~Abramovsky and O.~V.~Kancheli,
  Pisma Zh.\ Eksp.\ Teor.\ Fiz.\  {\bf 31} (1980) 566.
  
\bibitem{Abramovsky:1988zh}
  V.~A.~Abramovsky, E.~V.~Gedalin, E.~G.~Gurvich and O.~V.~Kancheli,
  JETP Lett.\  {\bf 47} (1988) 337
   [Pisma Zh.\ Eksp.\ Teor.\ Fiz.\  {\bf 47} (1988) 281].
  
\end{thebibliography}
\end{document}